\documentclass[prb,reprint,superscriptaddress]{revtex4-1}
\usepackage{graphicx}
\usepackage{amsmath}
\usepackage{amsfonts}
\usepackage{amssymb}
\usepackage{tabularx}
\usepackage{bbm}
\usepackage{color}
\usepackage[hidelinks=true]{hyperref}

	\newcommand{\ud}{\mathrm{d}}
	\newcommand{\sgn}[1]{\mathrm{sgn}(#1)}
	\renewcommand{\exp}[1]{\mathrm{e}^{#1}}
	
	\newcommand{\real}[1]{\mathrm{Re}\,#1}
	\newcommand{\imag}[1]{\mathrm{Im}\,#1}
	
	\newcommand{\Ams}{\mathrm{\AA}}
	\newcommand{\comm}[2]{\left[#1,#2 \right]}
	
	\newcommand{\kk}{\mathbf{k}}
	\newcommand{\qq}{\mathbf{q}}
	\newcommand{\QQ}{\mathbf{Q}}
	\newcommand{\GG}{\mathbf{G}}
	\newcommand{\KK}{\mathbf{K}}
	\newcommand{\bb}{\mathbf{b}}
	\newcommand{\RR}{\mathbf{R}}

	\newcommand{\rr}{\mathbf{r}}
	\newcommand{\xxi}{\boldsymbol{\xi}}

	\newcommand{\abs}[1]{\left| #1 \right|}
	\newcommand{\ket}[1]{\big| #1 \big>}

	\newcommand{\braoket}[3]{\left<  #1 \left| #2 \right| #3 \right>}
	\newcommand{\braoketf}[3]{\big<  #1 \big| #2 \big| #3 \big>}
	
\begin{document}
\title{Interlayer hybridization and moir\'e superlattice minibands for electrons and excitons in heterobilayers of transition-metal dichalcogenides}
\date{\today}

\author{David A.\ Ruiz-Tijerina}
\affiliation{National Graphene Institute, University of Manchester. Booth St.\ E.\, Manchester M13 9PL, United Kingdom}
\affiliation{Centro de Nanociencias y Nanotecnolog\'ia, Universidad Nacional Aut\'onoma de M\'exico. Apdo.\ postal 14, 22800, Ensenada, Baja California, M\'exico}

\author{Vladimir I.\ Fal'ko}
\affiliation{National Graphene Institute, University of Manchester. Booth St.\ E.\, Manchester M13 9PL, United Kingdom}
\affiliation{Henry Royce Institute for Advanced Materials, University of Manchester, Manchester M13 9PL, United Kingdom}

\begin{abstract}
Geometrical moir\'e patterns, generic for almost aligned bilayers of two-dimensional (2D) crystals with similar lattice structure but slightly different lattice constants, lead to zone folding and miniband formation for electronic states. Here, we show that moir\'e superlattice (mSL) effects in MoSe${}_2$/WS${}_2$ and MoTe${}_2$/MoSe${}_2$ heterobilayers that feature alignment of the band edges are enhanced by resonant interlayer hybridization, and anticipate similar features in twisted homobilayers of TMDs, including examples of narrow minibands close to the actual band edges. Such hybridization determines the optical activity of interlayer excitons in transition-metal dichalcogenide (TMD) heterostructures, as well as energy shifts in the exciton spectrum. We show that the resonantly hybridized exciton (hX) energy should display a sharp modulation as a function of the interlayer twist angle, accompanied by additional spectral features caused by umklapp electron-photon interactions with the mSL. We analyze the appearance of resonantly enhanced mSL features in absorption and emission of light by the interlayer exciton hybridization with both intralayer A and B excitons in MoSe${}_2$/WS${}_2$, MoTe${}_2$/MoSe${}_2$, MoSe${}_2$/MoS${}_2$, WS${}_2$/MoS${}_2$, and WSe${}_2$/MoSe${}_2$. 
\end{abstract}

\maketitle

\section{Introduction}\label{sec:introduction}
\begin{figure}[h!]
\begin{center}
\includegraphics[width=0.65\columnwidth]{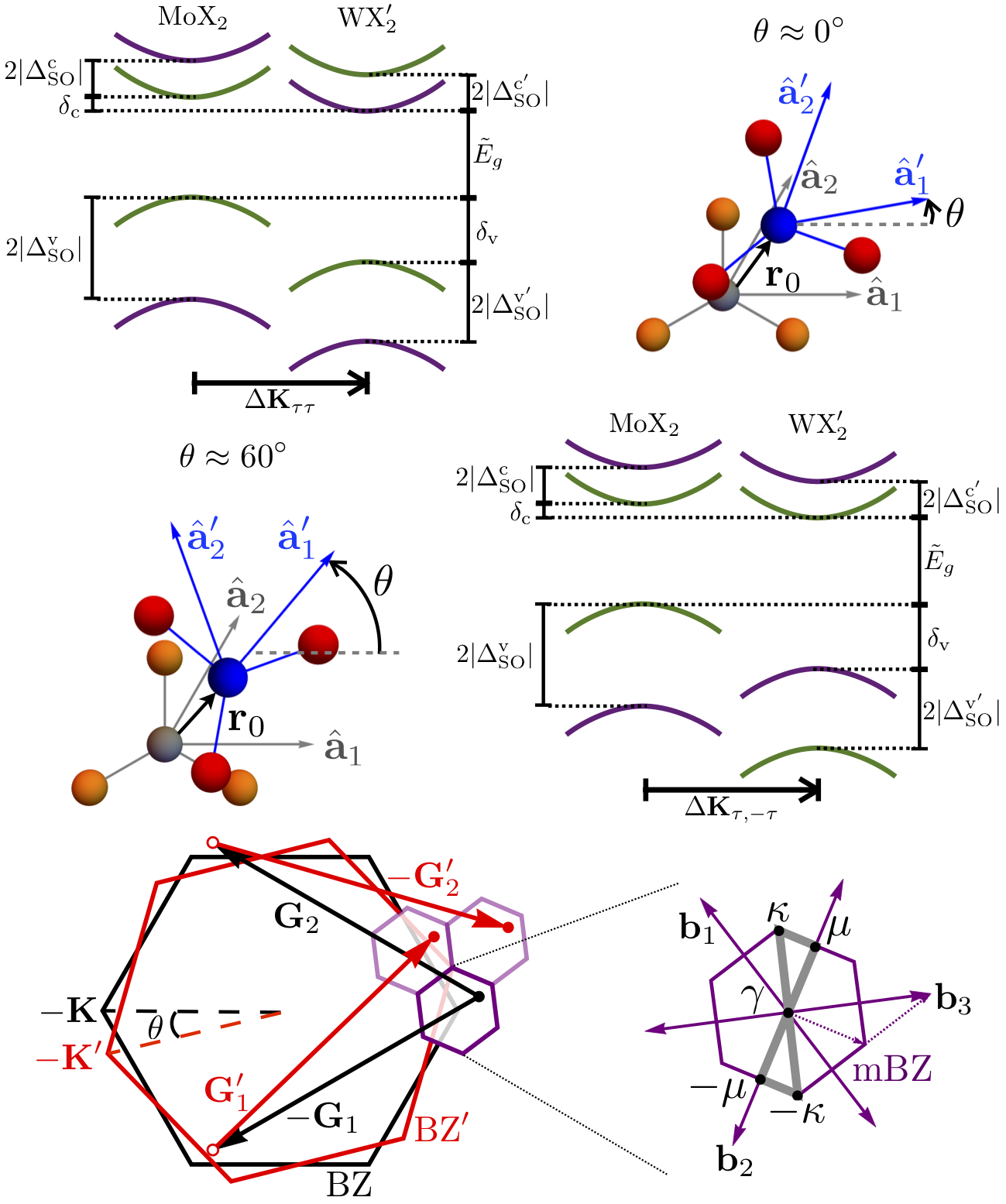}
\caption{Twisted bilayer of TMDs ${\rm MoX}_2$ and ${\rm WX_2'}$. Top-left: Band alignment, where almost resonant conduction-band states of same spin and valley quantum numbers are identified by the same color. Top-right: Atomic arrangement and bond orientation for almost parallel (P, $\theta \approx 0^\circ$) orientation of the two crystals. Basis Bravais vectors for each layer are indicated as $\hat{\mathbf{a}}_n$ and $\hat{\mathbf{a}}_n'$. Transition-metal atoms are shown in gray and blue, and chalcogens in orange and red. Center: Band alignment and atomic arrangement for the almost anti-parallel (AP, $\theta \approx 60^\circ$) heterobilayer. Bottom: Hexagonal Brillouin zones of the two crystals (${\rm BZ}$ and ${\rm BZ'}$), where the band edges appear at the Brillouin zone corners $\mathbf{K}$ and $\mathbf{K}'$. Interlayer hybridization produces electron Bragg scattering from the moir\'e superlattice, illustrated in the left sketch. This leads to band folding, and the formation of the superlattice mini Brillouin zone (mBZ) on the right.}
\label{fig:Figure1}
\end{center}
\end{figure}

Van der Waals (vdW) heterostructures consist of layers of atomically-thin two-dimensional (2D) crystals, vertically stacked and held together by vdW forces\cite{GeimNature2013,vdW_review}. The weak vdW interlayer bonding lifts the usual lattice-matching restrictions, allowing the formation of stable, high-quality heterostructures of incommensurate 2D crystals, both aligned and with an arbitrary mutual orientation. This has been demonstrated by recent experiments with graphene on boron nitride\cite{ponomarenko_GhBN_2013,dean_GhBN_2013,hunt_GhBN_2013}, where moir\'e superlattice minibands have been observed in scanning tunneling microscopy\cite{Li2010,Yankowitz2012}, magnetotransport\cite{roshan_2018}, capacitance\cite{capacitance_2014} and infrared spectroscopy\cite{IR_2015} measurements. Of particular interest for optoelectronics are vdW heterostructures of various transition-metal dichalcogenides (TMDs)\cite{ting_acsnano_2014,rigosi_nanolett_2015,hill_nanolett_2016,nayak_acsnano_2017,alexeev_nanolett_2017,Zhange1601459}, due to the gapped nature of these semiconducting 2D materials, which have a direct band gap in the monolayer form\cite{mak_MoS2_2010,splendiani_MoS2_2010}, strong coupling to light\cite{review_2018}, and valley-dependent optical selection rules\cite{wangyao_prb_2008,zeng_natnano_2012,mak_natnano_2012}. When combined into bilayers, the pair of 2D crystals acquires the band alignment shown in Fig.\ \ref{fig:Figure1}. The nearly identical lattice constants of TMDs with hexagonal lattices leads to the appearance of moir\'e patterns\cite{kuwabara_1990,Zhange1601459}, which have long periods in the case of almost aligned heterostructures. The resulting moir\'e superlattice (mSL) can generate flat minibands with high densities of states, potentially interesting from the point of view of strongly correlated states in TMD heterobilayers\cite{macdonald_hubbard}, analogous to recent observations in twisted bilayer graphene\cite{Cao2018}. It also has potential to modify the excitonic spectrum and change selection rules for optical transitions, due to electron-photon umklapp processes involving mSL reciprocal lattice vectors.

In this paper we study the interplay between relative interlayer orientation and band alignment in TMD heterobilayers and twisted homobilayers; in particular, in the regime of resonant interlayer hybridization. Based on the TMD work function data and band alignments available in the literature\cite{band_alignment,first_principles_2018,Kozawa2015}, we choose to focus this study  on mSLs in heterobilayers formed by TMD layers with nearly degenerate carrier bands: MoSe${}_2$/WS${}_2$ and MoTe${}_2$/MoSe${}_2$, which feature almost exact band alignment in undoped structures, and also on twisted homobilayers of TMDs, such as MoSe${}_2$. We study the dependence of hybridization and moir\'e effects on the misalignment angle $\theta$ of the 2D crystals in such heterostructures,  and find that, while superlattice effects are weak for arbitrary angles, they become dominant for close interlayer alignment near $\theta = 0$ and $\theta = 60^\circ$ (Fig.\ \ref{fig:Figure1}), producing narrow minibands near the actual band edges. We argue that, analogously to the case of twisted bilayer graphene\cite{macdonald_pnas}, these systems are highly non-perturbative, and their description must explicitly consider hybridization effects, rendering recent theoretical approaches based on harmonic moir\'e potentials\cite{macdonald_intra, hongyi_moire, Wu2017, macdonald_hubbard}---while applicable to heterobilayers with non-resonant band edges---unsuitable to describe this class of TMD heterostructures.

\begin{figure}[t!]
\begin{center}
\includegraphics[width=\columnwidth]{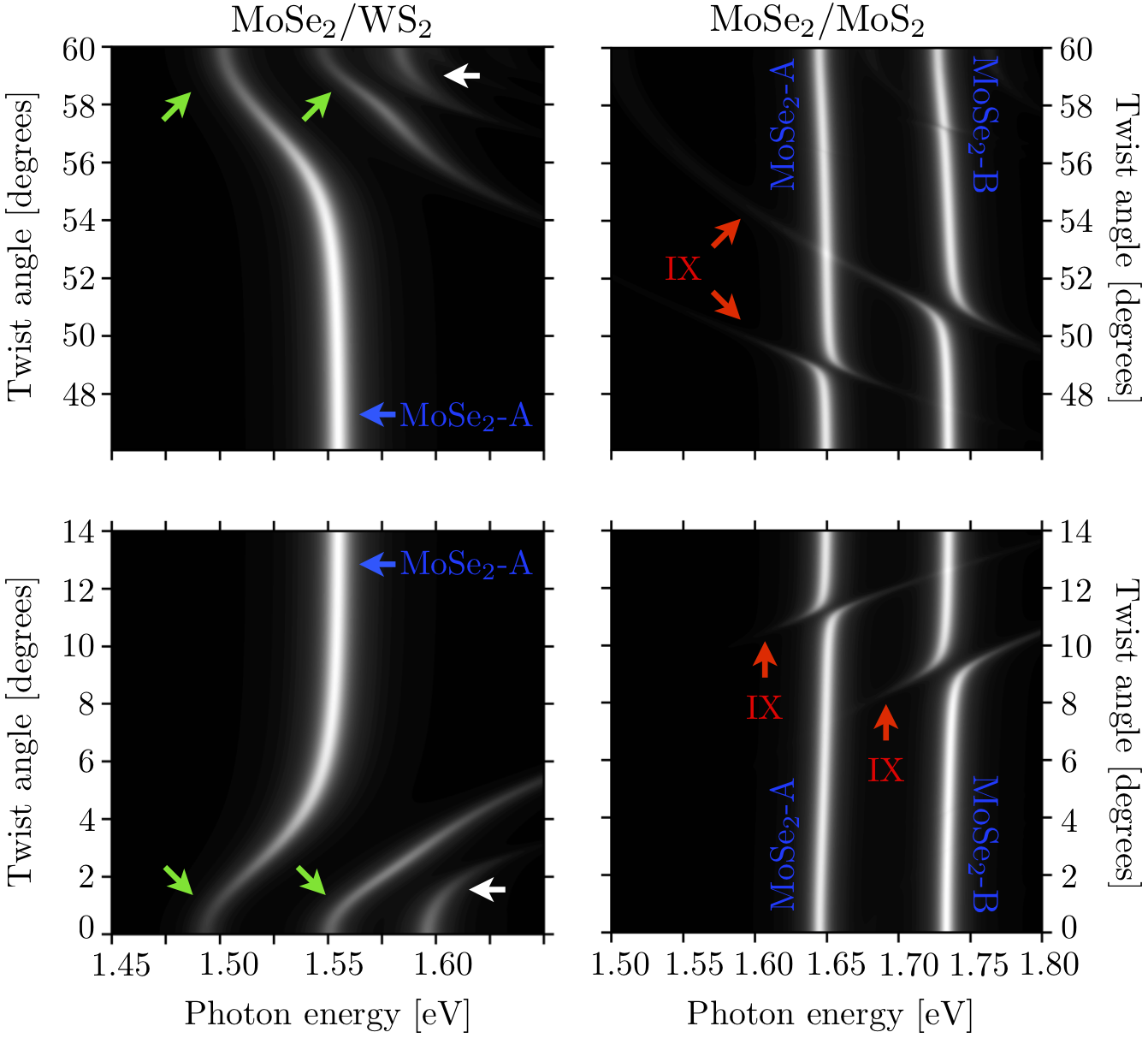}
\caption{Comparison between the low-energy absorption spectra (in arbitrary units) of TMD heterobilayers with resonant (MoSe${}_2$/WS${}_2$, left) and non-resonant (MoSe${}_2$/MoS${}_2$, right) conduction-band edges, as function of the interlayer twist angle. In MoSe${}_2$/WS${}_2$, hybridized excitons form near perfect alignment ($\theta = 0^\circ$) and anti-alignment ($\theta=60^\circ$), leading to the avoided crossings marked with green arrows. The white arrows point to absorption lines enabled by moir\'e umklapp processes, corresponding to higher exciton momentum states that become visible as they are folded onto zero momentum  by the moir\'e superlattice. For MoSe${}_2$/MoS${}_2$, the MoSe${}_2$ A and B excitons lie hundreds of ${\rm meV}$ above the lowest momentum-bright IX states, such that their hybridization is negligible, making the IX dark in our approximation. For large twist angles, the zero-momentum IX energies are raised toward the intralayer exciton energies, becoming semi-bright and eventually producing hX states.}
\label{fig:Summary}
\end{center}
\end{figure}

Also, we study the interplay between resonant hybridization of intra- and interlayer excitons and moir\'e superlattices in MoSe${}_2$/WS${}_2$ and MoTe${}_2$/MoSe${}_2$ heterobilayers, leading to the formation of hybridized excitons (hX) containing strongly mixed electron or hole states involved in the formation of intralayer (X) and interlayer (IX) excitons. Our estimates for the X and IX binding energies indicate that the weaker binding of the latter, due to the additional electron and hole out-of-plane separation, can significantly enhance the resonant condition between the two exciton species. We show that the optical spectra of hXs are dominated by their bright intralayer exciton component, resulting in identical selection rules as intralayer excitons in monolayer TMDs, in stark contrast to earlier predictions for IXs in non-resonant TMD heterostructures \cite{wang_yao_tvvtcc,hongyi_moire}. We present an analysis of the optical spectra of both resonant and non-resonant heterobilayers, compared in Fig.\ \ref{fig:Summary}. In the former case, we find that the energy and state composition of optically-active hybridized excitons varies sharply with interlayer orientation, producing a strong modulation of the corresponding absorption signatures with twist angle, marked with green arrows in Fig.\ \ref{fig:Summary}. For closely aligned resonant heterobilayers, the optical spectrum also displays a bright absorption line at higher energies enabled by moir\'e umklapp processes (white arrows in Fig.\ \ref{fig:Summary}), which fold finite-momentum exciton states onto zero momentum, allowing them to acquire a finite oscillator strength and providing direct experimental evidence for mSL minibands for the excitons in the system. These signatures are absent in closely aligned non-resonant heterobilayers, where the low-energy optical features correspond to IXs (red arrows in Fig.\ \ref{fig:Summary}) that only mix weakly with the bright intralayer exciton states. 
Finally, we show that hXs are sensitive to out-of-plane electric fields, due to their large IX components, and argue that vertical electrical bias can be used to tune the strength of mSL effects on excitons in TMD heterostructures.

For this purpose, in Sec.\ \ref{sec:model}, we introduce a general interlayer hybridization model for twisted TMD heterobilayers, parameterized using currently available \emph{ab initio} parameters of monolayer TMDs (Table \ref{tab:parameters}). In Sec.\ \ref{sec:harmonic}, we derive an effective low-energy Hamiltonian that incorporates moir\'e superlattice effects in terms of harmonic potentials specific to each of the monolayer bands\cite{macdonald_intra, hongyi_moire, Wu2017, macdonald_hubbard}, which we find are applicable to TMD heterostructures with large band-edge offsets, such as MoSe${}_2$/MoS${}_2$ bilayers. We discuss the shortcomings of this harmonic potential approach, and show that it breaks down in the case of resonant hybridization. In Sec.\ \ref{sec:homobilayer}, we study the opposite limit of perfect interlayer band-edge degeneracy in TMD homobilayers, and present results for their band structure features, such as the nature of the conduction-band edges and the appearance of van Hove singularities in the conduction bands. In Sec.\ \ref{sec:resonant}, we discuss two cases of TMD heterobilayers with nearly resonant band edges, namely, MoTe${}_2$/MoSe${}_2$ and MoSe${}_2$/WS${}_2$, and show that they constitute an intermediate case between typical TMD hetero- and homobilayers, which are exactly in the resonant hybridization regime. In Sec.\ \ref{sec:optical}, we study the effects of strong interlayer hybridization on the band structures of excitons in such heterostructures based on reported experimental values for the intralayer and interlayer exciton energies, and present theoretical predictions for the full optical spectra MoTe${}_2$/MoSe${}_2$, MoSe${}_2$/WS${}_2$, MoSe${}_2$/MoS${}_2$ and WSe${}_2$/MoS${}_2$ as functions of the interlayer alignment angle $\theta$, and electric field strength.

\section{Model}\label{sec:model}
\begin{figure*}[t!]
\begin{center}
\includegraphics[width=1.8\columnwidth]{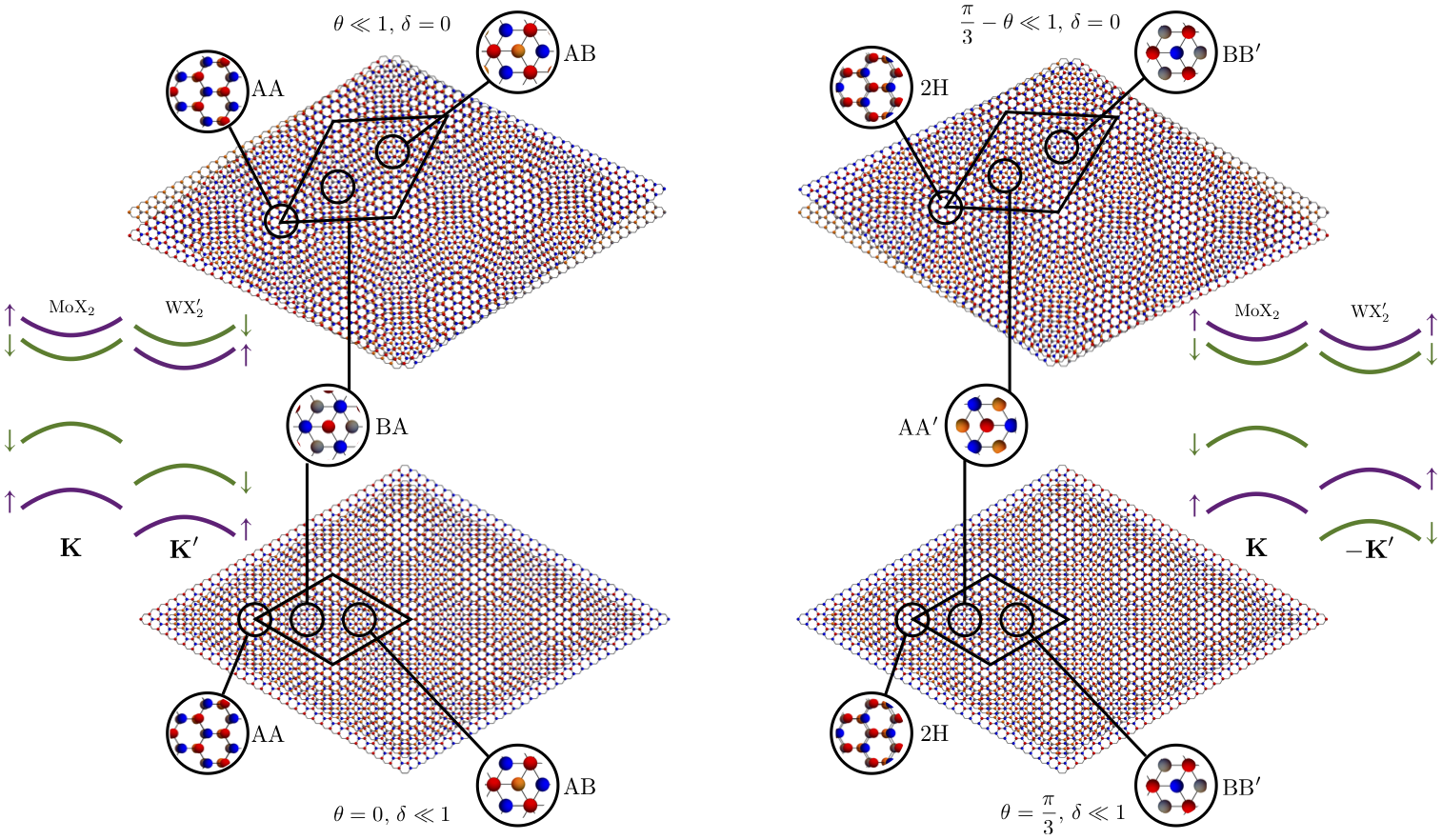}
\caption{Moir\'e patterns formed by parallel (P) and anti-parallel (AP) aligned TMD bilayers. Top- and bottom-layer metal atoms are shown in blue and gray, with corresponding chalcogens shown in red and orange. For each stacking type, the interlayer conduction- and valence-band alignment and spin ordering near the MX${}_2$ layer $\tau=+1$ valley is shown for ${\rm M}={\rm Mo}$ and ${\rm M'}={\rm W}$. The moir\'e unit cell is indicated by the black rhombus, and the encircled areas correspond to regions of the heterobilayer with local registry between the two lattices. For parallel alignment, these local registries correspond to AA, BA and AB stacking, analogous to the case of bilayer graphene. By contrast, for anti-parallel alignment we find regions with local 2H, AA' and BB' stacking, as shown in the insets.}
\label{fig:moire_stacking}
\end{center}
\end{figure*}

We describe electronic states in a TMD heterobilayer in terms of the monolayer conduction- and valence-band $\kk\cdot\mathbf{p}$ theory near the band edges of its two constituent layers. The band edges of the bottom ${\rm MX_2}$ layer, to which the highest valence band belongs, are located at the $\tau \KK$ valleys of its Brillouin zone, ${\rm BZ}$ ($\tau=\pm 1$). We set $\KK=(4\pi/a_{\rm MX_2})\hat{\mathbf{x}}$, according to the lattice vectors $\mathbf{a}_1=a_{\rm MX_2}[\hat{\mathbf{x}}/2 + \sqrt{3}\hat{\mathbf{y}}/2]$ and $\mathbf{a}_1=a_{\rm MX_2}[\hat{\mathbf{x}}/2 - \sqrt{3}\hat{\mathbf{y}}/2]$, where $a_{\rm MX_2}$ is the corresponding lattice constant. Similarly, for the top layer, ${\rm M'X_2'}$, containing the lowest conduction band, the band edges appear at the valleys $\tau' \KK'$ of the Brillouin zone ${\rm BZ}'$. Because of the lattice mismatch,
\begin{equation}
	\delta = 1 - a_{\rm M'X_2'}/a_{\rm MX_2},
\end{equation}
and the relative twist angle $\theta$ between the two crystals (Fig.\ \ref{fig:Figure1}, bottom), the valley momenta are related by $\KK'=(1+\delta)^{-1}R_{\theta}\KK$, where $R_{\theta}$ represents counterclockwise rotation by an angle $\theta$ about the $z$-axis. Henceforth, ${\rm M'X'_2}$-layer variables are identified  with a prime, and heterobilayers are labeled as ${\rm MX}_2/{\rm M'X_2'}$. We will discuss two inequivalent stacking types: parallel (P) stacking, for twist angles $|\theta| < 30^\circ$; and anti-parallel (AP) stacking, for $|\theta - 60^\circ|<30^\circ$. The two configurations are shown in Figs.\ \ref{fig:Figure1} and \ref{fig:moire_stacking}, and any twist angle outside the range $0^\circ \le \theta \le 60^\circ$ is related to one of these two stacking types by $120^\circ$ rotations or mirror reflection.\footnote{An alternative nomenclature is used, \emph{e.g.}, in Refs.\ \onlinecite{wang_yao_tvvtcc,wangyao_coupling}, where P and AP stacking configurations are referred to as R and H stacking, respectively. We choose the former convention to avoid confusion with standard nomenclature for commensurate stacking.}.

Locally, the exact heterostructure stacking is determined by $\theta$, $\delta$, and a unit-cell vector $\rr_0$, representing the shortest in-plane shift between transition metal atoms of the two layers, as illustrated in Fig.\ \ref{fig:Figure1}. For small twist angle and/or lattice mismatch, a superlattice structure emerges, known as a moir\'e pattern\cite{kuwabara_1990,Zhange1601459}, where the stacking determined by $\rr_0$, set to relate positions of ${\rm M}$ and ${\rm M}'$ atoms, is approximately preserved locally at the origin, and periodically along the heterostructure's surface, as shown in Fig.\ \ref{fig:moire_stacking}. Moreover, the interlayer registry varies inside the superlattice unit cell, producing two additional regions of approximately commensurate stacking, corresponding to local values of $\rr_0$ different than that at the origin. As an example, Fig.\ \ref{fig:moire_stacking} shows the case of $\rr_0=\boldsymbol{0}$, where the sequence of locally commensurate regions in the moir\'e unit cell is AA, BA and AB for P stacking; and 2H, AA' and BB' for AP stacking\cite{tmd_stacking_2013,tmd_stacking_2014}, contrasting the two stacking types. The local $\rr_0$ values for these commensurate stacking types are shown in Table \ref{tab:r0}.

The heterobilayer Hamiltonian has the general form
\begin{equation}\label{eq:fullH}
	H = H_0 + H_{\rm t},
\end{equation}
where
\begin{equation}\label{eq:H0}
\begin{split}
	H_0& = \sum_{s,\tau}\sum_{\alpha={\rm c,v,c',v'}}\sum_{\kk}E_{\tau s}^\alpha(k)c_{\alpha\tau s}^\dagger(\kk)c_{\alpha \tau s}(\kk).
\end{split}
\end{equation}
Here, the operators $c_{{\rm c}\tau s}(\kk)$ and $c_{{\rm v}\tau s}(\kk)$ [$c_{{\rm c}'\tau' s}(\kk)$ and $c_{{\rm v}'\tau' s}(\kk)$], annihilate electrons of spin quantum number $s=\uparrow,\,\downarrow$ and wave vector $\tau\KK+\kk$ ($\tau'\KK'+\kk$) in the conduction and valence bands of the ${\rm MX}_2$ (${\rm M'X'}_2$) layer. Setting the energy reference at the highest valence-band edge, the conduction and valence band dispersions can be approximated as
\begin{equation}\label{eq:dispersions}
\begin{split}
	&E_{\tau s}^{\rm v'}(k) =  -\delta_{\rm v}-[s\tau + \sgn{\Delta_{\rm SO}^{\rm v'}}]\Delta_{\rm SO}^{\rm v'}-\frac{\hbar^2k^2}{2m_{\rm v'}},\\
	&E_{\tau s}^{\rm v}(k) =  -[s\tau + \sgn{\Delta_{\rm SO}^{\rm v}}] \Delta_{\rm SO}^{\rm v}-\frac{\hbar^2k^2}{2m_{\rm v}},\\
	&E_{\tau s}^{\rm c'}(k) = \tilde{E}_g+[s\tau + \sgn{\Delta_{\rm SO}^{\rm c'}}]\Delta_{\rm SO}^{\rm c'} + \frac{\hbar^2k^2}{2m_{\rm c'}},\\
	&E_{\tau s}^{\rm c}(k) = \tilde{E}_g+\delta_{\rm c}+[s\tau + \sgn{\Delta_{\rm SO}^{\rm c}}]\Delta_{\rm SO}^{\rm c} + \frac{\hbar^2k^2}{2m_{\rm c}},
\end{split}
\end{equation}
where $\tilde{E}_g$ is the heterostructure band gap; $\delta_{\rm c}$ and $\delta_{\rm v}$ are the interlayer conduction and valence band edge detunings; $\Delta_{\rm SO}^{\alpha}$ is the spin-orbit splitting of band $\alpha={\rm v,\,c,\,v',\,c'}$; and $m_\alpha$ are effective masses. These model parameters, illustrated in Fig.\ \ref{fig:Figure1} and presented in Table \ref{tab:parameters}, are based on DFT\cite{kdotp,  komsa_2015, first_principles_2018} and $GW$\cite{band_alignment, Cheiwchanchamnangij2012} calculations recently reported in the literature.

\begin{table}[b!]
\caption{Interlayer in-plane translation vector $\rr_0$ corresponding to different commensurate stackings, for twist angles $\theta = 0^\circ$ (P) and $\theta = 60^\circ$ (AP), and perfect lattice matching, $\delta = 0$. We use lattice vectors $\mathbf{a}_1=a_0[\tfrac{\hat{\mathbf{x}}}{2} + \frac{\sqrt{3}}{2}\hat{\mathbf{y}}]$ and $\mathbf{a}_1=a_0[\tfrac{\hat{\mathbf{x}}}{2} - \frac{\sqrt{3}}{2}\hat{\mathbf{y}}]$, where $a_0$ is the lattice constant.}
\begin{center}
\begin{tabular}{c| c c}
\hline\hline
$\rr_0$ & P-stacking & AP-stacking\\
\hline\hline
 $\mathbf{0}$ & AA & ${\rm AA'}$\\
 $\frac{\mathbf{a}_1-\mathbf{a}_2}{3}=\frac{a_0}{\sqrt{3}}\hat{\mathbf{y}}$ & AB & ${\rm BB'}$\\
 $\frac{\mathbf{a}_2-\mathbf{a}_1}{3}=-\frac{a_0}{\sqrt{3}}\hat{\mathbf{y}}$ & BA & ${\rm 2H}$\\ 
\hline\hline
\end{tabular}
\end{center}
\label{tab:r0}
\end{table}%

The matrix elements for interlayer tunneling between the conduction- ($\alpha = {\rm c}$) or valence-bands ($\alpha = {\rm v}$) have the general form
\begin{equation}\label{eq:Ht_melem}
\begin{split}
&\braoket{\alpha',\tau'\KK'+\kk'}{H_{\rm t}}{\alpha, \tau\KK+\kk} =\\ 
&\frac{1}{\sqrt{NN'}}\sum_{\RR,\RR'}\exp{i(\tau\KK+\kk)\cdot \RR'}\exp{-i(\tau'\KK'+\kk')\cdot \RR} \braoket{\varphi_{\alpha',\RR'}}{H_{\rm t}}{\varphi_{\alpha,\RR}},
\end{split}
\end{equation}
where $H_{\rm t}$ is the tunneling Hamiltonian, $N$ and $N'$ are the numbers of unit cells in the MX${}_2$ and M'X'${}_2$ layers, and $\varphi_{\alpha,\rr}$ is a Wannier function centered at atomic site $\rr$. In the two-center approximation, the atomic matrix element $\braoket{\varphi_{\alpha',\RR'}}{H_{\rm t}}{\varphi_{\alpha,\RR}}$ can be Fourier transformed as
\begin{equation}
	\braoket{\varphi_{\alpha',\RR'}}{H_{\rm t}}{\varphi_{\alpha,\RR}} = \frac{1}{\sqrt{NN'}}\sum_{\qq}\exp{i\qq\cdot(\RR'-\RR)}t_{\alpha'\alpha}(\qq),
\end{equation}
which after substitution into \eqref{eq:Ht_melem} gives $H_{\rm t}$ in the form
\begin{equation}\label{eq:Ht}
\begin{split}
	&H_{\rm t}=\sum_{s,\tau,\tau'}\sum_{\kk \in {\rm BZ}}\sum_{\kk' \in {\rm BZ}'}\left[T^{\rm c}_{\tau'\tau}(\kk',\kk)c_{{\rm c}'\tau's}^\dagger(\kk')c_{{\rm c}\tau s}(\kk)\right. \\
	&\qquad\qquad\qquad+ \left.T^{\rm v}_{\tau'\tau}(\kk',\kk)c_{{\rm v}'\tau's}^\dagger(\kk')c_{{\rm v}\tau s}(\kk) \right],
\end{split}
\end{equation}
with interlayer hopping terms\cite{wangyao_coupling}
\begin{equation}\label{eq:hopping}
\begin{split}
	T_{\tau'\tau}^\alpha(\kk',\kk) = &\sum_{\GG,\GG'}\delta_{\kk-\kk',\GG'-\GG + (\tau'\KK'-\tau\KK)}\\
	&\times t_{\alpha}(\kk + \tau\KK + \GG)\,\exp{- i \GG\cdot \rr_0}.
\end{split}
\end{equation}
Here, $\GG$ and $\GG'$ are the reciprocal lattice vectors of the ${\rm MX_2}$ and ${\rm M'X_2'}$ layers, respectively, and $\tau'\KK'-\tau\KK$ gives the interlayer valley mismatch. As described in Ref.\ \onlinecite{wangyao_coupling}, the interlayer tunneling functions $t_{\rm c}(\qq)$ and $t_{\rm v}(\qq)$ are constrained, respectively, by the angular momentum quantum numbers $\ell_z$ of the conduction and valence bands at the $\tau$ and $\tau'$ valleys. Because the conduction-band states near the $K$ valleys are formed by in-plane-isotropic $\varphi_{\rm c} = d_{z^2}$ orbitals\cite{kdotp}, for both valleys we have $\ell_z=0$. By contrast, the valence-band states at the $\tau$ valley consist of  $\varphi_{\rm v}=(d_{x^2-y^2}+i\tau d_{xy})/\sqrt{2}$ orbitals, with $\ell_z=-\tau$. Therefore, under $C_3$ rotations we obtain
\begin{equation}\label{eq:t_sym}
	t_{\rm c}(C_3\qq) = t_{\rm c}(\qq),\qquad t_{\rm v}(C_3 \qq) = \exp{i\frac{2\pi}{3}(\tau' - \tau)}t_{\rm v}(\qq).
\end{equation}
To a good approximation\cite{wangyao_coupling}, we can set 
\begin{equation}\label{eq:qapprox}
	|t_\alpha(\qq)|=\left\{\begin{array}{ccc}
	|t_\alpha| & \text{for} & q = \KK,\,C_3\KK,\, C_3^2\KK \\
	0 & \text{for} & q > K
	\end{array}\right.,
\end{equation}
where $C_3^n$ represents rotation by $\tfrac{2n\pi}{3}$. Thus, we may define
\begin{equation}
\begin{split}
	&t_{\rm c}(\KK) = t_{\rm c}(C_3\KK) = t_{\rm c}(C_3^2\KK) = t_{\rm c},\\
	&t_{\rm v}(\KK) =  t_{\rm v},\quad t_{\rm v}(C_3\KK) = \exp{i\tfrac{2\pi}{3}(\tau'-\tau)}t_{\rm v},\\ 
	&t_{\rm v}(C_3^2\KK) = \exp{i\tfrac{4\pi}{3}(\tau'-\tau)}t_{\rm v}.
\end{split}
\end{equation}

The approximation \eqref{eq:qapprox} truncates the sum in Eq.\ \eqref{eq:hopping} to include only $\GG=\mathbf{0}$, and the two Bragg vectors
\begin{equation}
\begin{split}
	\GG_2=&(C_3-\mathbbm{1})\KK,\\
	-\GG_1=&(C_3^2-\mathbbm{1})\KK,
\end{split}
\end{equation}
which connect the three equivalent $K$ valleys. These reciprocal lattice vectors are shown with black arrows in the bottom panel of Fig.\ \ref{fig:Figure1}. At this point, the stacking type (P or AP) must be specified to determine which ${\rm M'X_2'}$-layer Bragg vectors give the dominant interlayer hopping terms. For closely aligned P-stacked structures, the Kronecker delta in Eq.\ \eqref{eq:hopping} couples states near the band edges only if $\tau'=\tau$, and for ${\rm M'X_2'}$-layer Bragg vectors  (red arrows in Fig.\ \ref{fig:Figure1}, bottom)
\begin{equation}\label{eq:Gp_P}
\begin{split}
	\GG_2'=&(C_3-\mathbbm{1})\KK',\\
	-\GG_1'=&(C_3^2-\mathbbm{1})\KK'.
\end{split}
\end{equation}
One can verify that, for these specific Bragg vectors, the generalized umklapp condition in Eq.\ \eqref{eq:hopping} becomes $\kk-\kk' = \tau C_3^\eta\Delta\KK$, where $\eta\in\{0,\,1,\,2\}$, and we have defined
\begin{equation}\label{eq:DKP}
	\Delta\KK \equiv \KK'-\KK  \quad (\text{P-stacking}).
\end{equation}
Alternatively, for AP stacking, we must set $\tau'=-\tau$, indicating that tunneling takes place between opposite $K$ valleys of the electron and hole layers, which are closely aligned in reciprocal space for this range of twist angles. The relevant ${\rm M'X_2'}$-layer Bragg vectors in this case are
\begin{equation}\label{eq:Gp_AP}
\begin{split}
	\GG_3'=&(C_3^2-C_3)\KK',\\ 
	-\GG_2'=&(\mathbbm{1}-C_3)\KK',
\end{split}
\end{equation}
leading to $\kk-\kk'=\tau C_3^{\eta}\Delta\KK$, with
\begin{equation}\label{eq:DKAP}
	\Delta\KK \equiv -\KK'-\KK  \quad (\text{AP-stacking}).
\end{equation}
This leads to the simplified hopping terms
\begin{equation}\label{eq:T_simpl}
\begin{split}
	T_{\tau'\tau}^{\rm c}(\kk',\kk) \approx& \sum_{\eta=0}^2\delta_{\kk-\kk',C_3^\eta\Delta\KK}\,t_{\rm c}\,\exp{ i\KK\cdot\rr_0}\exp{-i C_3^\eta\KK\cdot \rr_0},\\
	T_{\tau'\tau}^{\rm v}(\kk',\kk) \approx& \sum_{\eta=0}^2\delta_{\kk-\kk',C_3^\eta\Delta\KK}\,t_{\rm v}\exp{i\tfrac{2\eta\pi}{3}(\tau'-\tau)}\,\exp{ i\KK\cdot\rr_0}\exp{-i C_3^\eta\KK\cdot \rr_0}.
\end{split}
\end{equation}
Equation \eqref{eq:T_simpl} allows us to determine how the different locally commensurate regions shown in Fig.\ \ref{fig:moire_stacking} contribute to interlayer carrier tunneling. Exactly at the valley ($\kk=\kk'=0$), and neglecting the lattice mismatch within each region ($\Delta \KK=0$), we write
\begin{equation}\label{eq:local_t}
\begin{split}
	T_{\tau'\tau}^{\rm c}(0,0) \approx& t_{\rm c}\exp{ i\KK\cdot\rr_0}\left[ \exp{-i \KK\cdot \rr_0} + \exp{-i C_3\KK\cdot \rr_0} + \exp{-i C_3^2\KK\cdot \rr_0} \right],\\
	T_{\tau'\tau}^{\rm v}(0,0) \approx& t_{\rm v}\exp{ i\KK\cdot\rr_0}\left[ \exp{-i \KK\cdot \rr_0} + \exp{-i [C_3\KK\cdot \rr_0 - \tfrac{2\pi}{3}(\tau'-\tau)]} \right. \\
	 & \left. \qquad\qquad\quad +\, \exp{-i [C_3^2\KK\cdot \rr_0-\tfrac{4\pi}{3}(\tau'-\tau)]} \right].
\end{split}
\end{equation}
Table \ref{tab:local_t} summarizes the results of Eq.\ \eqref{eq:local_t} for the various locally commensurate regions of the moir\'e superlattice, obtained by substituting the appropriate $\rr_0$ values of Table \ref{tab:r0}. For P-stacked (AP-stacked) TMD heterostructures, conduction-band tunneling takes place in AA (AA') regions, whereas valence-band tunneling occurs in AA (2H) regions\cite{wang_yao_tvvtcc, kormanyos_kdotp_2018}. This result was first presented in Ref.\ \onlinecite{wang_yao_tvvtcc}, where it was also reported that the parameter $|t_{\rm v}|$ is somewhat larger for AP stacking, and $|t_{\rm c}| \lesssim |t_{\rm v}|$ for both stacking types, based on DFT calculations. In addition, it was shown that matrix elements $t_{\rm cv}$ and $t_{\rm vc}$ exist, representing electron hopping between the conduction and valence bands of different layers, which are significantly smaller than $t_{\rm c}$. As the latter couple states separated by energies comparable to the heterostructure's band gap, we neglect them in the following. Below, we assume this hierarchy for the interlayer hopping elements, setting $t_{\rm c}=26\,{\rm meV}$, based on recent experiments on MoSe${}_2$/WS${}_2$ heterobilayers\cite{twist_angle2018}, and $t_{\rm v}=2t_{\rm c}$. We use these values for all materials discussed, for the purpose of obtaining a general qualitative description of TMD heterobilayers, keeping in mind that these matrix elements are material-dependent.

\begin{table}[h!]
\caption{Interlayer conduction- and valence-band tunneling in the locally commensurate regions of moir\'e superlattices in P- and AP-stacked TMD heterobilayers.}
\begin{center}
\begin{tabular}{c|cc}
\hline
\hline
\textbf{P-stacking} & $|T_{+,+}^{\rm c}|$ & $|T_{+,+}^{\rm v}|$\\
\hline
\hline
AA                         & $3|t_{\rm c}|$  &  $3|t_{\rm v}|$ \\
AB                         & $0$                         &  $0$\\
BA                         & $0$                         &  $0$\\
\hline\hline
\end{tabular}
\begin{tabular}{c|cc}
\hline
\hline
\textbf{AP-stacking} & $|T_{-,+}^{\rm c}|$ & $|T_{-,+}^{\rm v}|$\\
\hline
\hline
AA'                         & $3|t_{\rm c}|$  &  $0$ \\
BB'                         & $0$                         &  $0$\\
2H                         & $0$                         &  $3|t_{\rm v}|$\\
\hline\hline
\end{tabular}
\end{center}
\label{tab:local_t}
\end{table}%

With Eq.\ \eqref{eq:T_simpl}, $H_{\rm t}$ periodically mixes electronic states of the two layers, whose wave vectors are separated by $\bb_{\pm n} \equiv \pm(C_3^{n-1}-C_3^{n-2})\Delta\KK$, where $n$ runs cyclically through $\{1,\,2,\, 3\}$ (Fig.\ \ref{fig:Figure1}, bottom). Note that $\bb_1$ and $\bb_2$ can be interpreted as the primitive vectors of the reciprocal lattice dual to the real-space moir\'e pattern shown in Fig.\ \ref{fig:moire_stacking}, and define the mini Brillouin zone (mBZ) presented in Fig.\ \ref{fig:Figure1}. Defining the reciprocal vectors $\bb_{mn} = m\bb_1+n\bb_2$, with $m$ and $n$ integers, the electron- and hole-layer dispersions can be folded into the mBZ to form a series of minibands with operators ($\alpha={\rm c,\,v,\,c',\,v'}$)
\begin{equation}\label{eq:foldings}
	c_{\alpha \tau s }^{mn}(\qq) \equiv c_{\alpha \tau s}(\qq + \bb_{mn})\quad;\quad \qq \in {\rm mBZ},
\end{equation}
which couple according to Eq.\ \eqref{eq:T_simpl} to produce what we henceforth call a moir\'e band structure. Then, the $n$th conduction and valence moir\'e bands have operators given by the linear combinations of the folded band operators,
\begin{equation}\label{eq:moire_band_operators}
\begin{split}
	C_{{\rm c}\tau s}^{n}(\qq) \equiv&\, \sum_{i,j}\mathcal{A}_{ij}^{n\tau s}(\qq)\,c_{{\rm c}\tau s}^{ij}(\qq) + \sum_{i,j}\mathcal{A}_{ij}^{n\tau s}{}'(\qq)\,c_{{\rm c'}\tau' s}^{ij}(\qq),\\
	C_{{\rm v}\tau s}^{n}(\qq) \equiv&\, \sum_{i,j}\mathcal{B}_{ij}^{n\tau s}(\qq)\,c_{{\rm v}\tau s}^{ij}(\qq) + \sum_{i,j}\mathcal{B}_{ij}^{n\tau s}{}'(\qq)\,c_{{\rm v'}\tau' s}^{ij}(\qq),
\end{split}
\end{equation}
where $\tau'=\pm \tau$ for P and AP stacking, respectively. In addition to the valley mismatch $\Delta\KK$, the spin-dependent amplitudes $\mathcal{A}_{ij}^{n\tau s}{}^{(')}$ and $\mathcal{B}_{ij}^{n\tau s}{}^{(')}$ depend on the spin ordering of the monolayer bands.  Fig.\ \ref{fig:moire_stacking} shows that the detuning between the highest spin-polarized MX${}_2$ valence band and the M'X'${}_2$ valence band of the same spin increases dramatically (by hundreds of meV) from P to AP stacking, consequence of the large intralayer valence-band spin-orbit splittings (see Table \ref{tab:parameters}). This leads to strong or weak interlayer valence-band mixing for P or AP stacking, respectively, leading to qualitatively different behaviors in the two stacking limits (see for example Figs.\ \ref{fig:P_BL_MoSe2} and \ref{fig:AP_BL_MoSe2}). This is not the case for the spin-polarized conduction bands, for which the spin-orbit splittings are much weaker, of only tens of meV.

Having defined the hybridization model \eqref{eq:fullH}, in the following Sections we study two important limit cases for the interlayer band alignment. First, in Sec.\ \ref{sec:harmonic} we look at MoSe${}_2$/MoS${}_2$ as a typical example of type-II semiconducting TMD heterostructures, with large band offsets $\delta_{\rm c},\,\delta_{\rm v} \gtrsim 100\,{\rm meV}$. Then, in Sec.\ \ref{sec:homobilayer}, we study the opposite limit of $\delta_{\rm c}=\delta_{\rm v}=0$, choosing bilayer MoSe${}_2$ as a case study.

\begin{table*}[t!]
\caption{\emph{Ab initio} parameters for the three heterobilayers discussed, an for bilayer MoSe${}_2$, extracted from Refs.\ \onlinecite{band_alignment, kdotp, mostaani_excitonic_prb_2017, Cheiwchanchamnangij2012, komsa_2015, first_principles_2018,mose2_d}. The effective masses are based on $GW$ or $G_0W_0$ calculations; heterostructure band gaps $\tilde{E}_g$ and conduction and valence band edge detunings $\delta_{\rm c}$ and $\delta_{\rm v}$ are based on the $G_0W_0$ approximation; spin-orbit couplings and momentum matrix elements at the valley ($\gamma$) are obtained from DFT (HSE and LDA); and the monolayer lattice constants $a_0$ and $a_0'$ and interlayer distances $d$ are based on DFT (HSE), or experimental values for bulk crystals. }
\begin{center}
\begin{tabular}{r | r @{.} l r @{.} l r @{.} l r @{.} l r @{.} l r @{.} l r @{.} l r @{.} l r @{.} l r @{.} l }
\hline \hline
\,&\multicolumn{2}{c}{$\tilde{E}_g$ [eV]}&\multicolumn{2}{c}{$\delta_{\rm v}$ [eV]} &\multicolumn{2}{c}{$\delta_{\rm c}$ [eV]}  & \multicolumn{2}{c}{$\Delta_{\rm SO}^{\rm e}$ [meV]}  & \multicolumn{2}{c}{$\Delta_{\rm SO}^{\rm e'}$ [meV]}  &  \multicolumn{2}{c}{$m_{\rm e}/m_0$} & \multicolumn{2}{c}{$m_{\rm e'}/m_0$} & \multicolumn{2}{c}{$\gamma\,[{\rm eV}\,\Ams]$} & \multicolumn{2}{c}{$a\,[\Ams]$}  & \multicolumn{2}{c}{$a'\,[\Ams]$} \\
\,&    \multicolumn{2}{c}{\,}           &               \multicolumn{2}{c}{\,}                  & \multicolumn{2}{c}{\,}                                & \multicolumn{2}{c}{$\Delta_{\rm SO}^{\rm h}$ [meV]}  & \multicolumn{2}{c}{$\Delta_{\rm SO}^{\rm h'}$ [meV]}  &  \multicolumn{2}{c}{$m_{\rm h}/m_0$} & \multicolumn{2}{c}{$m_{\rm h'}/m_0$} & \multicolumn{2}{c}{$\gamma'\,[{\rm eV}\,\Ams]$} & \multicolumn{2}{c}{$d\,[\Ams]$}  \\
\hline \hline
BL-MoSe${}_2$  & 1&330${}^{\rm a}$ & 0&0 & 0&0   & 11&0${}^{\rm b}$      &  11&0${}^{\rm b}$       & 0&38${}^{\rm c}$  & 0&38${}^{\rm d}$  & 2&20${}^{\rm f}$ & 3&289${}^{\rm b}$ & 3&289${}^{\rm b}$  \\
\,                                      & \multicolumn{2}{c}{\,}                  & \multicolumn{2}{c}{\,}                  & \multicolumn{2}{c}{\,}                    & 93&0${}^{\rm b}$     &   93&0${}^{\rm b}$       & 0&44${}^{\rm c}$  & 0&44${}^{\rm d}$ & 2&20${}^{\rm f}$ & 6&463${}^{\rm g}$ \\ 
MoSe${}_2$/MoS${}_2$  & 0&960${}^{\rm a}$ & 0&630${}^{\rm a}$ & 0&370${}^{\rm a}$   & 11&0${}^{\rm b}$      &  1&5${}^{\rm b}$       & 0&38${}^{\rm c}$  & 0&35${}^{\rm d}$  & 2&20${}^{\rm f}$ & 3&289${}^{\rm b}$ & 3&157${}^{\rm b}$  \\
\,                                      & \multicolumn{2}{c}{\,}                  & \multicolumn{2}{c}{\,}                  & \multicolumn{2}{c}{\,}                    & 93&0${}^{\rm b}$     &   74&0${}^{\rm b}$       & 0&44${}^{\rm c}$  & 0&43${}^{\rm d}$ & 2&22${}^{\rm f}$ & 6&972${}^{\rm f}$ \\ 
MoTe${}_2$/MoSe${}_2$ & 0&860${}^{\rm a}$ & 0&470${}^{\rm a}$ & 0&070${}^{\rm a}$   & 18&0${}^{\rm b}$      &  11&0${}^{\rm b}$        & 0&69${}^{\rm e}$  & 0&38${}^{\rm c}$  & 2&16${}^{\rm f}$ & 3&516${}^{\rm b}$ & 3&289${}^{\rm b}$    \\
\,                                      & \multicolumn{2}{c}{\,}                   & \multicolumn{2}{c}{\,}                 & \multicolumn{2}{c}{\,}                    & 109&5${}^{\rm b}$ &  93&0${}^{\rm b}$        & 0&66${}^{\rm e}$  & 0&44${}^{\rm c}$ & 2&20${}^{\rm f}$ & 7&421${}^{\rm f}$ \\
MoSe${}_2$/WS${}_2$    & 1&270${}^{\rm a}$ & 0&270${}^{\rm a}$ & 0&060${}^{\rm a}$   & 11&0${}^{\rm b}$      & -16&0${}^{\rm b}$        & 0&38${}^{\rm c}$   & 0&27${}^{\rm c}$ & 2&20${}^{\rm f}$ & 3&289${}^{\rm b}$ & 3&16${}^{\rm b}$    \\
\,                                      & \multicolumn{2}{c}{\,}                   & \multicolumn{2}{c}{\,}                  & \multicolumn{2}{c}{\,}                   &  93&0${}^{\rm b}$     & 241&5${}^{\rm b}$    & 0&44${}^{\rm c}$   & 0&32${}^{\rm c}$ & 2&59${}^{\rm f}$ & 6&913${}^{\rm f}$\\
\hline \hline
\end{tabular}\\
a.\ [\onlinecite{band_alignment}];\quad b.\ [\onlinecite{kdotp}];\quad c.\ [\onlinecite{mostaani_excitonic_prb_2017}];\quad d.\ [\onlinecite{Cheiwchanchamnangij2012}]; \quad e.\ [\onlinecite{komsa_2015}];\quad f.\ [\onlinecite{first_principles_2018}];\quad g.\ [\onlinecite{mose2_d}].
\end{center}
\label{tab:parameters}
\end{table*}%

\section{Perturbation theory for non-resonant interlayer hybridization and harmonic potential approximation for moir\'e superlattices}\label{sec:harmonic}
The importance of interlayer hybridization depends crucially on the ratio between the interlayer tunneling matrix elements $t_{\alpha}$ and the band edge detunings $\delta_\alpha$. When these ratios are small, one can treat $H_{\rm t}$ perturbatively, in terms of the $\kk$-dependent energy corrections produced by the tunneling processes, which in real space form a periodic potential\cite{uchoa,wallbank}. This approach to describing the effects of a moir\'e superlattice on the electronic states has been used in Ref.\ \onlinecite{macdonald_hubbard}, and for excitons in Refs.\ \onlinecite{macdonald_intra, hongyi_moire, Wu2017}, where the potential was estimated from \emph{ab initio} calculations. In this section, we derive the tunneling contribution to this potential from the microscopic Hamiltonian \eqref{eq:fullH}, based on a perturbative treatment of the elementary excitations in the heterobilayer (conduction-band electrons and valence-band holes). For clarity, the final result is presented in terms of the conduction- and valence-band dispersions.

We apply the unitary transformation $\mathcal{U}=\exp{iS}$ to the Hamiltonian \eqref{eq:fullH}, with $S$ an anti-Hermitian operator. The resulting rotated Hamiltonian $\tilde{H} = \mathcal{U}H\mathcal{U}^\dagger$ is given to second order in $S$ as
\begin{equation}\label{eq:BHexp}
	\tilde{H}=H_0+H_{\rm t} + i\comm{S}{H_0+H_{\rm t}} - \frac{1}{2!}\comm{S}{\comm{S}{H_0+H_{\rm t}}}.
\end{equation}
We eliminate $H_{\rm t}$ to first order by choosing\cite{schrieffer_wolff} $i[S,\,H_0]=-H_{\rm t}$, and keep only terms up to second order in $S$ to get the effective model $\tilde{H}=\tilde{H}_0+H_{\rm m}$. The first term corresponds to Eq.\ \eqref{eq:H0}, with the renormalized dispersions
\begin{equation}\label{eq:Erenorm}
\begin{split}
	\tilde{E}_{\tau' s}^{\rm v'}(\kk)  =& E_{\tau' s}^{\rm v'}(k) - \sum_{\eta=0}^2\frac{\abs{t_{\rm v}}^2}{E_{\tau s}^{\rm v}(\kk-C_3^\eta\Delta\KK)-E_{\tau' s}^{\rm v'}(k)},\\
	\tilde{E}_{\tau s}^{\rm v}(\kk)  =& E_{\tau s}^{\rm v}(k) + \sum_{\eta=0}^2\frac{\abs{t_{\rm v}}^2}{E_{\tau s}^{\rm v}(k) - E_{\tau' s}^{\rm v'}(\kk+C_3^\eta\Delta\KK)},\\
	\tilde{E}_{\tau' s}^{\rm c'}(\kk)  =& E_{\tau' s}^{\rm c'}(k) - \sum_{\eta=0}^2\frac{\abs{t_{\rm c}}^2}{E_{\tau s}^{\rm c}(\kk+C_3^\eta\Delta\KK) -E_{\tau' s}^{\rm c'}(k)},\\
	\tilde{E}_{\tau s}^{\rm c}(\kk)  =& E_{\tau s}^{\rm c}(k) + \sum_{\eta=0}^2\frac{\abs{t_{\rm c}}^2}{E_{\tau s}^{\rm c}(k) -E_{\tau' s}^{\rm c'}(\kk-C_3^\eta\Delta\KK)},
\end{split}
\end{equation}
whereas the second term gives ($n=\pm 1,\,\pm2,\,\pm3$)
\begin{widetext}
\begin{equation}\label{eq:harmonic}
{\small
\begin{split}
	H_{\rm m} = &\frac{1}{2}\sum_{s,\tau,n}\sum_{\kk}\Bigg[ \frac{\abs{t_{\rm c}}^2\exp{i\GG_n\cdot\rr_0}c_{{\rm c}\tau s}^\dagger(\kk+\bb_n)c_{{\rm c}\tau s}(\kk)}{E_{\tau s}^{\rm c}(k) -E_{\tau' s}^{\rm c'}(\kk-C_3^{\sgn{n}(n+1)}\Delta\KK)}+ \frac{\abs{t_{\rm v}}^2\exp{i\GG_n\cdot\rr_0}\exp{i\tfrac{2\pi}{3}\sgn{n}\sgn{|n|-2}}c_{{\rm v}\tau s}^\dagger(\kk+\bb_n)c_{{\rm v}\tau s}(\kk)}{E_{\tau's}^{\rm v'}(\kk-C_3^{\sgn{n}(n+1)}\Delta\KK) -E_{\tau s}^{\rm v}(\kk)} \\
&-\frac{\abs{t_{\rm c}}^2\exp{i\GG_n\cdot\rr_0}c_{{\rm c}'\tau's}^\dagger(\kk+\bb_n)c_{{\rm c}'\tau's}(\kk)}{E_{\tau s}^{\rm c}(\kk+C_3^{\sgn{n}(n-1)}\Delta\KK)-E_{\tau's}^{\rm c'}(k)} - \frac{\abs{t_{\rm v}}^2\exp{i\GG_n\cdot\rr_0}\exp{i\tfrac{2\pi}{3}\sgn{n}\sgn{|n|-2}}c_{{\rm v}'\tau's}^\dagger(\kk)c_{{\rm v'}\tau' s}(\kk+\bb_n)}{E_{\tau' s}^{\rm v'}(k)-E_{\tau s}^{\rm v}(\kk+C_3^{\sgn{n}(n-1)}\Delta\KK)}  \Bigg]+{\rm H.c.}
\end{split}
}
\end{equation}
\end{widetext}

\begin{figure}[t!]
\begin{center}
\includegraphics[width=\columnwidth]{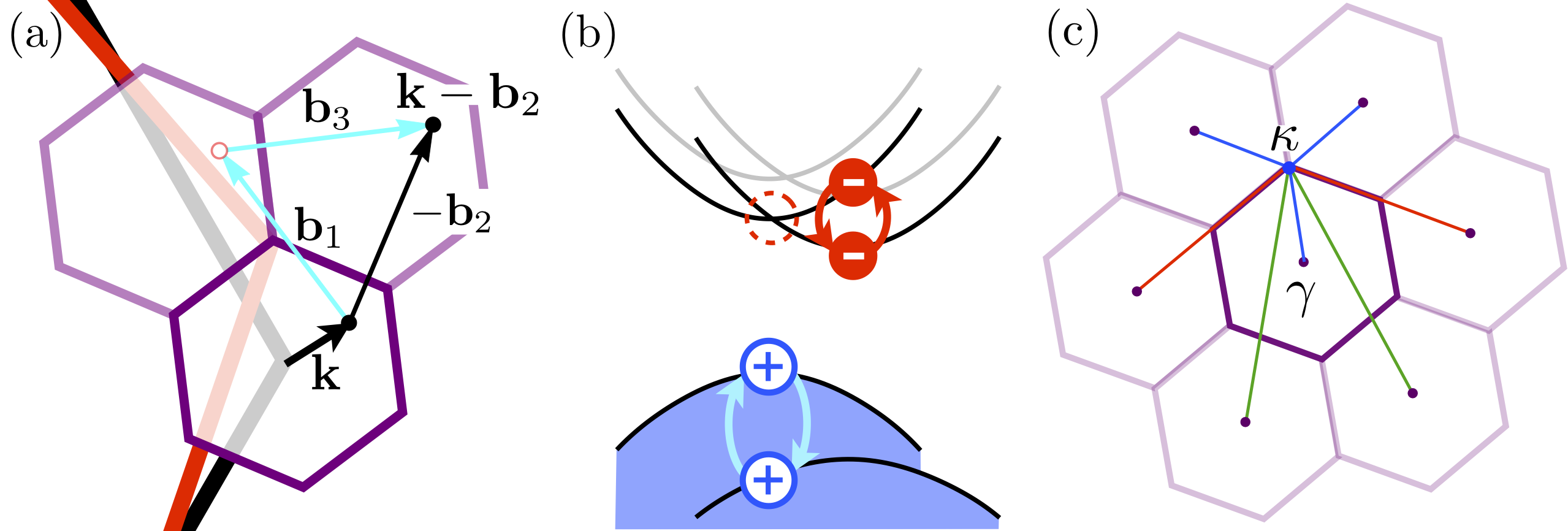}
\caption{(a) Virtual processes giving rise to the intralayer scattering terms \eqref{eq:harmonic}. A state in band ${\rm c}$ with wave vector $\kk$ (black dot) tunnels to the band ${\rm c}'$ state $\kk'=-\Delta\KK+\kk+\bb_1$ (faint red circle), subsequently hopping back to band ${\rm c}$, with final wave vector $\kk-\bb_2$. (b) Energy diagram of the virtual processes, for $\theta \approx 0^\circ$ or $\theta \approx 60^\circ$. The effective model \eqref{eq:harmonic_potential} breaks down at the crossings between bands ${\rm c}$ and ${\rm c}'$ (red dashed circle). A small detuning $\delta_{\rm c}$ allows a crossing at or near the band edges, such that the model cannot describe low-energy electrons. Because of larger values of $\delta_{\rm v}$, the model correctly describes low-energy holes. See also Figs.\ \ref{fig:R_MoTe2_MoSe2}--\ref{fig:H_MoSe2_WS2}. (c) Extended mBZ scheme, representing the top valence minibands that couple through \eqref{eq:harmonic} at the $\kappa$ point. Same color lines represent degenerate levels at the $\kappa$ point, with the three blue ones being closest to the band edge, and mixing through the term \eqref{eq:crossing}.}
\label{fig:mBZpert}
\end{center}
\end{figure}

$H_{\rm m}$ represents scattering of electrons and holes by moir\'e vectors $\bb_{n}$, produced by two sequential interlayer tunneling processes. Fig.\ \ref{fig:mBZpert}(a) shows an electron near the $\tau=1$ valley of band ${\rm c}$ tunnel into band ${\rm c}'$ through one of the processes depicted in the botom panel of Fig.\ \ref{fig:Figure1}, followed by a second tunneling process back into band ${\rm c}$. The net result is a scattering process of the initial state by a moir\'e Bragg vector $\bb_n$. An inverse Fourier transform of Eq.\ \eqref{eq:harmonic}, taking $k \rightarrow 0$ in the dispersions, gives simple real-space harmonic potentials for each of the bands, of the form ($\alpha = {\rm c},\,{\rm c'},\,{\rm v},\,{\rm v'}$)
\begin{equation}\label{eq:harmonic_potential}
	V_{\alpha}(\rr) = \sum_{n=1}^3\left( V_{\alpha}^n\exp{i\bb_n\cdot\rr} + V_{\alpha}^n{}^*\exp{-i\bb_n\cdot\rr} \right).
\end{equation}
This is the same type of harmonic potential, as used in Refs.\ \onlinecite{macdonald_intra, hongyi_moire, Wu2017, macdonald_hubbard} for both carriers and excitons in TMD heterobilayers. Whereas in those cases the coefficients $V_\alpha^n$ were determined by fitting to the spatial variation of the heterostructure band gap, as determined by DFT calculations, in our analysis they are determined from a microscopic model. We point out, however, that our approach is based purely on interlayer tunneling, and neglects lattice relaxation in the regions of commensurate stacking.

Whether Eqs.\ \eqref{eq:harmonic_potential} constitute a valid low-energy theory for carriers near the band edges in a heterostructure with twist angle $\theta$ depends on the band alignment. To illustrate this, we take the first term of Eq.\ \eqref{eq:harmonic} near the $\tau=1$ valley ($k \rightarrow 0$), and note the divergence when $\delta_{\rm c}+s(\tau\Delta_{\rm SO}^{\rm e}-\tau'\Delta_{\rm SO}^{\rm e'})+(|\Delta_{\rm SO}^{\rm c}|-|\Delta_{\rm SO}^{\rm c'}|)=\tfrac{\hbar^2\Delta K_{\tau\tau'}^2}{2m_{\rm c'}}$. As shown in Fig.\ \ref{fig:mBZpert}(b), this is due to a crossing of the two conduction bands, which can occur at or near the bottom of the higher-energy band for some values of $\Delta\KK(\theta)$. The resulting strong interlayer mixing of electronic states near the higher band edge leads to the breakdown of perturbation theory. Turning to band ${\rm c}'$, the third term in Eq.\ \eqref{eq:harmonic} does not show a divergence, reflecting the fact that a higher parabolic band can never cross the bottom of a lower one. This, however, does not guarantee the validity of the harmonic-potential approximation. To make this statement precise, we define perturbative parameters
\begin{equation}\label{eq:perturbative_parameter}
\begin{split}
	\mathcal{P}_{{\rm c}s}^\tau=&\abs{\frac{t_{\rm c}}{E_{\tau s}^{\rm c}(0)-E_{\tau' s}^{\rm c'}(\Delta K)}},\\
	\mathcal{P}_{{\rm c'}s}^{\tau'}=&\abs{\frac{t_{\rm c}}{E_{\tau s}^{\rm c}(\Delta K)-E_{\tau' s}^{\rm c'}(0)}},\\
	\mathcal{P}_{{\rm v}s}^{\tau}=&\abs{\frac{t_{\rm v}}{E_{\tau' s}^{\rm v'}(\Delta K)-E_{\tau s}^{\rm v}(0)}},\\
	\mathcal{P}_{{\rm v'}s}^{\tau'}=&\abs{\frac{t_{\rm v}}{E_{\tau' s}^{\rm v'}(0)-E_{\tau s}^{\rm v}(\Delta K)}},
\end{split}
\end{equation}
for each band, where $\tau$ and $\tau'$ are determined by the twist angle, as discussed in Sec.\ \ref{sec:model}: $\tau'=\tau$ for P stacking, and $\tau'=-\tau$ for AP stacking. The effective potential \eqref{eq:harmonic_potential} correctly describes low-energy carriers in valley $\tau$ of band $\alpha$ when $\mathcal{P}_\alpha^\tau \ll 1$, and interlayer band mixing is weak. This condition may not be met if the interlayer detunings $\delta_{\rm c}$ or $\delta_{\rm v}$ are small, as illustrated in Fig.\ \ref{fig:mBZpert}(b).
\begin{figure}[t!]
\begin{center}
\includegraphics[width=0.98\columnwidth]{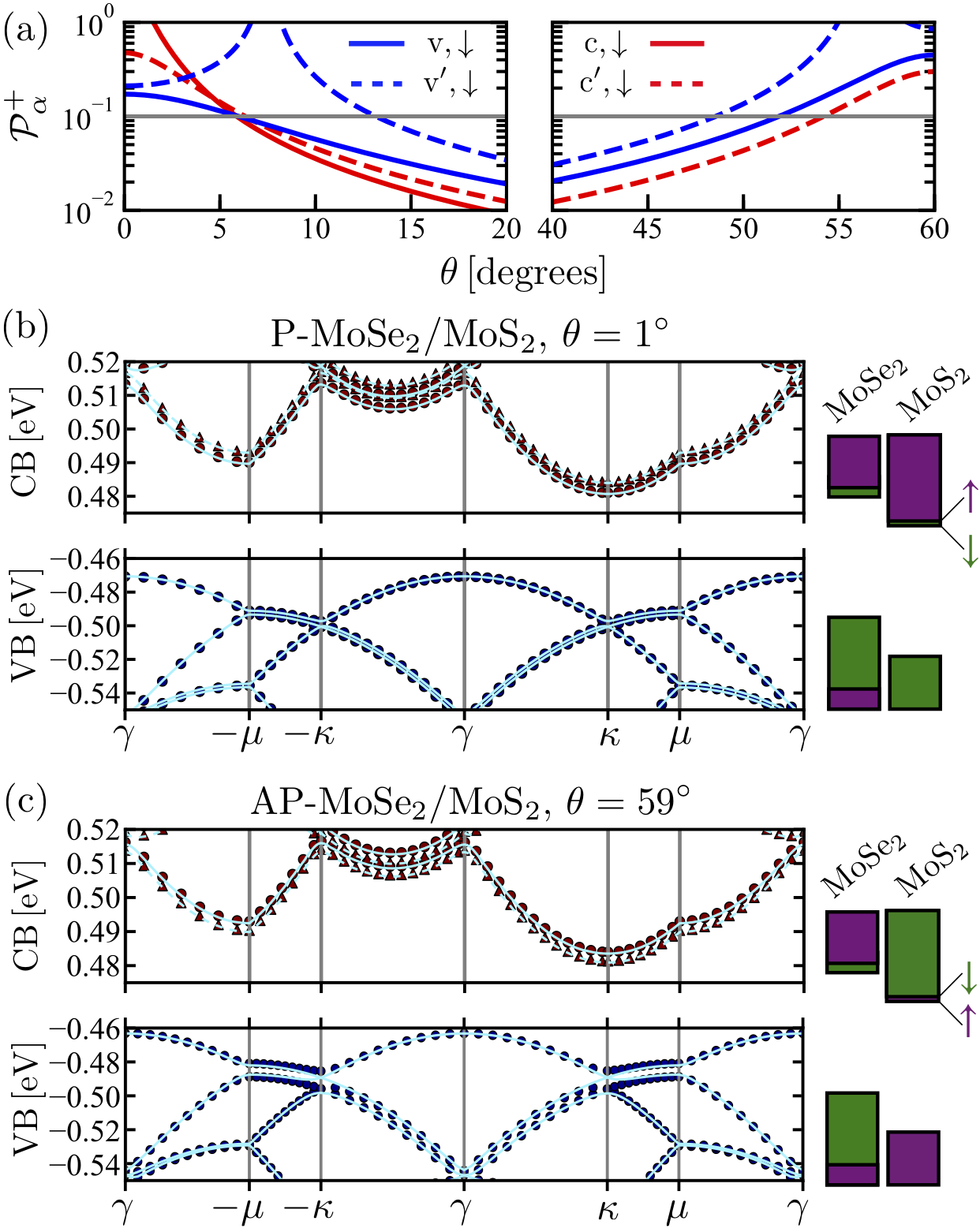}
\caption{(a) Perturbative parameters for all four $\tau=1$ monolayer band edges of MoSe${}_2$/MoS${}_2$, as functions of twist angle. The harmonic-potential \eqref{eq:harmonic_potential} is valid below the gray line, representing $\mathcal{P}_{\alpha  s}^+=0.1$. (b) and (c) Moir\'e miniband structure of P-stacked MoSe${}_2$/MoS${}_2$ at twist angle $\theta = 1^\circ$ and AP-stacked MoSe${}_2$/MoS${}_2$ at $\theta = 59^\circ$, respectively, obtained from direct diagonalization of both the full hybridization Hamiltonian and the effective harmonic-potential model. For the full Hamiltonian, spin-up (down) bands are shown with triangles (circles). Spin-up and -down minibands of the harmonic-potential model are shown with dashed and solid cyan lines, respectively. In each case, the band diagrams on the right indicate the spin ordering of the hybridizing bands. We use \cite{twist_angle2018} $t_{\rm c}=t_{\rm v}/2=26\,{\rm meV}$; all other parameters are listed in Table \ref{tab:parameters}.}
\label{fig:BS_MoSe2_MoS2}
\end{center}
\end{figure}

Using MoSe${}_2$/MoS${}_2$ as an example, we take \emph{ab initio}\cite{band_alignment,first_principles_2018} results for the monolayer band parameters (Table \ref{tab:parameters}). The perturbative parameters plotted in Fig.\ \ref{fig:BS_MoSe2_MoS2}(a) as functions of the twist angle suggest that the harmonic-potential picture holds for angles $|\theta| < 5^\circ$ and $|\theta - 60^\circ| < 5^\circ$. Although the small-twist-angle approximations leading to Eqs.\ \eqref{eq:Gp_P} and \eqref{eq:Gp_AP} are not applicable for $\theta \sim 30^\circ$, Fig.\ \ref{fig:BS_MoSe2_MoS2}(a) shows that all perturbative parameters are negligible for large twist angles, and interlayer tunneling effects can be neglected, as expected for strongly misaligned heterobilayers. Thus, our model can be applied safely for $0^\circ \le \theta \le 60^\circ$. We numerically diagonalized both the full hybridization Hamiltonian \eqref{eq:fullH}, and the harmonic-potential effective model \eqref{eq:harmonic_potential}, using a large basis of moir\'e bands \cite{wallbank_annalen_2015}. Dispersions with valley quantum number $\tau=1$ near the main conduction and valence band edges are shown in Figs.\ \ref{fig:BS_MoSe2_MoS2}(b) and \ref{fig:BS_MoSe2_MoS2}(c), for P- and AP-stacked configurations, respectively, along the mBZ path defined in Fig.\ \ref{fig:Figure1}. Their $\tau=-1$ counterparts can be obtained by time-reversal symmetry, and are not explicitly shown. The figures show quantitative agreement between the full Hamiltonian and the harmonic approximation near the band edges for MoSe${}_2$/MoS${}_2$. A shortcoming of the model \eqref{eq:perturbative_parameter} is visible in the valence bands, however, where the avoided crossings at $\pm \kappa$ are not captured by the harmonic approximation, and instead a Dirac cone appears. This crossing is not accidental, but exact; it appears for all material pairs (see Figs.\ \ref{fig:R_MoTe2_MoSe2}-\ref{fig:H_MoSe2_WS2}), and can be understood as follows: the three lowest minibands, with indices $(00)$, $(01)$ and $(0,-1)$, become degenerate at the $\pm \kappa$-points, as sketched in Fig.\ \ref{fig:mBZpert}(c). Evaluating the corresponding coefficients, given by the second term of Eq.\ \eqref{eq:harmonic}, we find that the three minibands couple through the $C_3$-symmetric Hamiltonian
\begin{equation}\label{eq:crossing}
	h=\begin{pmatrix}
	\varepsilon & t\,\exp{-i\GG_1\cdot \rr_0} & t\,\exp{i\GG_2\cdot \rr_0}\\
	t\,\exp{i\GG_1\cdot \rr_0} & \varepsilon & t\,\exp{-i\GG_3\cdot \rr_0}\\
	t\,\exp{-i\GG_2\cdot \rr_0} & t\,\exp{i\GG_3\cdot \rr_0} & \varepsilon
	\end{pmatrix},
\end{equation}
where
\begin{equation}
	\varepsilon=\tilde{E}_{+\downarrow}^{\rm v}(\Delta K),\quad t=\frac{\abs{t_{\rm v}}^2}{E_{+\downarrow}^{\rm v'}(\Delta K) -E_{+ \downarrow}^{\rm v}(\Delta K)}.
\end{equation}
The resulting eigenvalues are $\varepsilon+2t$, and a doubly degenerate level $\varepsilon-t$, responsible for the spurious level crossing. By comparison, the harmonic potentials proposed in Refs.\ \onlinecite{macdonald_intra, hongyi_moire, Wu2017, macdonald_hubbard} give the simpler but less symmetric form
\begin{equation*}
	h=\begin{pmatrix}
	\varepsilon & V & V^*\\
	V^* & \varepsilon & V\\
	V & V^* & \varepsilon
	\end{pmatrix}
\end{equation*} 
with eigenvalues $\varepsilon + 2\real{V}$ and $\varepsilon -\real{V} \pm\sqrt{3}|\imag{V}|$, which allow a gap opening at $\kappa$.

\section{Resonant interlayer hybridization in twisted TMD homobilayers}\label{sec:homobilayer}
\begin{figure}[t!]
\begin{center}
\includegraphics[width=\columnwidth]{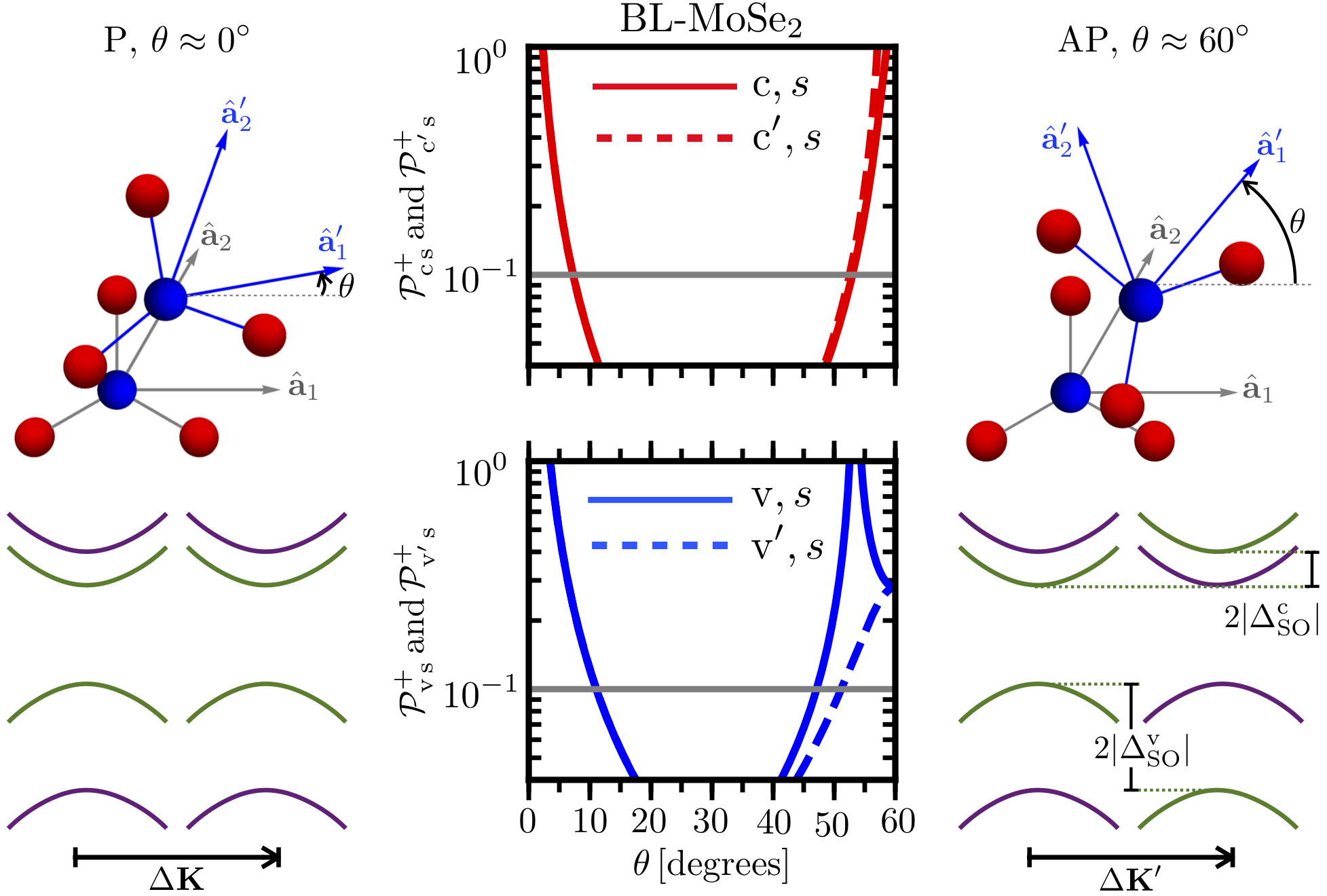}
\caption{Perturbative parameters for the conduction- and valence-band edges of twisted bilayer MoSe${}_2$, as functions of the twist angle. For all bands, the perturbative approach described in Sec.\ \ref{sec:harmonic} is valid only for strong misalignment angles $10^\circ \lesssim \theta \lesssim 50^\circ$, and breaks down near close alignment and anti-alignment. Left and right panels show schematics of the homobilayer's atomic arrangement and band alignment for P and AP stacking, respectively.}
\label{fig:PP_BL_MoSe2}
\end{center}
\end{figure}

Whereas the large band-edge offsets $\delta_{\rm v}$ and $\delta_{\rm c}$ guarantee the validity of the harmonic-potential model for most TMD heterobilayers, the opposite limit of $\delta_{\rm v}=\delta_{\rm c}=0$ can be found in TMD (homo)bilayers. Fig.\ \ref{fig:PP_BL_MoSe2} shows that the perturbative parameters for twisted bilayer MoSe${}_2$ are small only for strongly misaligned configurations ($10^\circ < \theta < 50^\circ$), whereas for close alignment or anti-alignment, interlayer hybridization cannot be treated as a perturbation.

\begin{figure}[t!]
\begin{center}
\includegraphics[width=\columnwidth]{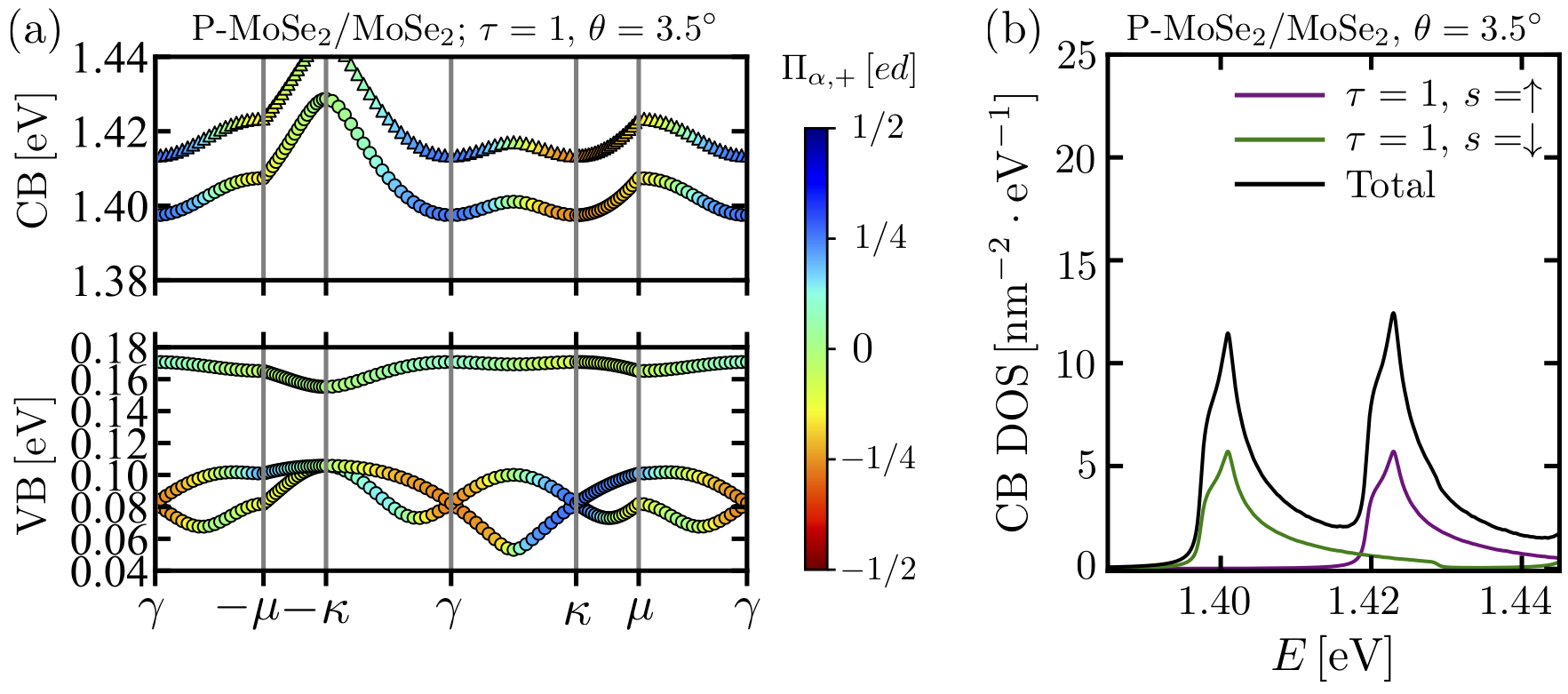}
\caption{(a) Moir\'e conduction and valence minibands of P-type twisted bilayer MoSe${}_2$, for twist angle $\theta=3.5^\circ$. Spin-up (spin-down) bands are shown with triangles (circles), and the symbol color represents the out-of-plane electric dipole moment of the state. (b) Corresponding density of states near the conduction-band edge.  We use \cite{twist_angle2018} $t_{\rm c}=t_{\rm v}/2=26\,{\rm meV}$; all other parameters are listed in Table \ref{tab:parameters}.}
\label{fig:P_BL_MoSe2}
\end{center}
\end{figure}

\begin{figure}[t!]
\begin{center}
\includegraphics[width=\columnwidth]{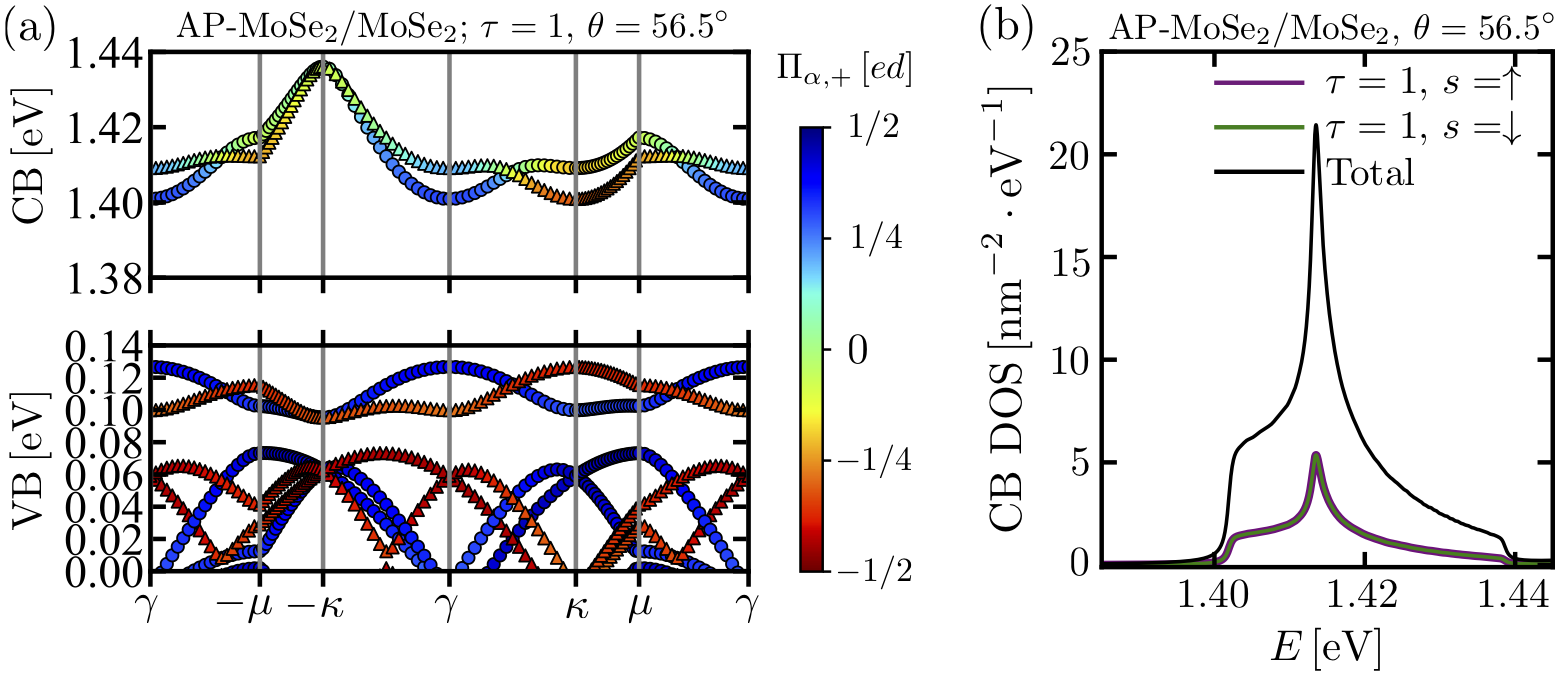}
\caption{(a) Moir\'e conduction and valence minibands of AP-type twisted bilayer MoSe${}_2$, for twist angle $\theta=56.5^\circ$. Spin-up (spin-down) bands are shown with triangles (circles), and the symbol color represents the out-of-plane electric dipole moment of the state. (b) Corresponding density of states near the conduction-band edge. We use \cite{twist_angle2018} $t_{\rm c}=t_{\rm v}/2=26\,{\rm meV}$; all other parameters are listed in Table \ref{tab:parameters}.}
\label{fig:AP_BL_MoSe2}
\end{center}
\end{figure}

We present the band structures of P- and AP-stacked bilayer MoSe${}_2$ in Figs.\ \ref{fig:P_BL_MoSe2} and \ref{fig:AP_BL_MoSe2}, respectively. We point out that, in the case of homobilayers, a more symmetric mBZ can be defined by shifting our chosen mBZ (Fig.\ \ref{fig:Figure1}) by $-\Delta \KK$. This transforms $\gamma \rightarrow \kappa$ and $\kappa \rightarrow \kappa'$, up to a moir\'e Bragg vector, corresponding to the convention followed in, \emph{e.g.}, Refs.\ \onlinecite{wallbank_annalen_2015,koshino_incommensurate}. Here, however, we will use the mBZ convention of Fig.\ \ref{fig:Figure1}, for the sake of consistency. To show the degree of interlayer state mixing, we color-code the plot symbols in Figs.\ \ref{fig:P_BL_MoSe2}(a) and \ref{fig:AP_BL_MoSe2}(a), according to the expectation value of the out-of-plane electric polarization, given by [see Eq.\ \eqref{eq:moire_band_operators}]
\begin{equation}\label{eq:polarization}
\begin{split}
	\Pi_{{\rm c}\tau s}^n(\qq) =\frac{ed}{2}&\sum_{i,j}\big[ \abs{\mathcal{A}_{ij}^{n\tau s}(\qq)}^2 - \abs{\mathcal{A}_{ij}^{n\tau s}{}'(\qq)}^2\big],\\
	\Pi_{{\rm v}\tau s}^n(\qq) =\frac{ed}{2}&\sum_{i,j}\big[ \abs{\mathcal{B}_{ij}^{n\tau s}(\qq)}^2 - \abs{\mathcal{B}_{ij}^{n\tau s}{}'(\qq)}^2\big],
\end{split}
\end{equation}
for each moir\'e band, at every wave vector $\qq \in {\rm mBZ}$. In Eq.\ \eqref{eq:polarization}, we have assumed, without loss of generality, that the ${\rm M'X_2'}$ (${\rm MX_2}$) layer is at the top (bottom) of the heterostructure; $e$ is the elementary charge, and $d$ is the interlayer distance (see Table \ref{tab:parameters}). The polarizations take values from $-1/2$ (blue) to $1/2$ (red), in units of $ed$, for electron states fully localized in the ${\rm MX_2}$ and ${\rm M'X_2'}$ layer, respectively. These values correspond to the state's out-of-plane electric dipole moment, measured with respect to the central plane of the stack.

Figs.\ \ref{fig:P_BL_MoSe2}(a) and \ref{fig:AP_BL_MoSe2}(a) show weak polarization of the electronic states in several regions of mBZ, indicating an even spatial distribution in the out-of-plane direction between the two MoSe${}_2$ layers, caused by the strong interlayer mixing. The highest valence states in the case of AP stacking are the exception, however, as seen in the bottom panel of Fig.\ \ref{fig:AP_BL_MoSe2}(b). This is because, as illustrated in the right panel of Fig.\ \ref{fig:PP_BL_MoSe2}, in AP-type bilayers the interlayer tunneling takes place between states of opposite valley quantum number ($\tau'=-\tau$), which due to spin-valley locking in the monolayers\cite{wangyao_spin_valley}, have opposite ordering of the spin-polarized bands. Therefore, bands of same spin quantum number in opposite layers are separated by a large spin-orbit splitting, typical of TMD valence bands (see Table \ref{tab:parameters}), and hybridize only weakly.

For the lowest conduction bands, however, a modest spin-orbit coupling strength of order $10\, {\rm meV}$ allows for strong interlayer hybridization also in the case of AP stacking (Fig.\ \ref{fig:PP_BL_MoSe2}, right), producing stark qualitative differences between moir\'e band structures for P- and AP-MoSe${}_2$ bilayers, seen in the top panels of Figs.\ \ref{fig:P_BL_MoSe2}(a) and \ref{fig:AP_BL_MoSe2}(a). For P-type bilayers, the lowest conduction bands of a given valley $\tau$ have the same spin quantum number in both monolayers, and mix to form the  moir\'e band edge shown in Fig.\ \ref{fig:P_BL_MoSe2}(a). The miniband edge has two branches, located at mBZ points $\gamma$ and $\kappa$, which belong to the same spin-polarized mixed miniband, and are separated only by a shallow saddle point, which produces a van Hove singularity close to the band edge, as shown in Fig.\ \ref{fig:P_BL_MoSe2}(b).  For AP-type bilayers, the two branches of the conduction-band edge belong to minibands of opposite spin quantum number, as shown in Fig.\ \ref{fig:AP_BL_MoSe2}(a). Note that above each band minimum at $\gamma$ or $\kappa$, the opposite-spin miniband flattens significantly, producing the van Hove singularity shown in Fig.\ \ref{fig:AP_BL_MoSe2}(b).

\section{Interlayer hybridization and moir\'e superlattice minibands for electrons in ${\rm MoTe_2/MoSe_2}$ and ${\rm MoSe_2/WS_2}$}\label{sec:resonant}
\begin{figure}[t!]
\begin{center}
\includegraphics[width=\columnwidth]{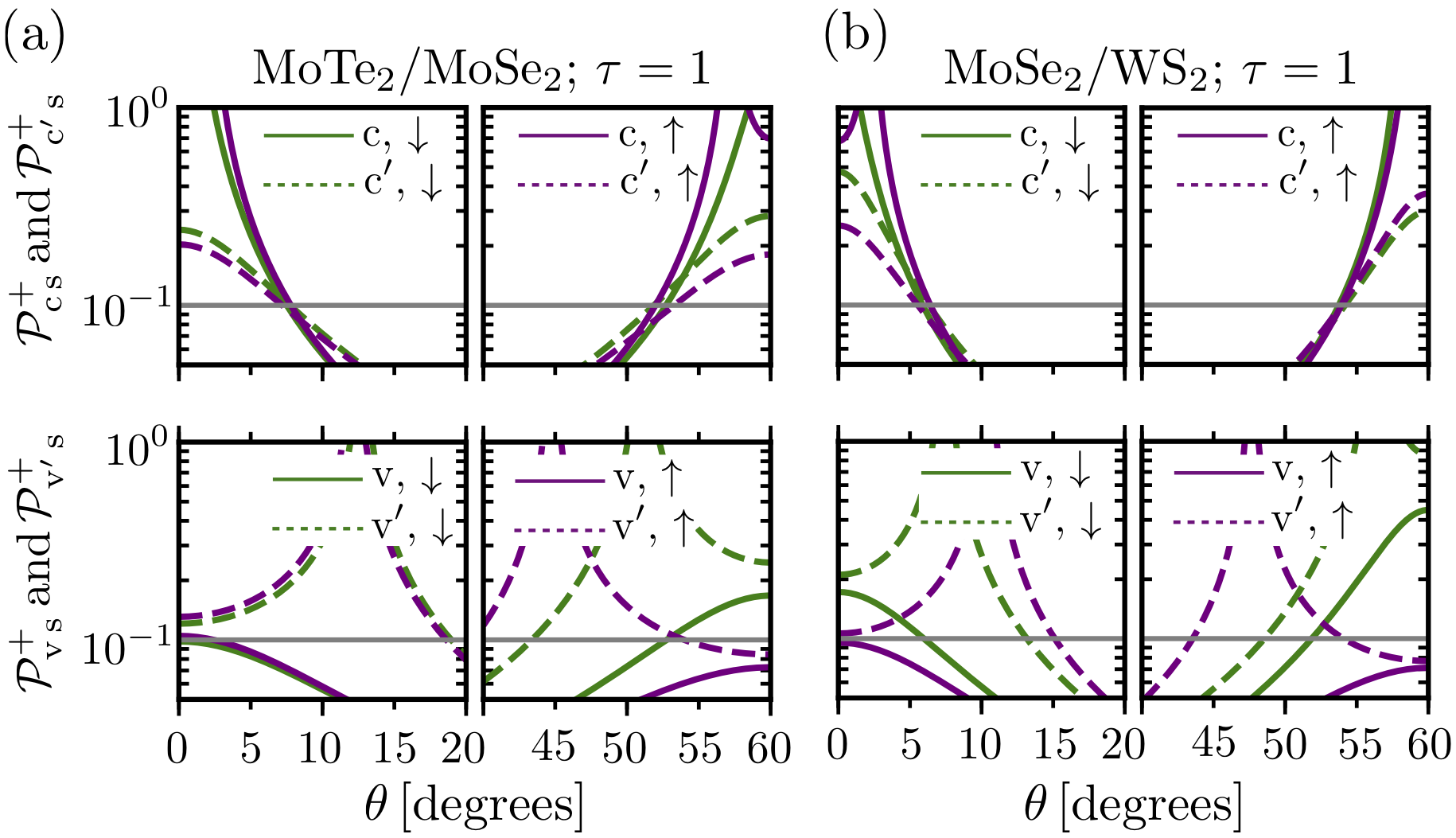}
\caption{Perturbative parameters for the conduction and valence bands of twisted (a) MoTe${}_2$/MoSe${}_2$ and (b) MoSe${}_2$/WS${}_2$ heterostructures. In both cases, the highest valence bands show weak interlayer mixing in close alignment and anti-alignment, and a perturbative treatment of interlayer hybridization is appropriate. By contrast, the lowest conduction bands mix resonantly in the moir\'e regime, and hybridization effects must be treated exactly.}
\label{fig:PP_resonant}
\end{center}
\end{figure}

\begin{figure*}[t!]
\begin{center}
\includegraphics[width=1.6\columnwidth]{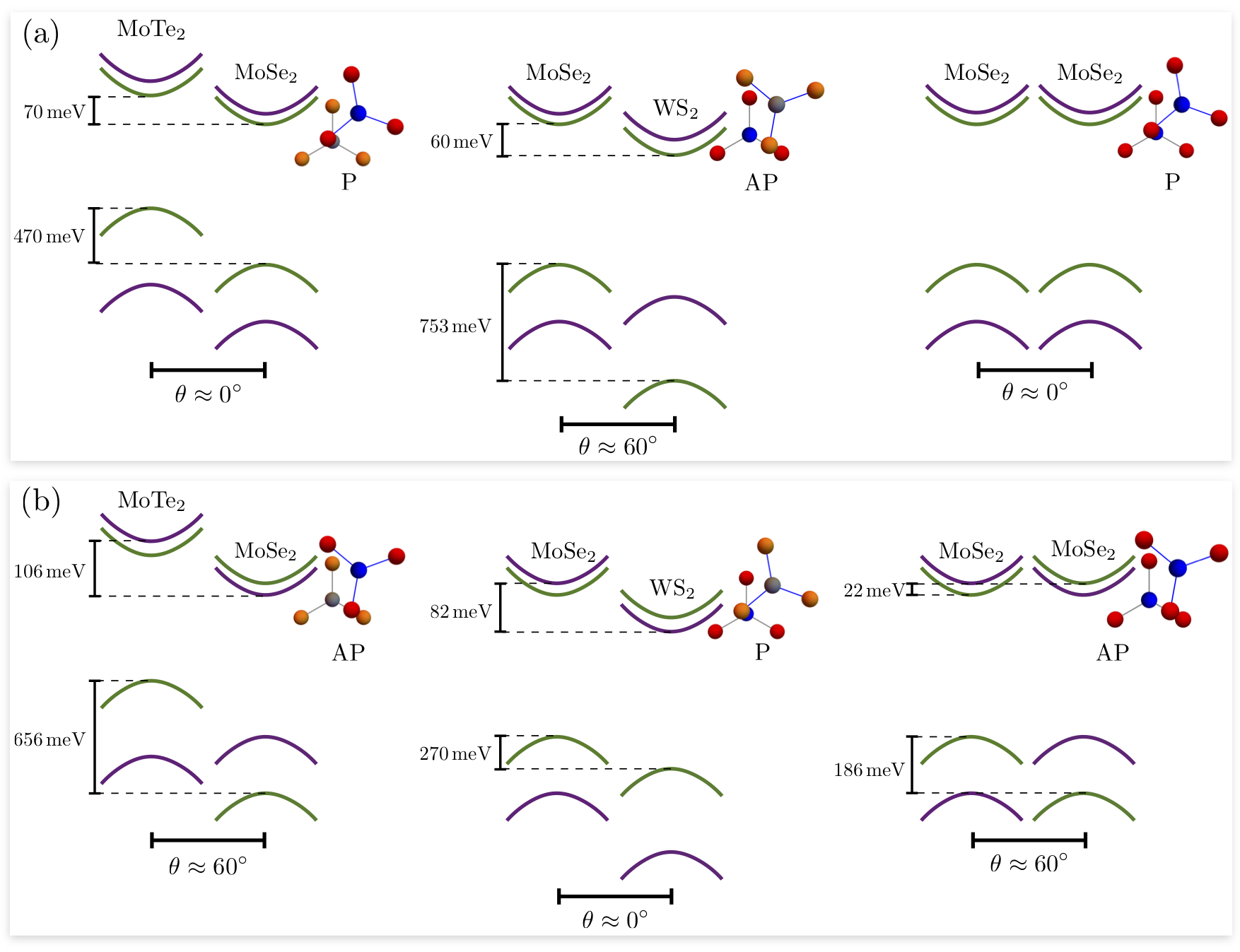}
\caption{Band alignment schematics for P- and AP-stacked twisted MoTe${}_2$/MoSe${}_2$ and MoSe${}_2$/WS${}_2$ heterostructures, and bilayer MoSe${}_2$. Respective atomic arrangements are also shown, for reference. (a) The spin ordering of the conduction bands in P-MoTe${}_2$/MoSe${}_2$ and AP-MoSe${}_2$/WS${}_2$ corresponds to the case of P-type bilayer MoSe${}_2$. (b) A similar correspondence is found between AP-MoTe${}_2$/MoSe${}_2$, P-MoSe${}_2$/WS${}_2$ and AP-type bilayer MoSe${}_2$.}
\label{fig:all_cases}
\end{center}
\end{figure*}

In this Section we discuss TMD heterobilayers in which the interlayer band alignment produces accidental near-resonant hybridization between either the conduction-  or valence-band edges of the two constituting TMD layers. This situation has been predicted\cite{band_alignment,Kozawa2015, first_principles_2018} for the conduction bands of MoSe${}_2$/WS${}_2$ heterostructures, and recently confirmed by photoluminescence experiments\cite{twist_angle2018}. Moreover, first principles estimates based on $G_0W_0$ calculations\cite{band_alignment} also point toward near-resonant conduction bands in MoTe${}_2$/MoSe${}_2$ (see Table \ref{tab:parameters} and Fig.\ \ref{fig:all_cases}). Similarly to the case of TMD homobilayers, for this class of heterostructures, the harmonic-potential approximation breaks down precisely for closely aligned and anti-aligned configurations, where effects of the moir\'e superlattice are most prominent. This is shown by the perturbative parameters  presented in Fig.\ \ref{fig:PP_resonant}, which indicate that, for $\theta \approx 0^\circ$ and $60^\circ$, it is necessary to treat the interlayer tunneling term \eqref{eq:Ht} exactly, due to near-resonant interlayer hybridization at the conduction-band edges.

\subsection{P-stacked ${\rm MoTe}_2/{\rm MoSe}_2$}\label{sec:R_MoTe2MoSe2}
\begin{figure}[t!]
\begin{center}
\includegraphics[width=\columnwidth]{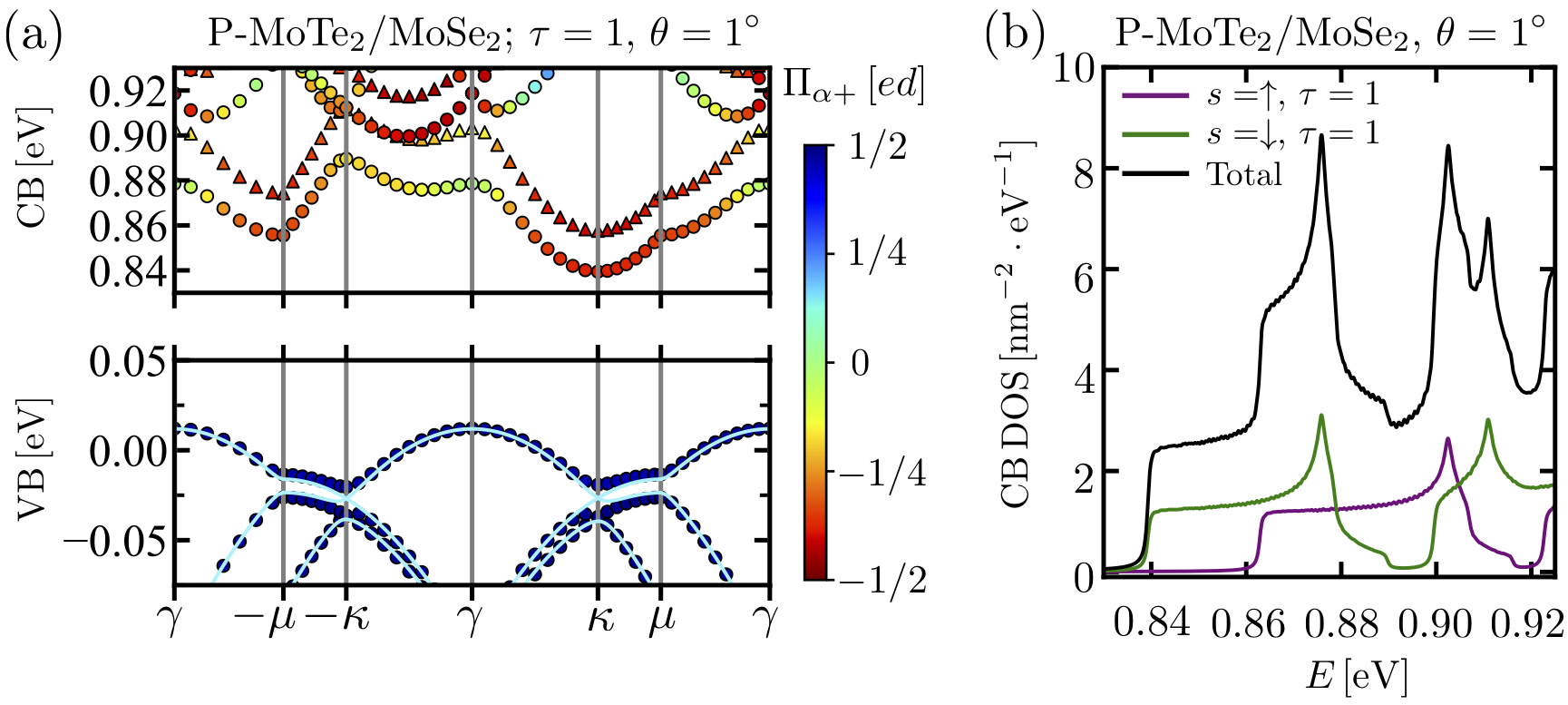}
\caption{(a) Moir\'e conduction and valence minibands of twisted P-MoTe${}_2$/MoSe${}_2$, with twist angle $\theta = 1^\circ$. Spin-up (-down) bands are shown with triangles (circles), with symbol color representing the state's out-of-plane electric dipole moment. The cyan curves in the lower panel show the harmonic-potential approximation to the highest valence bands, which hybridize weakly between layers. (b) Corresponding density of states near the conduction-band edge.}
\label{fig:R_MoTe2_MoSe2}
\end{center}
\end{figure}

Fig.\ \ref{fig:all_cases}(a) sketches the atomic arrangement and interlayer band alignment of P-stacked twisted MoTe${}_2$/MoSe${}_2$. The corresponding moir\'e conduction and valence miniband structures for $\theta=1^\circ$ are presented in Fig.\ \ref{fig:R_MoTe2_MoSe2}(a), as obtained by direct diagonalization of the full Hamiltonian \eqref{eq:fullH}. All bands shown have valley quantum number $\tau=1$, and the $\tau=-1$ bands can be obtained by a time reversal transformation. As expected from Fig.\ \ref{fig:all_cases}(a), we find that P-MoTe${}_2$/MoSe${}_2$ is an indirect-band-gap semiconductor, with the valence and conduction band edges located at the $\gamma$ and $\kappa$ points. However, note that, whereas the highest valence bands are well localized in the main hole layer, similarly to the case of MoSe${}_2$/MoS${}_2$ [Fig.\ \ref{fig:BS_MoSe2_MoS2}(b)], the lowest conduction bands show significant depolarization across the mBZ. This is due to the strong interlayer mixing caused by the relatively small detuning of $77\,{\rm meV}$ between hybridizing band edges, which distributes the miniband states between the two layers, similarly to the case of P-type TMD homobilayers [Fig.\ \ref{fig:P_BL_MoSe2}(a)]. Moreover, a comparison between the conduction-miniband density of states of P-MoTe${}_2$/MoSe${}_2$ and of P-type bilayer MoSe${}_2$ [Figs.\ \ref{fig:R_MoTe2_MoSe2}(b) and \ref{fig:P_BL_MoSe2}(b)] shows important parallels between the two cases; in particular, the formation of two van Hove singularities above the band edge.

\subsection{AP-stacked ${\rm MoTe_2}/{\rm MoSe_2}$}\label{sec:H_MoTe2MoSe2}
\begin{figure}[t!]
\begin{center}
\includegraphics[width=\columnwidth]{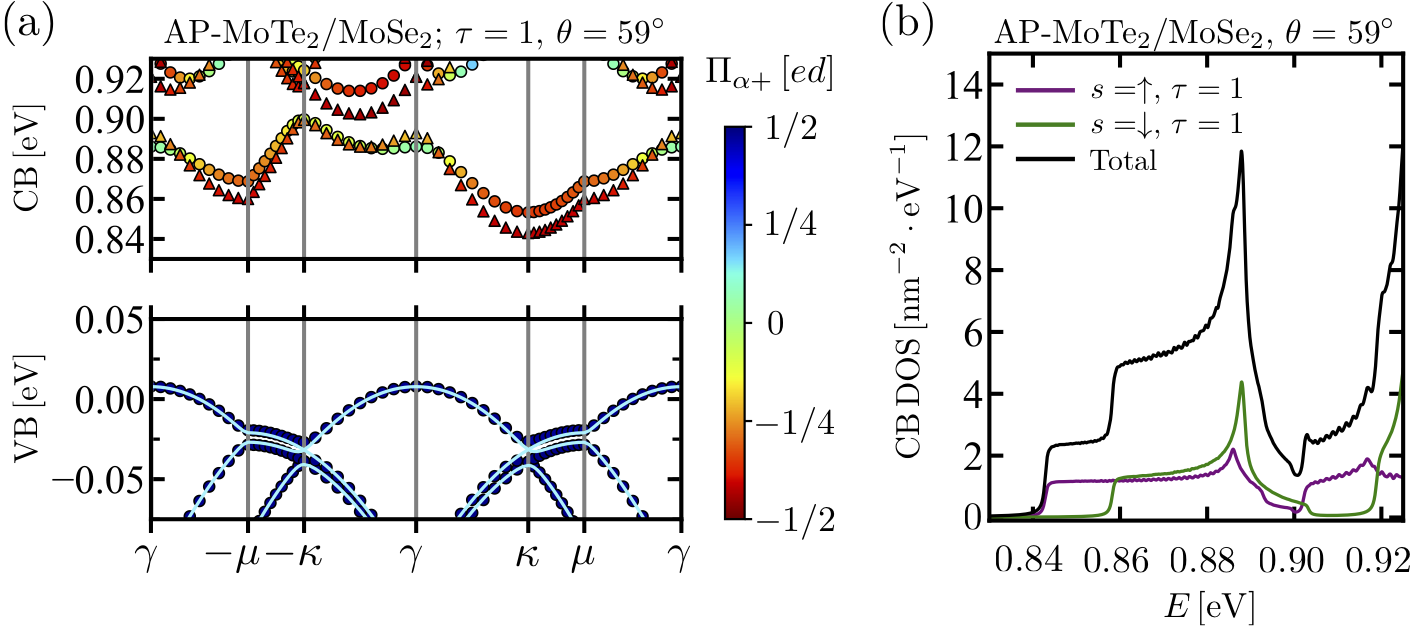}
\caption{(a)  Moir\'e conduction and valence minibands of twisted AP-MoTe${}_2$/MoSe${}_2$, with twist angle $\theta = 59^\circ$. Spin-up (-down) bands are shown with triangles (circles), with symbol color representing the state's out-of-plane electric dipole moment. The cyan curves in the lower panel show the harmonic-potential approximation to the highest valence bands. (b) Corresponding density of states near the conduction-band edge.}
\label{fig:H_MoTe2_MoSe2}
\end{center}
\end{figure}

Fig.\ \ref{fig:all_cases}(b) shows the band alignment of AP-stacked MoTe${}_2$/MoSe${}_2$, where states near the MoTe${}_2$ $\tau$ valley hybridize with those at the MoSe${}_2$ $\tau'=-\tau$ valley, which have the opposite spin ordering in both the conduction and valence bands. The corresponding moir\'e miniband structure, for twist angle $\theta = 59^\circ$, is shown in Fig.\ \ref{fig:H_MoTe2_MoSe2}(a), and the conduction-miniband density of states is presented in Fig.\ \ref{fig:H_MoTe2_MoSe2}(b). As in the case of P stacking, for AP stacking we find an indirect gap semiconductor, whose highest valence bands are largely confined to the MoTe${}_2$ layer, whereas the conduction minibands show different degrees of interlayer mixing throughout the mBZ. The conduction-band alignment in this case is more closely related to AP TMD homobilayers (Fig.\ \ref{fig:AP_BL_MoSe2}), due to the opposite spin ordering of the bands, which somewhat diminishes resonant hybridization of the band edges. Similar qualitative features are apparent in the density of states [Figs.\ \ref{fig:AP_BL_MoSe2}(b) and \ref{fig:H_MoTe2_MoSe2}(b)], which show a single van Hove singularity near the band edge.

\subsection{P-stacked ${\rm MoSe}_2/{\rm WS}_2$}\label{sec:R_MoSe2WS2}
MoSe${}_2$/WS${}_2$ heterostructures are different from all the cases discussed so far, due to the presence of different transition metal atoms in the two TMD layers. In particular, tungsten-based TMDs are known\cite{kdotp} to display a negative spin-orbit coupling constant for the conduction band, as opposed to the positive one found in molybdenum-based TMDs (see Table \ref{tab:parameters}). This results in opposite ordering of the spin-polarized conduction bands of MoSe${}_2$ and WS${}_2$ in a P-MoSe${}_2$/WS${}_2$ heterostructure, as illustrated in Fig.\ \ref{fig:all_cases}(b)---an analogous situation to AP-type homobilayers.

\begin{figure}[t!]
\begin{center}
\includegraphics[width=\columnwidth]{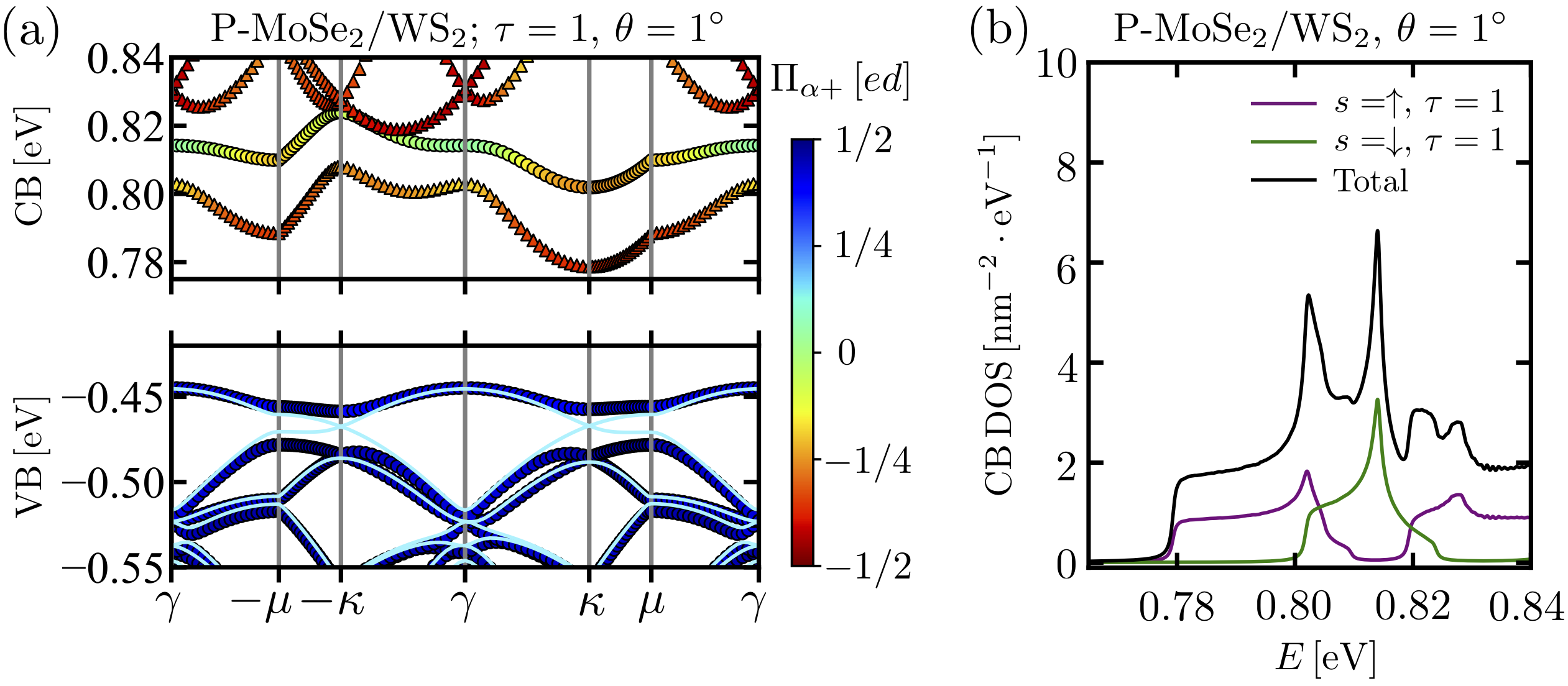}
\caption{(a)  Moir\'e conduction and valence minibands of twisted P-MoSe${}_2$/WS${}_2$, with twist angle $\theta = 1^\circ$. Spin-up (-down) bands are shown with triangles (circles), with symbol color representing the state's out-of-plane electric dipole moment. The cyan curves in the lower panel show the harmonic-potential approximation to the highest valence bands. (b) Corresponding density of states near the conduction band edge.}
\label{fig:R_MoSe2_WS2}
\end{center}
\end{figure}
Fig.\ \ref{fig:R_MoSe2_WS2}(a) shows the moir\'e band structure of P-MoSe${}_2$/WS${}_2$ at twist angle $\theta=1^\circ$, with the density of states corresponding to the conduction band shown in Fig.\ \ref{fig:R_MoSe2_WS2}(b). Notice the strong electric dipole moment imbalance between the spin-up and spin-down conduction bands, indicated by the symbol colors in Fig.\ \ref{fig:R_MoSe2_WS2}(a). This strong spin asymmetry is caused by the combination of opposite spin splittings of the monolayer conduction bands, and the small interlayer offset $\delta_{\rm c}=60\,{\rm meV}$, which produces almost perfect alignment between the ($\tau=1$) spin-down band edges ($28\,{\rm meV}$) and a much larger detuning of the spin-up bands ($82\,{\rm meV}$), as illustrated in Fig.\ \ref{fig:all_cases}(b). This leads to different levels of interlayer mixing for the two spin-polarized minibands.

\subsection{AP-stacked ${\rm MoSe}_2/{\rm WS}_2$}\label{sec:H_MoSe2WS2}
\begin{figure}[t!]
\begin{center}
\includegraphics[width=\columnwidth]{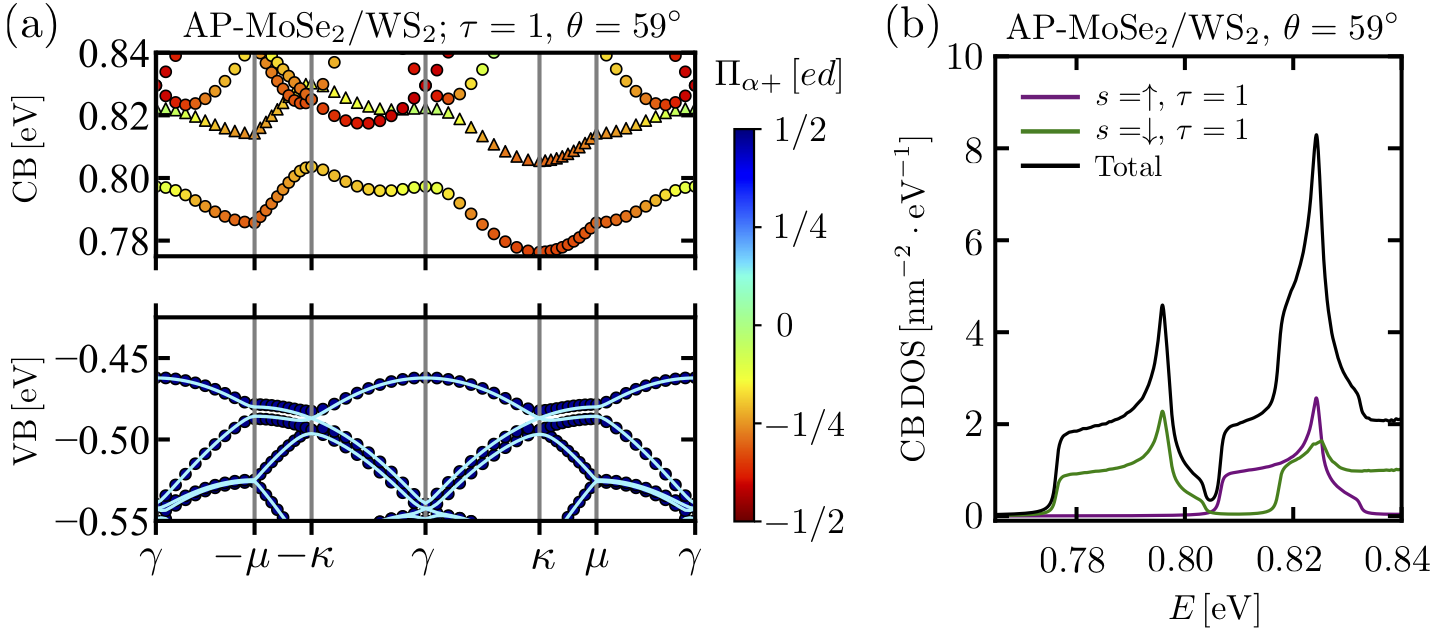}
\caption{(a)  Moir\'e conduction and valence minibands of twisted AP-MoSe${}_2$/WS${}_2$, with twist angle $\theta = 59^\circ$. Spin-up (-down) bands are shown with triangles (circles), with symbol color representing the state's out-of-plane electric polarization. The cyan curves in the lower panel show the harmonic-potential approximation to the highest valence bands. (b) Corresponding density of states near the conduction-band edge.}
\label{fig:H_MoSe2_WS2}
\end{center}
\end{figure}

Finally, Fig.\ \ref{fig:H_MoSe2_WS2} shows the moir\'e band structure and conduction-miniband DOS for AP-type MoSe${}_2$/WS${}_2$ at twist angle $\theta=59^\circ$, corresponding to the schematic shown in Fig.\ \ref{fig:all_cases}(a). For this range of angles, where MoSe${}_2$ $\tau$ valley states hybridize with WS${}_2$ states of valley quantum number $\tau'=-\tau$, both layers show the same spin ordering in the conduction bands, and the situation is qualitatively similar to the case of P-type homobilayers. Some similarities between the two cases can be found in Fig.\ \ref{fig:H_MoSe2_WS2}, such as the spin polarization of the bottom band across the mBZ, and a rough two-peak structure in the density of states near the conduction-band edge, reminiscent of the van Hove singularities shown in Fig.\ \ref{fig:P_BL_MoSe2}(b).

\subsection{Electrical control of moir\'e superlattice effects}\label{sec:stark}
The band structure calculations presented in Figs.\ \ref{fig:R_MoTe2_MoSe2}--\ref{fig:H_MoSe2_WS2} show that P- and AP-type MoTe${}_2$/MoSe${}_2$ and MoSe${}_2$/WS${}_2$ heterostructures are type II semiconductors, with a $\gamma$--$\kappa$ indirect band gap. However, it is possible to reduce the offset between the conduction band edges by application of a positive interlayer bias voltage\cite{MoSe2_WSe2_2015,klein_backgate_2016,wang_backgate_2017} $V_{\rm B}$. In fact, recent experiments\cite{klein_backgate_2016} have demonstrated that interlayer voltages of up to approximately $200\,{\rm mV}$ can be produced in TMD bilayers by means of metallic gates. The resulting potential gradient along the heterostructure's out-of-plane axis will lower the ${\rm MX_2}$ (bottom) layer's ${\rm c}$ band while raising the ${\rm M'X_2'}$ (top) layer's ${\rm c}'$ band, such that a suitable value of $V_{\rm B}$ can impose a degeneracy between the two minima at $\gamma$ and $\kappa$.
\begin{figure}[b!]
\begin{center}
\includegraphics[width=\columnwidth]{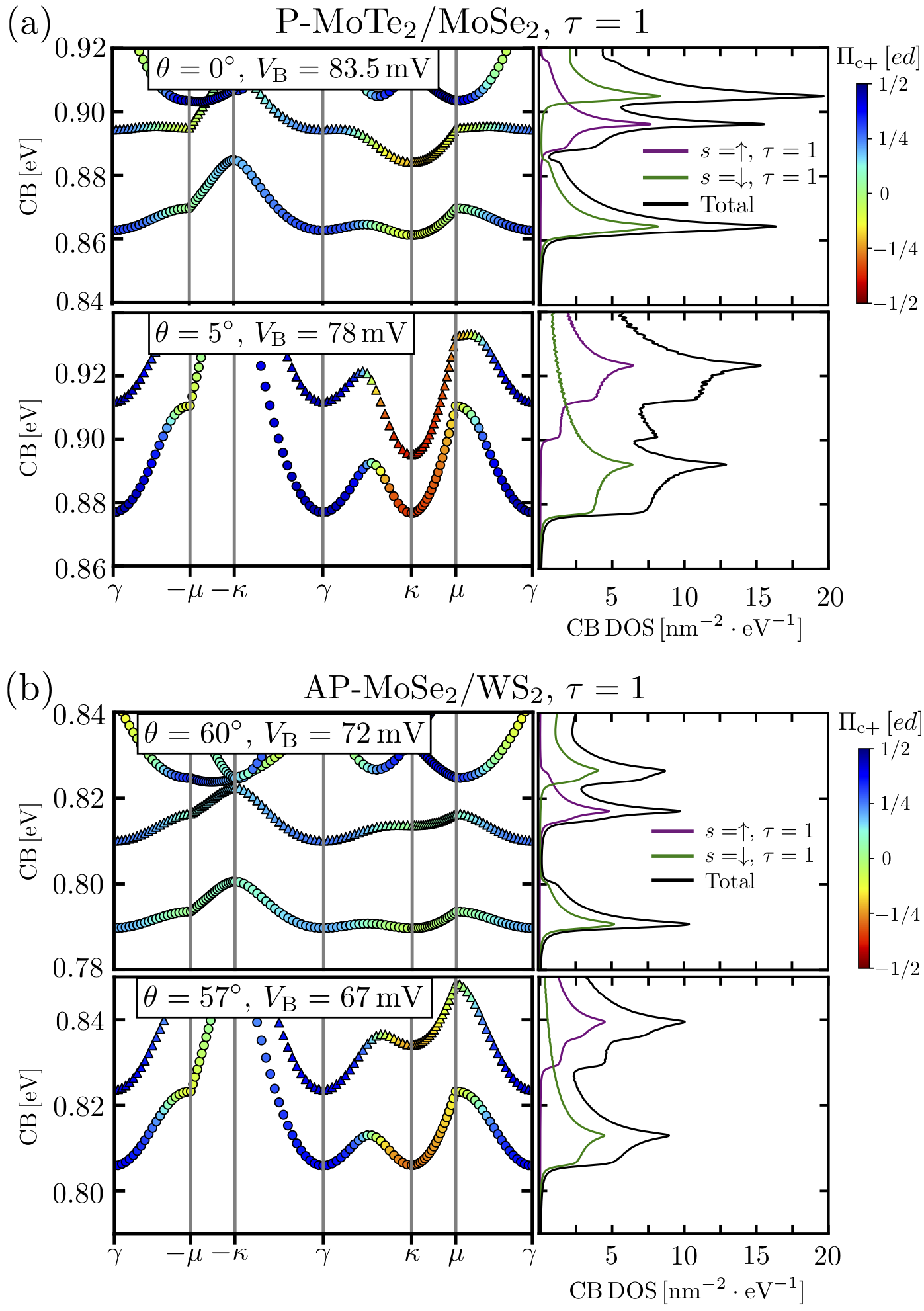}
\caption{Conduction minibands and DOS at critical bias voltage $V_{\rm B}$, for (a) P-MoTe${}_2$/MoSe${}_2$ and (b) AP-MoSe${}_2$/WS${}_2$ heterostructures with different degrees of alignment. In both cases, $V_{\rm B}$ is weakly twist-angle dependent. Spin-up (-down) bands are shown with triangles (circles). The two branches of the band edge belong to a single spin-polarized band, analogously to the case of P-type TMD homobilayers (Fig.\ \ref{fig:P_BL_MoSe2}).}
\label{fig:CBS_Ez_MoTe2MoSe2}
\end{center}
\end{figure}

Figs.\ \ref{fig:CBS_Ez_MoTe2MoSe2}(a) and \ref{fig:CBS_Ez_MoTe2MoSe2}(b) show the conduction miniband structures of P-type MoTe${}_2$/MoSe${}_2$ and AP-type MoSe${}_2$/WS${}_2$, under the critical bias voltages $V_{\rm B}$ that establish a band-edge degeneracy at the $\gamma$ and $\kappa$ points. In both cases, we find that the critical bias is twist-angle dependent, giving values of $83.5\,{\rm mV}$ for perfectly aligned P-MoTe${}_2$/MoSe${}_2$, and a lower value of $78\,{\rm mV}$ for twist angle $\theta=5^\circ$. Note that both of these values correspond to energies greater than the band offset of $70\,{\rm meV}$, indicated in Fig.\ \ref{fig:all_cases}(a). The same trend is found for AP-MoSe${}_2$/WS${}_2$, where the critical bias for perfect anti-alignment ($\theta = 60^\circ$) is $72\,{\rm mV}$, compared to $67\,{\rm mV}$ for $\theta = 57^\circ$, and to the $60\,{\rm meV}$ offset shown in Fig.\ \ref{fig:all_cases}(a). The lowest conduction minibands shown in Fig.\ \ref{fig:CBS_Ez_MoTe2MoSe2} bear striking resemblance to those of P-type TMD homobilayers (Fig.\ \ref{fig:P_BL_MoSe2}), with two branches of the band edge at the $\gamma$ and $\kappa$ points belonging to the same spin-polarized band, and a van Hove singularity forming just above the band edge.

\begin{figure}[t!]
\begin{center}
\includegraphics[width=\columnwidth]{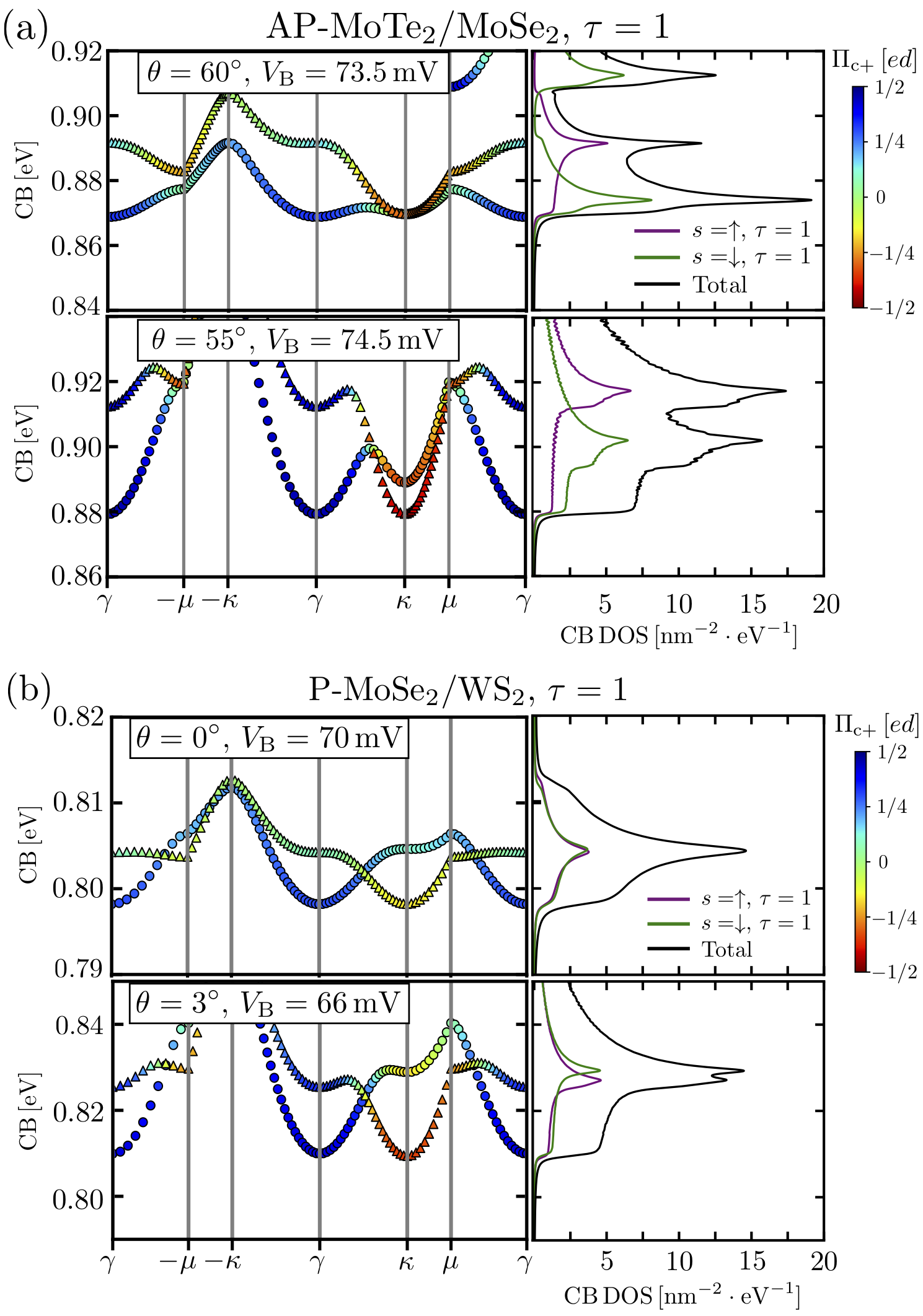}
\caption{Conduction minibands and DOS at critical bias voltage $V_{\rm B}$, for (a) AP-MoTe${}_2$/MoSe${}_2$ and (b) P-MoSe${}_2$/WS${}_2$ heterostructures with different degrees of alignment. Spin-up (-down) bands are shown with triangles (circles). The two branches of the band edge belong to different minibands of opposite spin polarization, similarly to the case of AP-type TMD homobilayers (Fig.\ \ref{fig:AP_BL_MoSe2}).}
\label{fig:CBS_Ez_MoSe2WS2}
\end{center}
\end{figure}

Similarly, Figs.\ \ref{fig:CBS_Ez_MoSe2WS2}(a) and \ref{fig:CBS_Ez_MoSe2WS2}(b) show the corresponding cases of critical interlayer bias for AP-MoTe${}_2$/MoSe${}_2$ and P-MoSe${}_2$/WS${}_2$, both for perfect (anti) alignment and for a finite misalignment angle, showing also a weak twist-angle dependence of the critical bias voltage. The conduction minibands in these cases show direct correspondence to those of AP-type TMD homobilayers (Fig.\ \ref{fig:AP_BL_MoSe2}), with the $\gamma$- and $\kappa$-point branches of the band edge belonging to bands of opposite spin polarization, and a flattening of the second conduction miniband above each minimum that is particularly clear in P-MoSe${}_2$/WS${}_2$, as a result of the similar effective electron masses of the two layers (Table \ref{tab:parameters}).

In the case of TMD homobilayers, the band-edge degeneracy at the $\gamma$ and $\kappa$ points is a direct consequence of the identical dispersions of the two layers, as well as the perfect band alignment in the case of P stacking. However, note that this is not the case for P- or AP-type MoTe${}_2$/MoSe${}_2$ and MoSe${}_2$/WS${}_2$, since the predicted critical $V_{\rm B}$ are twist-angle dependent, and can, in fact, be larger than the actual offsets between the hybridizing bands. The reason behind this discrepancy is the asymmetry between the conduction-band dispersions of the two layers, parametrized by their different effective electron masses (Table \ref{tab:parameters}). Thus, when these bands are folded into the mBZ, different miniband configurations are obtained at $\gamma$ and $\kappa$, producing an asymmetry between the band minima at those points in the mBZ.

\section{${\rm m}$SL minibands for resonantly hybridized intra- and interlayer excitons}\label{sec:optical}
\begin{table}[t!]
\caption{Theoretical binding energies and Bohr radii of the four types of intralayer and interlayer excitons shown in Eq.\ \eqref{eq:XandY}, for different TMD heterostructures on a SiO${}_2$ substrate. The screening lengths were obtained from Refs.\ \onlinecite{berkelbach_variational,mostaani_excitonic_prb_2017,kumar_physicab_2012}, and we assume an average dielectric constant $\epsilon=2.45$ for the SiO${}_2$/vacuum environment, following Ref.\ \onlinecite{complexes2018}.}
\begin{center}
\begin{tabular}{c | c c r @{.} l r @{.} l r @{.} l r @{.} l}
\hline\hline
\, & \, & \, & \multicolumn{2}{c}{$\varepsilon_{\rm X}\,[{\rm eV}]$} & \multicolumn{2}{c}{$\varepsilon_{\rm IX}\,[{\rm eV}]$} & \multicolumn{2}{c}{$a_{\rm X}\,[\Ams]$} & \multicolumn{2}{c}{$a_{\rm IX}\,[\Ams]$}\\
\raisebox{1.5ex}[0pt]{${\rm MX_2/M'X_2'}$} & \raisebox{1.5ex}[0pt]{$r_*\,[\Ams]$} & \raisebox{1.5ex}[0pt]{$r_*'\,[\Ams]$} & \multicolumn{2}{c}{$\varepsilon_{\rm X' }\,[{\rm eV}]$} & \multicolumn{2}{c}{$\varepsilon_{\rm IX'}\,[{\rm eV}]$} & \multicolumn{2}{c}{$a_{\rm X'}\,[\Ams]$} & \multicolumn{2}{c}{$a_{\rm IX'}\,[\Ams]$}\\
\hline\hline
\,                                                                                                                                                               & \, & \, & 0&176 & 0&164 & 21&4 & 21&4\\
\raisebox{1.5ex}[0pt]{WS${}_2$/MoS${}_2$}      & \raisebox{1.5ex}[0pt]{37.89} & \raisebox{1.5ex}[0pt]{38.62} & 0&194 & 0&163 & 18&2 & 21&5\\
\,                                                                                                                                                               & \, & \, & 0&170 & 0&157 & 21&4 & 21&7\\
\raisebox{1.5ex}[0pt]{WSe${}_2$/MoS${}_2$}    & \raisebox{1.5ex}[0pt]{45.11} & \raisebox{1.5ex}[0pt]{38.62} & 0&183 & 0&158 & 18&9 & 21&7\\
\,                                                                                                                                                              & \, & \, & 0&195 & 0&170 & 17&8 & 19&8\\
\raisebox{1.5ex}[0pt]{MoSe${}_2$/MoS${}_2$} & \raisebox{1.5ex}[0pt]{39.79} & \raisebox{1.5ex}[0pt]{38.62} & 0&191 & 0&172 & 18&4 & 19&4\\
\,                                                                                                                                                              & \, & \, & 0&196 & 0&162 & 17&7 & 21&5\\
\raisebox{1.5ex}[0pt]{MoSe${}_2$/WS${}_2$}   & \raisebox{1.5ex}[0pt]{39.79} & \raisebox{1.5ex}[0pt]{37.89} & 0&174 & 0&164 & 21&5 & 21&1\\
\,                                                                                                                                                              & \, & \, & 0&169 & 0&158 & 21&5 & 21&3\\
\raisebox{1.5ex}[0pt]{WSe${}_2$/MoSe${}_2$} & \raisebox{1.5ex}[0pt]{45.11} & \raisebox{1.5ex}[0pt]{39.79} & 0&185 & 0&157 & 18&4 & 21&7\\
\,                                                                                                                                                               & \, & \, & 0&177 & 0&147 & 15&7 & 20&1\\
\raisebox{1.5ex}[0pt]{MoTe${}_2$/MoSe${}_2$} & \raisebox{1.5ex}[0pt]{73.61} & \raisebox{1.5ex}[0pt]{39.79} & 0&152 & 0&151 & 20&9 & 18&9\\
\hline\hline
\end{tabular}
\end{center}
\label{tab:excitons}
\end{table}%

Much of the current interest in the properties of TMD systems stems from their outstanding monolayer optical properties\cite{mak_MoS2_2010,splendiani_MoS2_2010,review_2018}, produced by their direct band gap at the $K$ points, and dominated by the formation of strongly-bound 2D intralayer excitons (X). By contrast, in twisted TMD heterobilayers such as those discussed in Secs.\ \ref{sec:harmonic} to \ref{sec:resonant}, the ground-state excitons are formed by electron and hole states confined to opposite layers, known as interlayer excitons (IXs)\cite{MoSe2_WSe2_2015,Heo2015,nayak_acsnano_2017,charge_separation_2017}. The wave vector mismatch between the electron and hole band edges shown in, \emph{e.g.}, Fig.\ \ref{fig:BS_MoSe2_MoS2}, means that the center-of-mass momentum of low-energy IXs is finite, and energy-momentum conservation forbids radiative recombination, unless mediated by some compensating mechanism, such as phonon or impurity scattering. In other words, the lowest-energy IX is momentum-dark.

Until recently\cite{twist_angle2018}, Xs and IXs have been mostly discussed as independent objects; however, it is clear that the band hybridization effects predicted in Secs.\ \ref{sec:homobilayer} and \ref{sec:resonant} must lead to mixing of IX and X states\cite{Thygesen_large_optical_2018}. Indeed, as the (positive) binding energy $\varepsilon_{\rm IX}$ of IXs is smaller than that of Xs, $\varepsilon_{\rm X}$, due to the additional out-of-plane distance between the electron and hole, the detuning between the lowest X and IX energies must be approximately $\delta_{\rm c}-(\varepsilon_{\rm X} - \varepsilon_{\rm IX})$, which improves the resonant condition. To show this, we estimate the binding energies of all possible species of X and IX for different TMD heterobilayers, by solving the two-body problem for electrons and holes using the finite elements method, considering the Keldysh-type\cite{keldysh} long-range intra- and interlayer interactions
\begin{equation}\label{eq:keldysh}
\begin{split}
  \varphi_{\rm intra}({\bf q}) \approx& \frac{2\pi}
  {\epsilon q\left[1+(r_{*}+r_{*}')q\right]},\\
  \varphi_{\rm inter}({\bf q}) \approx& \frac{2\pi}
  {\epsilon q\left[1+(r_{*}+r_{*}'+d)q\right]},
\end{split}
\end{equation}
derived in Ref.\ \onlinecite{complexes2018}. The results are presented in Table \ref{tab:excitons}. In Eq.\ \eqref{eq:keldysh}, $\qq$ represents momentum transfer; $d$ is the interlayer distance; $r_*=2\pi \kappa/\epsilon$ ($r_*'=2\pi \kappa'/\epsilon$) is the screening length of layer ${\rm MX_2}$ (${\rm M'X_2'}$), with $\kappa$ ($\kappa'$) the in-plane dielectric susceptibility, and $\epsilon$ is the average dielectric constant of the environment. The corresponding exciton Bohr radii were estimated as the RMS width of the numerically obtained lowest bound-state wave function, which is of $s$ type. From Tables \ref{tab:parameters} and \ref{tab:excitons}, we can see that the difference in binding energies of the intra- and interlayer excitons, $\varepsilon_{\rm X} - \varepsilon_{\rm IX}$, is comparable to $\delta_{\rm c}$ for materials with near-resonant band edges, such as MoTe${}_2$/MoSe${}_2$ and MoSe${}_2$/WS${}_2$. This leads to enhanced hybridization between these Xs and IXs, as compared to electrons and holes, resulting in hybridized excitons (hXs) formed by resonantly mixed X and IX states. As we show below, similar resonant conditions can arise also for higher-energy intra- and interlayer excitons, such that signatures of hXs can appear all throughout the optical spectra of TMD heterobilayers. These strongly mixed states have a large intralayer component that allows them to recombine radiatively, making hXs semi-bright, whereas the out-of-plane electric dipole moment, inherited from their interlayer component, makes hXs sensitive to the electrostatic environment of the heterobilayer, through the Stark effect.

\begin{figure}[h!]
\begin{center}
\includegraphics[width=0.8\columnwidth]{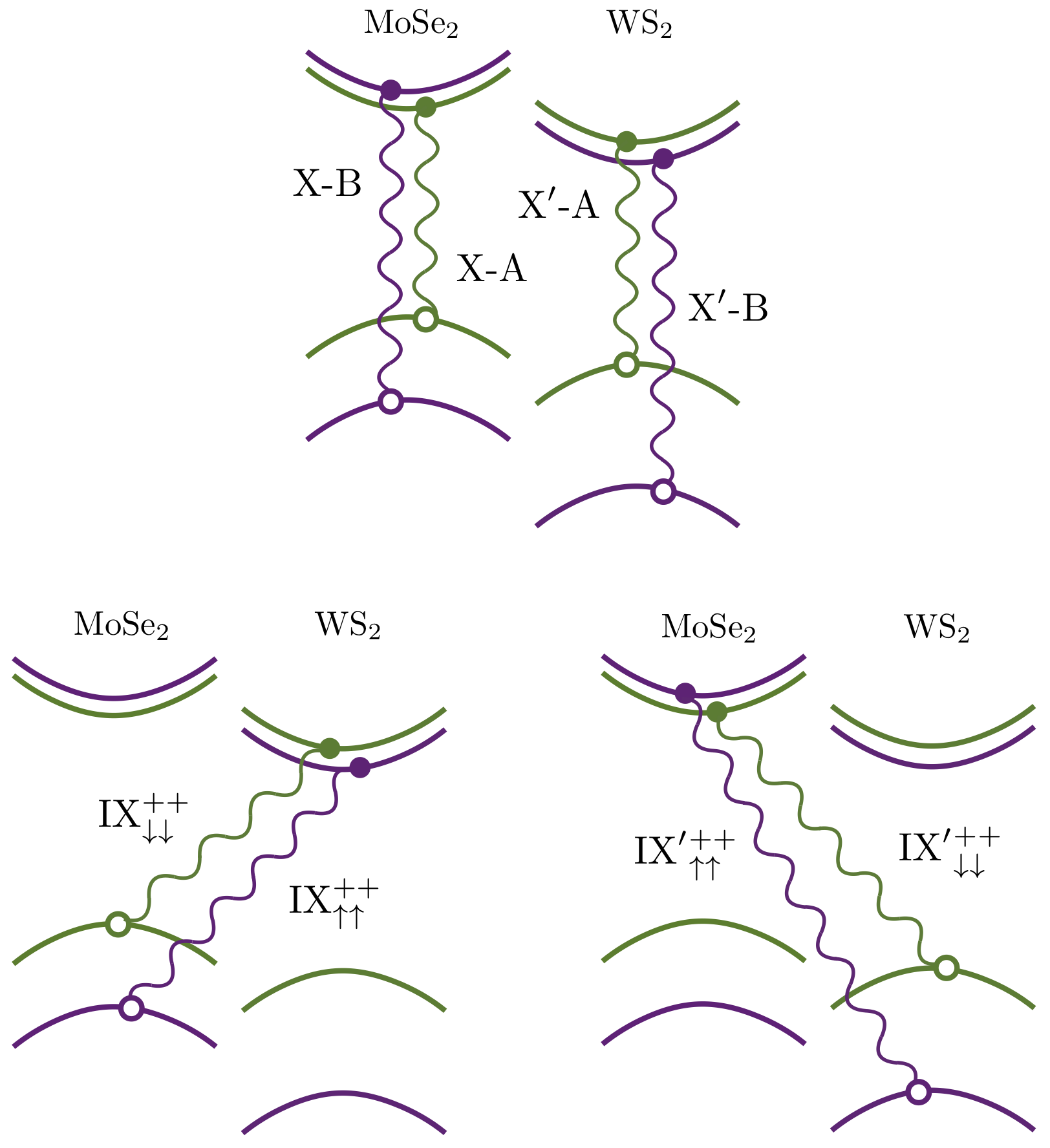}
\caption{Schematics of the different intra- and interlayer excitons in Eq.\ \eqref{eq:XandY} with valley quantum number $\tau=1$, for P-stacked MoSe${}_2$/WS${}_2$. Wavy lines indicate binding of the electron and hole by electrostatic interactions. The MoSe${}_2$ A and B excitons correspond to the states ${\rm X}_{\downarrow\downarrow}^{++}$ and ${\rm X}_{\uparrow\uparrow}^{++}$, whereas the WS${}_2$ A and B excitons are ${\rm X}'{}_{\downarrow\downarrow}^{++}$ and ${\rm X}'{}_{\uparrow\uparrow}^{++}$, respectively.}
\label{fig:XXpYZ}
\end{center}
\end{figure}

Consider the X states of ${\rm MX_2}$ and ${\rm M'X_2'}$, and all possible IX states of the ${\rm MX_2/M'X_2'}$ heterobilayer, given respectively by\cite{moskalenko2000bose,wangyao_interlayer_X}
\begin{equation}\label{eq:XandY}
\begin{split}
	\ket{{\rm X}^{\tau\bar{\tau}}_{s\bar{s}}(\QQ)}=&\frac{1}{\sqrt{S}}\sum_{\qq}X_{\qq}\,c_{{\rm c}\tau s}^\dagger(\tfrac{m_{\rm c}}{m_{\rm c}+m_{\rm v}}\QQ+\qq)\\
	&\times c_{{\rm v}\bar{\tau} \bar{s}}(-\tfrac{m_{\rm v}}{m_{\rm c}+m_{\rm v}}\QQ+\qq)\ket{\Omega},\\
	\ket{{\rm X}'{}^{\tau'\bar{\tau}'}_{s'\bar{s}'}(\QQ)}=&\frac{1}{\sqrt{S}}\sum_{\qq}X_{\qq}'\,c_{{\rm c'}\tau' s'}^\dagger(\tfrac{m_{\rm c'}}{m_{\rm c'}+m_{\rm v'}}\QQ+\qq)\\
	&\times c_{{\rm v'}\bar{\tau}' \bar{s}'}(-\tfrac{m_{\rm v'}}{m_{\rm c'}+m_{\rm v'}}\QQ+\qq)\ket{\Omega},\\
	\ket{{\rm IX}^{\tau'\bar{\tau}}_{s'\bar{s}}(\QQ)}=&\frac{1}{\sqrt{S}}\sum_{\qq}Y_{\qq}\,c_{{\rm c}'\tau' s'}^\dagger(\tfrac{m_{\rm c'}}{m_{\rm c'}+m_{\rm v}}\QQ+\qq)\\
	&\times c_{{\rm v}\bar{\tau} \bar{s}}(-\tfrac{m_{\rm v}}{m_{\rm c'}+m_{\rm v}}\QQ+\qq)\ket{\Omega},\\
	\ket{{\rm IX}'{}^{\tau\bar{\tau}'}_{s\bar{s}'}(\QQ)}=&\frac{1}{\sqrt{S}}\sum_{\qq}Y_{\qq}'\,c_{{\rm c}\tau s}^\dagger(\tfrac{m_{\rm c}}{m_{\rm c}+m_{\rm v'}}\QQ+\qq)\\
	&\times c_{{\rm v}'\bar{\tau}' \bar{s}'}(-\tfrac{m_{\rm v'}}{m_{\rm c}+m_{\rm v'}}\QQ+\qq)\ket{\Omega}.
\end{split}
\end{equation}
In each case, the exciton center-of-mass  momentum is represented by $\QQ$; $X_{\qq}$, $X_{\qq}'$, $Y_{\qq}$ and $Y_{\qq}'$ are the corresponding electron-hole relative motion wave functions in reciprocal space; $| \Omega \rangle$ is the neutral ground state of the heterobilayer in the absence of interactions; and $S$ is the heterostructure's surface area. The Xs and IXs have parabolic dispersions given by
\begin{equation}\label{eq:Xdisp}
\begin{split}
	E_{{\rm X},s\bar{s}}^{\tau \bar{\tau}}(Q) =& E_{{\rm X},s\bar{s}}^{\tau \bar{\tau}} + \frac{\hbar^2Q^2}{2(m_{\rm c}+m_{\rm v})},\\
	E_{{\rm X}',s'\bar{s}'}^{\tau' \bar{\tau}'}(Q) =& E_{{\rm X}',s'\bar{s}'}^{\tau' \bar{\tau}'} + \frac{\hbar^2Q^2}{2(m_{\rm c'}+m_{\rm v'})},\\
	E_{{\rm IX},s'\bar{s}}^{\tau' \bar{\tau}}(Q) =& E_{{\rm IX},s'\bar{s}}^{\tau' \bar{\tau}} + \frac{\hbar^2Q^2}{2(m_{\rm c'}+m_{\rm v})},\\
	E_{{\rm IX}',s\bar{s}'}^{\tau \bar{\tau}'}(Q) =& E_{{\rm IX}',s\bar{s}'}^{\tau \bar{\tau}'} + \frac{\hbar^2Q^2}{2(m_{\rm c}+m_{\rm v'})}.
\end{split}
\end{equation}
We will focus on intravalley X and ${\rm X}'$ states, formed by electrons and holes that can recombine in the absence of intervalley scattering. For IX and ${\rm IX}'$, we consider only the set of low-energy states that can hybridize with intravalley X and ${\rm X}'$ according to Eq.\ \eqref{eq:T_simpl}, such that we must set $\tau'=\pm \tau$ for P or AP stacking, respectively. We further assume that no spin scattering mechanisms are present in the system, and excitons can recombine only if its constituting electron and hole have opposite spin quantum numbers. We call such excitons spin-bright, and in the opposite case, the exciton is called spin-dark. The eight spin-bright exciton species are represented schematically in Fig.\ \ref{fig:XXpYZ} for the case of P-MoSe${}_2$/WS${}_2$, and the intralayer exciton states are labeled according to the usual nomenclature of A and B excitons.

In principle, the exciton energies at zero momentum can be obtained from the \emph{ab initio} band alignment parameters of Table \ref{tab:parameters}, together with the binding energies of Table \ref{tab:excitons}, as
\begin{equation}
\begin{split}
	E_{{\rm X},s\bar{s}}^{\tau \bar{\tau}} =& E_{\tau s}^{\rm c}(0) - E_{\bar{\tau} \bar{s}}^{\rm v}(0) - \varepsilon_{\rm X} ,\\
	E_{{\rm X}',s'\bar{s}'}^{\tau' \bar{\tau}'} =& E_{\tau' s'}^{\rm c'}(0) - E_{\bar{\tau}' \bar{s}'}^{\rm v'}(0) - \varepsilon_{{\rm X}'},\\
	E_{{\rm IX},s'\bar{s}}^{\tau' \bar{\tau}} =& E_{\tau' s'}^{\rm c'}(0) - E_{\bar{\tau} \bar{s}}^{\rm v}(0) - \varepsilon_{\rm IX},\\
	E_{{\rm IX}',s\bar{s}'}^{\tau \bar{\tau}'} =& E_{\tau s}^{\rm c}(0) - E_{\bar{\tau}' \bar{s}'}^{\rm v'}(0) - \varepsilon_{{\rm IX}'}.
\end{split}
\end{equation}
However, the former quantities depend directly on the intra- and interlayer band gaps, which may be underestimated by \emph{ab initio} methods. Instead, we use experimental values available in the literature, presented in Table \ref{tab:iX_exp}. These correspond to the A- and B-Xs of each layer, and the lowest-energy IX excitons of the heterostructure. In our present notation, these states are respectively ($\tau=+1$) ${\rm X}_{\downarrow\downarrow}^{++}$, ${\rm X}'{}_{\downarrow\downarrow}^{\tau'\tau'}$ and ${\rm IX}_{\downarrow\downarrow}^{\tau' +}$, where $\tau'=\pm$ for P and AP stacking. 

We estimate the energies of the higher states ${\rm IX}_{\uparrow\uparrow}^{\tau'+}$ and ${\rm IX}'{}_{ss}^{+\tau'}$ by combining the experimental values of Table \ref{tab:iX_exp} with the \emph{ab initio} spin-orbit splittings and conduction-band-edge offsets of Table \ref{tab:parameters}, and the binding energies of Table \ref{tab:excitons}. This same strategy is followed for all IXs in the case of MoTe${}_2$/MoSe${}_2$, for which no experimental results are available at the moment, using the experimental monolayer X energies in Table \ref{tab:excitons_exp}.

\begin{table}[t!]
\caption{Experimental values for the intralayer and interlayer A exciton energies in TMD heterobilayers, extracted from Refs.\ \onlinecite{twist_angle2018, MoSe2_WSe2_2015, WS2_MoS2_2014, MoSe2_WSe2_2017, MoSe2_MoS2_2018}. In our calculations for material pairs with more than one reported value, we take the average of the values shown.}
\begin{center}
\begin{tabular}{c | r @{.} l r @{.} l c} 
\hline\hline
\, & \multicolumn{4}{c}{Intralayer} & \, \\
\raisebox{1.5ex}[0pt]{${\rm MX_2/M'X_2'}$} & \multicolumn{2}{c}{${\rm MX_2}$ [eV]} & \multicolumn{2}{c}{${\rm M'X_2'}$ [eV]} & \raisebox{1.5ex}[0pt]{Interlayer [eV]} \\
\hline\hline
WS${}_2$/MoS${}_2$     & 1&97$^{\rm a}$      & 1&82$^{\rm a}$      & 1.42$^{\rm a}$ \\
MoSe${}_2$/MoS${}_2$  & 1&65${}^{\rm b}$    & 1&95${}^{\rm b}$    & 1.33$^{\rm b}$ \\
MoSe${}_2$/WS${}_2$   & 1&56${}^{\rm c}$    & 1&96${}^{\rm c}$    & 1.57$^{\rm c}$ \\
WSe${}_2$/MoSe${}_2$ & 1&65${}^{\rm d}$     & 1&57${}^{\rm d}$   & 1.35$^{\rm d}$, 1.39$^{\rm e}$ \\
\hline\hline
\end{tabular}\\
a.\ [\onlinecite{WS2_MoS2_2014}];\, 
b.\ [\onlinecite{MoSe2_MoS2_2018}];\, c.\ [\onlinecite{twist_angle2018}];\, d.\  [\onlinecite{MoSe2_WSe2_2017}];\, e.\ [\onlinecite{MoSe2_WSe2_2015}]
\end{center}
\label{tab:iX_exp}
\end{table}%
\begin{table}[h!]
\caption{Experimental values for the intralayer A and B exciton energies in monolayer TMDs, extracted from Refs.\ \onlinecite{X_in_main_four_2014,X_MoTe2_2014,X_MoTe2_2015,X_MoTe2_2016,X_WS2_and_WSe2_2013}. In our calculations for materials with more than one reported value, we take the average of the values shown.}
\begin{center}
\begin{tabularx}{0.95\columnwidth}{c|XX}
\hline\hline
\, & A exciton energy [eV] & B exciton energy [eV] \\
\hline\hline
MoS${}_2$     & 1.84$^{\rm a}$ & 2.00$^{\rm a}$ \\
MoSe${}_2$   & 1.58$^{\rm a}$ & 1.76$^{\rm a}$ \\
MoTe${}_2$   &  1.10--1.20$^{\rm b,c,d}$ & 1.35$^{\rm b}$, 1.44$^{\rm d}$ \\
WS${}_2$      &  2.01$^{\rm a}$, 1.99$^{\rm e}$ & 2.39$^{\rm a}$, 2.26$^{\rm e}$ \\
WSe${}_2$    &  1.66$^{\rm a}$, 1.63$^{\rm e}$ & 2.10$^{\rm a}$, 2.07$^{\rm e}$ \\
\hline\hline
\end{tabularx}\\
a.\ [\onlinecite{X_in_main_four_2014}];\,  b.\ [\onlinecite{X_MoTe2_2014}];\, c.\ [\onlinecite{X_MoTe2_2015}] ;\, d.\ [\onlinecite{X_MoTe2_2016}];\, e.\ [\onlinecite{X_WS2_and_WSe2_2013}]
\end{center}
\label{tab:excitons_exp}
\end{table}%

Similarly to carrier states, different intralayer and interlayer exciton species are mixed by the tunneling term $H_{\rm t}$ in Eq.\ \eqref{eq:fullH}. Using the simplified form \eqref{eq:T_simpl}, we obtain the matrix elements
\begin{equation}\label{eq:melemX}
\begin{split}
	\braoketf{{\rm IX}_{s\bar{s}}^{\tau'\tau}(\bar{\QQ})}{H_{\rm t}}{{\rm X}_{s\bar{s}}^{\tau\tau}(\QQ)} =& \sum_{\eta=0}^2\delta_{\QQ-\bar{\QQ},C_3^\eta\Delta\KK} M_{\rm IX\,X}^{\eta}(\QQ),\\
	\braoketf{{\rm IX}'{}_{s\bar{s}}^{\tau\tau'}(\bar{\QQ})}{H_{\rm t}}{{\rm X}_{s\bar{s}}^{\tau\tau}(\QQ)} =& \sum_{\eta=0}^2\delta_{\QQ-\bar{\QQ},C_3^\eta\Delta\KK} M_{\rm IX'\,X}^{\eta}(\QQ),\\
	\braoketf{{\rm IX}_{s\bar{s}}^{\tau'\tau}(\bar{\QQ})}{H_{\rm t}}{{\rm X}'{}_{s\bar{s}}^{\tau'\tau'}(\QQ)} =& \sum_{\eta=0}^2\delta_{\QQ-\bar{\QQ},C_3^\eta\Delta\KK} M_{\rm IX\,X'}^{\eta}(\QQ),\\
	\braoketf{{\rm IX}'{}_{s\bar{s}}^{\tau\tau'}(\bar{\QQ})}{H_{\rm t}}{{\rm X}'{}_{s\bar{s}}^{\tau'\tau'}(\QQ)} =& \sum_{\eta=0}^2\delta_{\QQ-\bar{\QQ},C_3^\eta\Delta\KK} M_{\rm IX'\,X'}^{\eta}(\QQ),
\end{split}
\end{equation}
with any other matrix elements between the relevant bright excitons being equal to zero. In Eq.\ \eqref{eq:melemX}, $\Delta\KK$ is chosen according to Eqs.\ \eqref{eq:DKP} and \eqref{eq:DKAP}, and we have defined
\begin{widetext}
\begin{equation}\label{eq:MMelem}
\begin{split}
	M_{\rm IX\,X}^{\eta}(\QQ) \equiv\quad& t_{\rm c}\exp{i\KK\cdot\rr_0}\exp{-iC_3^\eta\KK\cdot\rr_0} \int d^2r\,\exp{-i\left[\tfrac{m_{\rm c}}{m_{\rm c}+m_{\rm v}} - \tfrac{m_{\rm c'}}{m_{\rm c'}+m_{\rm v}} \right]\QQ\cdot\rr}\exp{-i\left[\tfrac{m_{\rm v}}{m_{\rm c'}+m_{\rm v}}\right]C_3^\eta\Delta\KK\cdot\rr}Y^*(\rr)X(\rr),\\
	M_{\rm IX'\,X}^{\eta}(\QQ) \equiv -&t_{\rm v}^*\exp{-i\KK\cdot\rr_0}\exp{iC_3^\eta\KK\cdot\rr_0} \int d^2r\,\exp{-i\left[\tfrac{m_{\rm c}}{m_{\rm c}+m_{\rm v}} - \tfrac{m_{\rm c}}{m_{\rm c}+m_{\rm v'}} \right]\QQ\cdot\rr}\exp{-i\left[\tfrac{m_{\rm c}}{m_{\rm c}+m_{\rm v'}}\right]C_3^\eta\Delta\KK\cdot\rr}Y'{}^*(\rr)X(\rr),\\
	M_{\rm IX\,X'}^{\eta}(\QQ) \equiv -& t_{\rm v}\exp{i\KK\cdot\rr_0}\exp{-iC_3^\eta\KK\cdot\rr_0} \int d^2r\,\exp{-i\left[\tfrac{m_{\rm c'}}{m_{\rm c'}+m_{\rm v'}} - \tfrac{m_{\rm c'}}{m_{\rm c'}+m_{\rm v}} \right]\QQ\cdot\rr}\exp{-i\left[\tfrac{m_{\rm c'}}{m_{\rm c'}+m_{\rm v}}\right]C_3^\eta\Delta\KK\cdot\rr}Y^*(\rr)X'(\rr),\\
	M_{\rm IX'\,X'}^{\eta}(\QQ) \equiv\quad& t_{\rm c}^*\exp{-i\KK\cdot\rr_0}\exp{iC_3^\eta\KK\cdot\rr_0} \int d^2r\,\exp{-i\left[\tfrac{m_{\rm v'}}{m_{\rm c}+m_{\rm v'}} - \tfrac{m_{\rm v'}}{m_{\rm c'}+m_{\rm v'}} \right]\QQ\cdot\rr}\exp{-i\left[\tfrac{m_{\rm v'}}{m_{\rm c}+m_{\rm v'}}\right]C_3^\eta\Delta\KK\cdot\rr}Y'{}^*(\rr)X'(\rr),
\end{split}
\end{equation}
\end{widetext}
with $X(\rr)$, $X'(\rr)$, $Y(\rr)$ and $Y'(\rr)$ the real-space relative motion wave functions, given by the inverse Fourier transforms of $X_{\qq}$, $X_{\qq}'$, $Y_{\qq}$ and $Y_{\qq}'$.
\begin{figure*}[t!]
\begin{center}
\includegraphics[width=1.95\columnwidth]{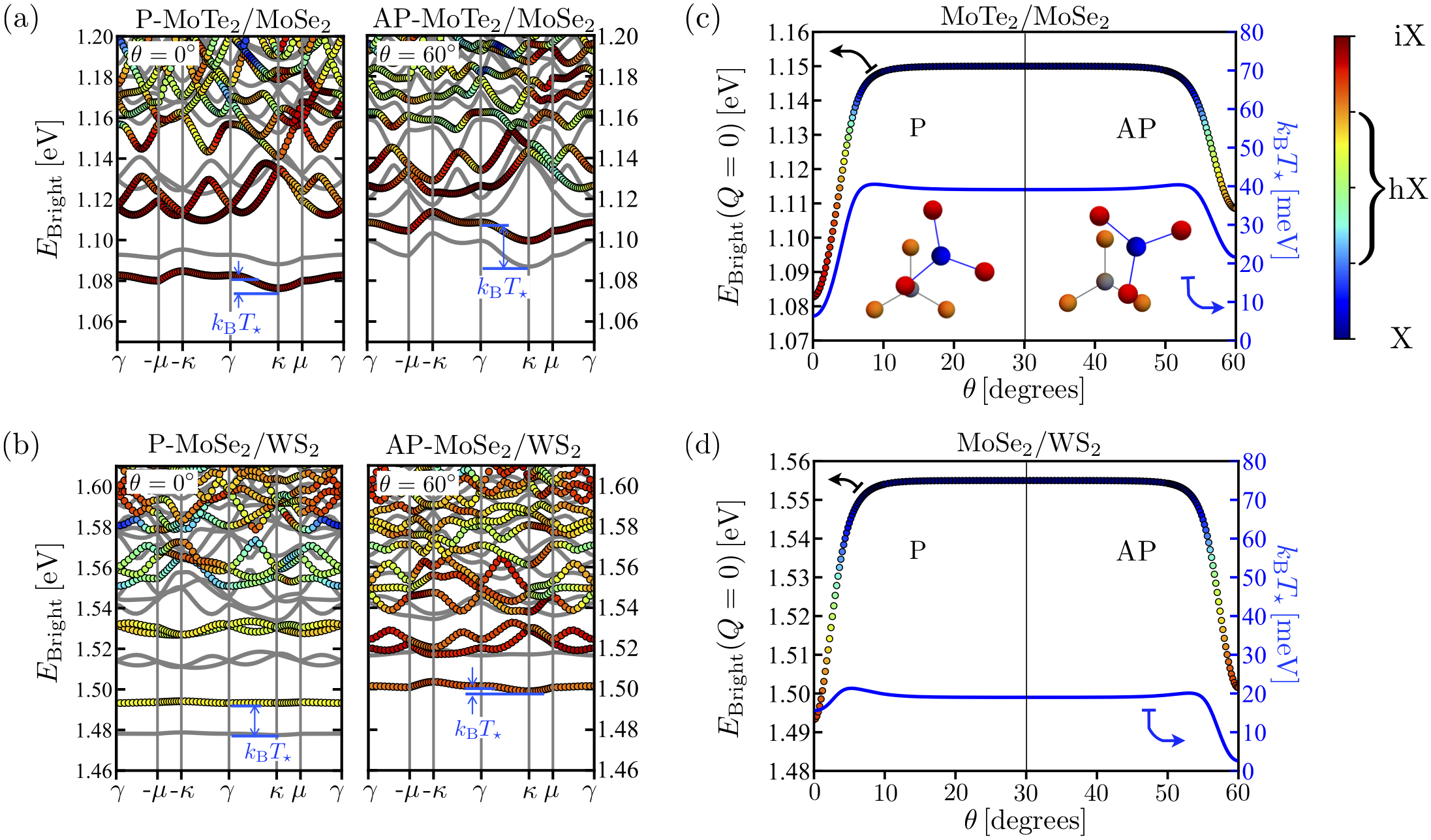}
\caption{(a) and (b) Main ($\tau=1$) bright exciton moir\'e bands for perfectly aligned and anti-aligned MoTe${}_2$/MoSe${}_2$ and MoSe${}_2$/WS${}_2$. The dark exciton bands are also shown in gray. For reference, the atomic arrangements of P- and AP-stacking are presented in the panel (c) insets. (c) and (d) Energy and activation temperature $T_\star$ of the main optically-active ($\QQ=0$) bright exciton state, as functions of the twist angle. In all panels, the symbol color represents the state composition, with red (blue) corresponding to pure interlayer (${\rm MX_2}$-intralayer) excitons. Intermediate colors indicate strong mixing of IX and X species, corresponding to hybridized excitons, hX.}
\label{fig:Xdispersions}
\end{center}
\end{figure*}

The expressions in \eqref{eq:MMelem} can be simplified by noting that, in each case, the difference of mass ratios appearing in the argument of the first exponential is much smaller than one for the heterostructures discussed (see Table \ref{tab:parameters}). Then, assuming two-dimensional $s$-states for the X and IX wave functions\cite{berkelbach_variational}, with corresponding Bohr radii $a_{\rm X}$, $a_{\rm X'}$, $a_{\rm IX}$ and $a_{\rm IX'}$, yields the momentum-independent expressions
\begin{widetext}
\begin{equation}\label{eq:Xmelem}
\begin{split}
	M_{\rm IX\,X}^{\eta} &\approx\quad \frac{4\,t_{\rm c}\exp{i\KK\cdot\rr_0}\exp{-iC_3^\eta\KK\cdot\rr_0}}{a_{\rm X}a_{\rm IX}}\left(\frac{a_{\rm X}+a_{\rm IX}}{a_{\rm X}a_{\rm IX}} \right)\,\, \left[\left(\frac{a_{\rm X}+a_{\rm IX}}{a_{\rm X}a_{\rm IX}} \right)^2 + \frac{m_{\rm v}^2 \Delta K ^2}{(m_{\rm c'}+m_{\rm v})^2} \right]^{-3/2},\\
	M_{\rm IX'\,X}^{\eta} &\approx -\frac{4\,t_{\rm v}^*\exp{-i\KK\cdot\rr_0}\exp{iC_3^\eta\KK\cdot\rr_0}}{a_{\rm X}a_{\rm IX'}}\,\left(\frac{a_{\rm X}+a_{\rm IX'}}{a_{\rm X}a_{\rm IX'}} \right)\,\left[\left(\frac{a_{\rm X}+a_{\rm IX'}}{a_{\rm X}a_{\rm IX'}} \right)^2 + \frac{m_{\rm c}^2 \Delta K^2}{(m_{\rm c}+m_{\rm v'})^2} \right]^{-3/2},\\
	M_{\rm IX\,X'}^{\eta} &\approx -\frac{4\,t_{\rm v}\exp{i\KK\cdot\rr_0}\exp{-iC_3^\eta\KK\cdot\rr_0}}{a_{\rm X'}a_{\rm IX}}\left(\frac{a_{\rm X'}+a_{\rm IX}}{a_{\rm X'}a_{\rm IX}} \right)\,\,\left[\left(\frac{a_{\rm X'}+a_{\rm IX}}{a_{\rm X'}a_{\rm IX}} \right)^2 + \frac{m_{\rm c'}^2 \Delta K^2}{(m_{\rm c'}+m_{\rm v})^2} \right]^{-3/2},\\
	M_{\rm IX'\,X'}^{\eta} &\approx\quad \frac{4\,t_{\rm c}^*\exp{-i\KK\cdot\rr_0}\exp{iC_3^\eta\KK\cdot\rr_0}}{a_{\rm X'}a_{\rm IX'}}\left(\frac{a_{\rm X'}+a_{\rm IX'}}{a_{\rm X'}a_{\rm IX'}} \right) \left[\left(\frac{a_{\rm X'}+a_{\rm IX'}}{a_{\rm X'}a_{\rm IX'}} \right)^2 + \frac{m_{\rm v'}^2 \Delta K^2}{(m_{\rm c}+m_{\rm v'})^2} \right]^{-3/2}.
\end{split}
\end{equation}
\end{widetext}
Analogously to Eq.\ \eqref{eq:T_simpl}, Eq.\ \eqref{eq:melemX} defines a mBZ for excitons, where X and IX states with center-of-mass momenta separated by moir\'e Bragg vectors $\bb_{mn}$ mix. The resulting intralayer-interlayer hybridization model can be solved by direct diagonalization within the mBZ defined in Fig.\ \ref{fig:Figure1}.

\begin{figure*}[t!]
\begin{center}
\includegraphics[width=2\columnwidth]{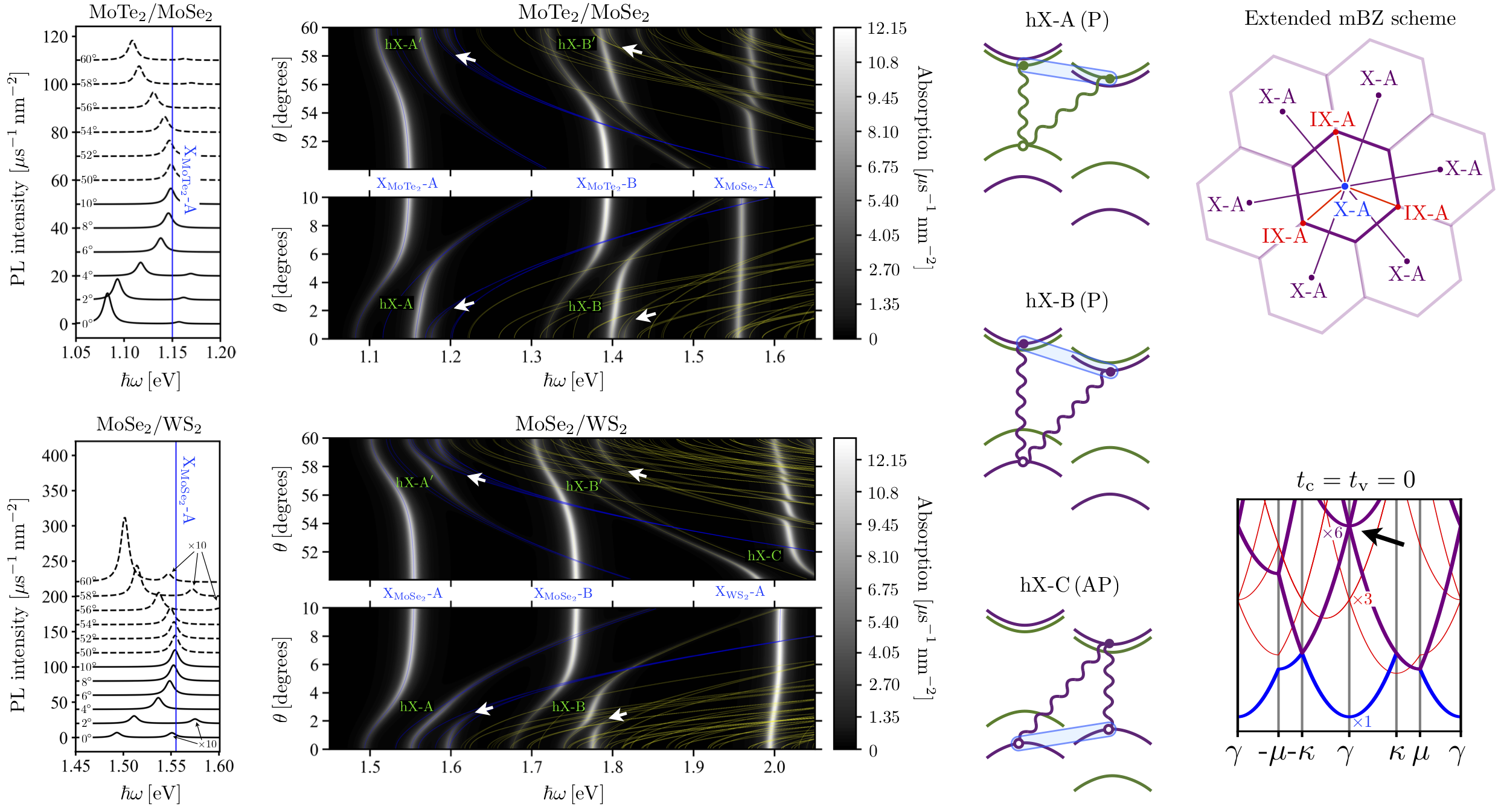}
\caption{Photoluminescence (PL) at room temperature (left) and absorption spectra (center) as functions of twist angle for MoTe${}_2$/MoSe${}_2$ (top) and MoSe${}_2$/WS${}_2$ (bottom), close to parallel and anti-parallel alignment. PL curves for different twist angles are offset for clarity, and the decoupled ${\rm MX_2}$-A exciton energy, obtained at $\theta = 30^\circ$, is indicated by the vertical blue line. For MoTe${}_2$/MoSe${}_2$, a second PL feature is clearly visible at room temperature, for $\theta \approx 0^\circ$ and $|\theta - 60^\circ |\approx 0^\circ$. A faint second PL peak appears also in MoSe${}_2$/WS${}_2$, shown magnified (multiplied by a factor of 10) in the figure. These secondary spectral features are thermally activated, and disappear at low temperatures. The full low-energy exciton spectrum is overlaid on top of each absorption map (blue and yellow curves), showing multiple momentum-dark exciton states. Blue curves correspond to the 10 minibands obtained from the first ${\rm MX_2}$-A exciton band (1 line), the lowest ${\rm IX}$ bands (3 lines), and the first folding of the ${\rm MX_2}$-A band into the mBZ (6 lines), shown schematically in the top- and bottom-right panels. In the top-right panel, same-color lines indicate degenerate exciton states at the $\gamma$ point. Multiple hX states are identified by avoided crossings of the exciton minibands close to (anti-)alignment, and sketched on the center-right panels. White arrows indicate absorption signatures from photon umklapp proceses associated to one of the latter 6 exciton bands, which is maximally hybridized with the first ${\rm MX_2}$-A exciton, thus becoming bright. This absorption line is direct evidence of the moir\'e superlattice. All PL and absorption line shapes are assumed of Lorentzian form, with broadening of $5\,{\rm meV}$.}
\label{fig:AbsVsTh}
\end{center}
\end{figure*}

\subsection{hX formed by IX hybridization with the interlayer A exciton}\label{sec:AX}
Fig.\ \ref{fig:Xdispersions} shows the moir\'e band structures for the ($\tau=1$) X and IX states in perfectly aligned (P-type) and anti-aligned (AP-type) MoTe${}_2$/MoSe${}_2$ and MoSe${}_2$/WS${}_2$. For both material pairs, the flatness of the lowest exciton bands is a consequence of the resonant condition between the intralayer A exciton and the ${\rm IX}_{\downarrow\downarrow}^{\tau'+}$ (detunings are approximately $10\,{\rm meV}$ for MoSe${}_2$/WS${}_2$ and $40\,{\rm meV}$ for MoTe${}_2$/MoSe${}_2$), combined with the reduced size of the mBZ at perfect alignment or anti-alignment. The symbol colors in Fig.\ \ref{fig:Xdispersions} indicate the composition of the exciton state, with blue or red representing a large X or IX component, and intermediate colors corresponding to hXs. Note that multiple high-energy $\gamma$-point hXs appear in both materials, some of which are optically active and should contribute to the heterostructure's absorption spectrum, as discussed in Figs.\ \ref{fig:AbsVsTh}, \ref{fig:hXAbsEz_MoTe2_MoSe2} and \ref{fig:hXAbsEz_MoSe2_WS2}.

The evolution with twist angle of the lowest bright $\gamma$ exciton energy and state composition are shown in Figs.\ \ref{fig:Xdispersions}(c) and \ref{fig:Xdispersions}(d), displaying sharp variations of approximately $50$ to $70\,{\rm meV}$ in both material pairs, when $\theta$ departs from $0^\circ$ or $60^\circ$. For those twist angle ranges, the lowest bright $\gamma$ exciton is hX-A, formed by resonant hybridization of an IX with the A intralayer exciton, whereas for strong misalignment angles $10^\circ \lesssim \theta \lesssim 50^\circ$, it is the fully bright X-A state. The slight asymmetry of the curve, especially visible for MoTe${}_2$/MoSe${}_2$ [Fig.\ \ref{fig:Xdispersions}(c)], is caused by the opposite conduction-band spin-orbit splitting in the $\tau'=\pm1$ valleys of the ${\rm M'X_2'}$ layer, resulting in different IX energies for P and AP configurations. The blue curves in Figs.\ \ref{fig:Xdispersions}(c) and \ref{fig:Xdispersions}(d) show the $\gamma$ exciton activation energy $k_{\rm B}T_\star$ as a function of twist angle, indicating what temperature $T_\star$ is required to populate these exciton states, and produce photoluminescence. Note that for the case of MoSe${}_2$/WS${}_2$, this is always  lower than room temperature ($k_{\rm B}T_\star < 25.8\,{\rm meV}$), reaching values as low as $\sim 1\,{\rm meV}$ at $\theta=60^\circ$. Based on these results, we have evaluated the photoluminescence spectra of both heterostructures (Appendix \ref{app:PL}), which we present for different twist angles in Fig.\ \ref{fig:AbsVsTh}.

The latter Figure also shows the MoTe${}_2$/MoSe${}_2$ and MoSe${}_2$/WS${}_2$ absorption spectra for varying interlayer twist angle (Appendix \ref{app:Abs}), which captures the full optical spectrum of the heterostructure. For both alignment cases, the lowest hX pair, labeled hX-A, is formed by resonant hybridization of an IX with the ${\rm MX}_2$ intralayer A exciton (${\rm IX}_{\downarrow\downarrow}^{+\pm}$ and ${\rm X}_{\downarrow\downarrow}^{++}$, respectively), driven by interlayer electron hopping. The IX component of each hX-A state is formed by an electron and hole residing in different layers, and thus separated by a distance of approximately $6\,\Ams$ (Table \ref{tab:parameters}). This IX component possesses an electric dipole moment of $30$ to $50\,{\rm D}$ ($\sim10^{-18}\,{\rm C} \cdot \Ams$), and will couple to out-of-plane electric fields $E_z$, modifying the hX energies (Stark shift) and state compositions, and ultimately splitting them into pure Xs and IXs. This is shown in Figs.\ \ref{fig:hXAbsEz_MoTe2_MoSe2} and \ref{fig:hXAbsEz_MoSe2_WS2}, where we present the absorption spectra of P- and AP-type MoTe${}_2$/MoSe${}_2$ and MoSe${}_2$/WS${}_2$, respectively, for fixed twist angle and varying field strength $E_z$. In each case, the IX can be easily identified within the lower-energy multiplet of lines by its Stark shift, and direct correspondence with the reference, free IX line, shown in red. X-A, on the other hand, can be identified by its lack of a Stark shift, recovering its unperturbed value at large, negative $E_z$ (blue line).

\begin{figure}[h!]
\begin{center}
\includegraphics[width=0.86\columnwidth]{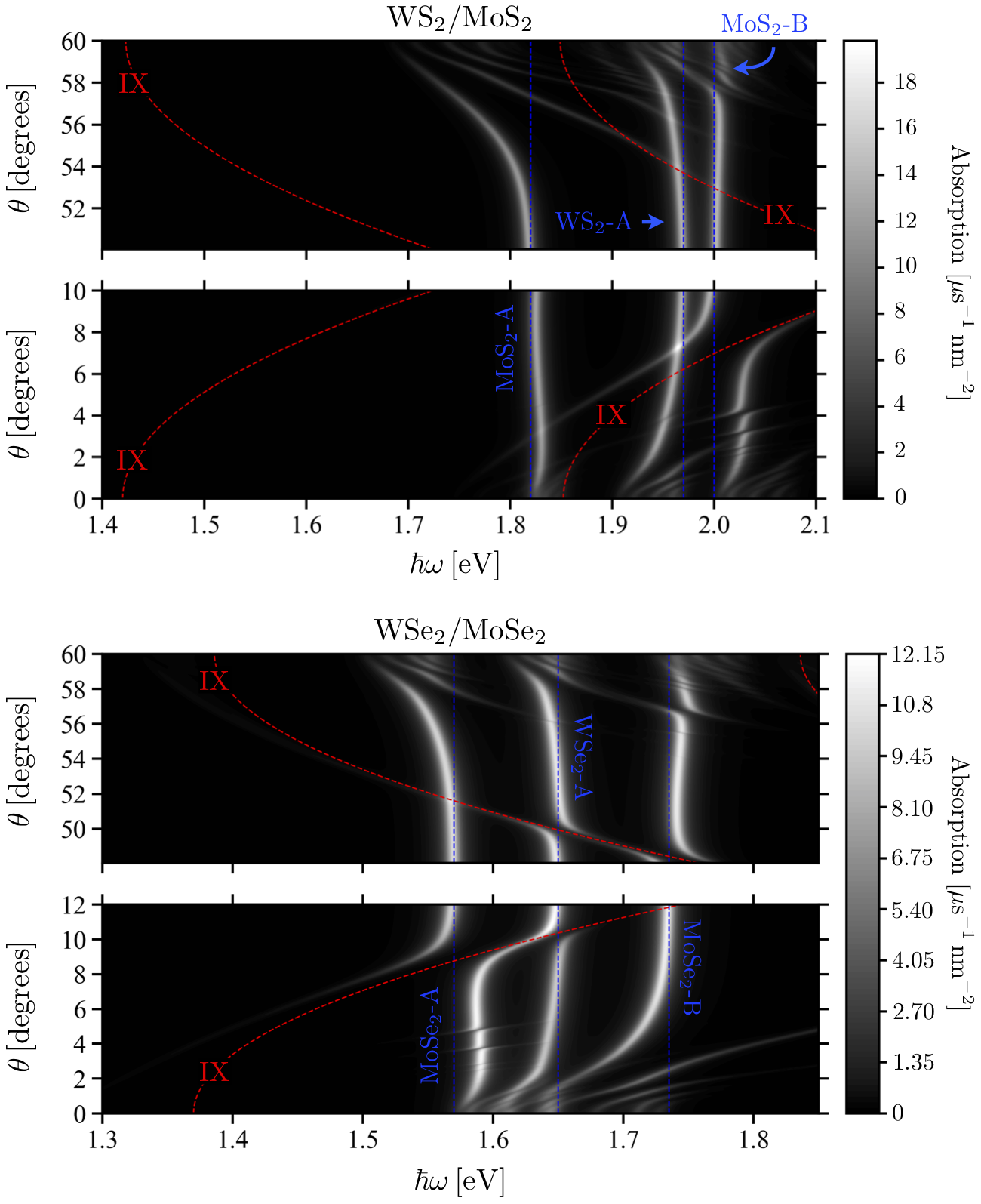}
\caption{Absorption spectra of WS${}_2$/MoS${}_2$ (top) and WSe${}_2$/MoSe${}_2$ (bottom). For reference, blue and red curves show the energies of different intra- and interlayer excitons, respectively, in the limit of $t_{\rm c}=t_{\rm v}=0$. Whereas for P-WSe${}_2$/MoSe${}_2$ a faint $\mathrm{IX}_{\downarrow\downarrow}^{++}$ line is visible at low energies, labeled IX, the corresponding line is absent in P-WS${}_2$/MoS${}_2$ due to its larger detuning with the MoS${}_2$-A exciton. For both material pairs, the intralayer exciton lines display complex fine structures close to perfect alignment and anti-alignment, coming from moir\'e exciton minibands that mix strongly with the main A or B exciton.}
\label{fig:NonRes}
\end{center}
\end{figure}

The optical spectra of hXs are dominated by their large intralayer-exciton component, leading to identical optical selection rules as the monolayers\cite{wangyao_spin_valley}. This is in stark contrast to IXs in TMD heterostructures with non-resonant band edges, whose optical selection rules are determined by the local stacking\cite{wangyao_interlayer_X,hongyi_moire, Wu2017} (see Fig.\ \ref{fig:moire_stacking}), and dominated by the weak interlayer tunneling matrix elements $t_{\rm cv}$ and $t_{\rm vc}$, discussed in Sec.\ \ref{sec:model}. In addition to the case of MoSe${}_2$/MoS${}_2$, shown in Fig.\ \ref{fig:Summary}, Fig.\ \ref{fig:NonRes} shows the absorption spectra of WS${}_2$/MoS${}_2$ and WSe${}_2$/MoSe${}_2$, where intra- and interlayer excitons are strongly off resonance, as reported in Table \ref{tab:iX_exp}. For WS${}_2$/MoS${}_2$, the lowest-energy interlayer exciton line ($\mathrm{IX}_{\downarrow\downarrow}^{\pm+}$ for P and AP stacking, respectively) is completely absent in our approximation ($t_{\rm vc}=t_{\rm cv}=0$) due to negligible hybridization with the bright intralayer WS${}_2$-A and MoS${}_2$-A excitons. In WSe${}_2$/MoSe${}_2$, the lowest absorption peak visible corresponds to $\mathrm{IX}_{\downarrow\downarrow}^{\pm+}$, which in our approximation gains oscillator strength only for large twist angles as it approaches the energy of the MoSe${}_2$-A exciton, even showing an avoided crossing at photon energies between $1.5$ and $1.6\,{\rm eV}$ for $\theta \approx 8^\circ$. Note that, due to the ordering of the intralayer exciton energies in these heterobilayers, this avoided crossing is produced by interlayer hole tunneling, as opposed to electron tunneling, discussed below in the context of MoSe${}_2$/WS${}_2$ and MoTe${}_2$/MoSe${}_2$ heterostructures.  Surprisingly, the intralayer exciton lines of perfectly aligned P- and AP-stacked WS${}_2$/MoS${}_2$ and WSe${}_2$/MoSe${}_2$ display intricate fine structures, corresponding to higher intralayer exciton minibands, which should be discernible at low temperatures, and appear as anomalous broadening of the intralayer exciton lines in high-temperature experiments.

\subsection{hX formed by IX hybridization with the intralayer B exciton}\label{sec:BX}
\begin{figure}[t!]
\begin{center}
\includegraphics[width=0.9\columnwidth]{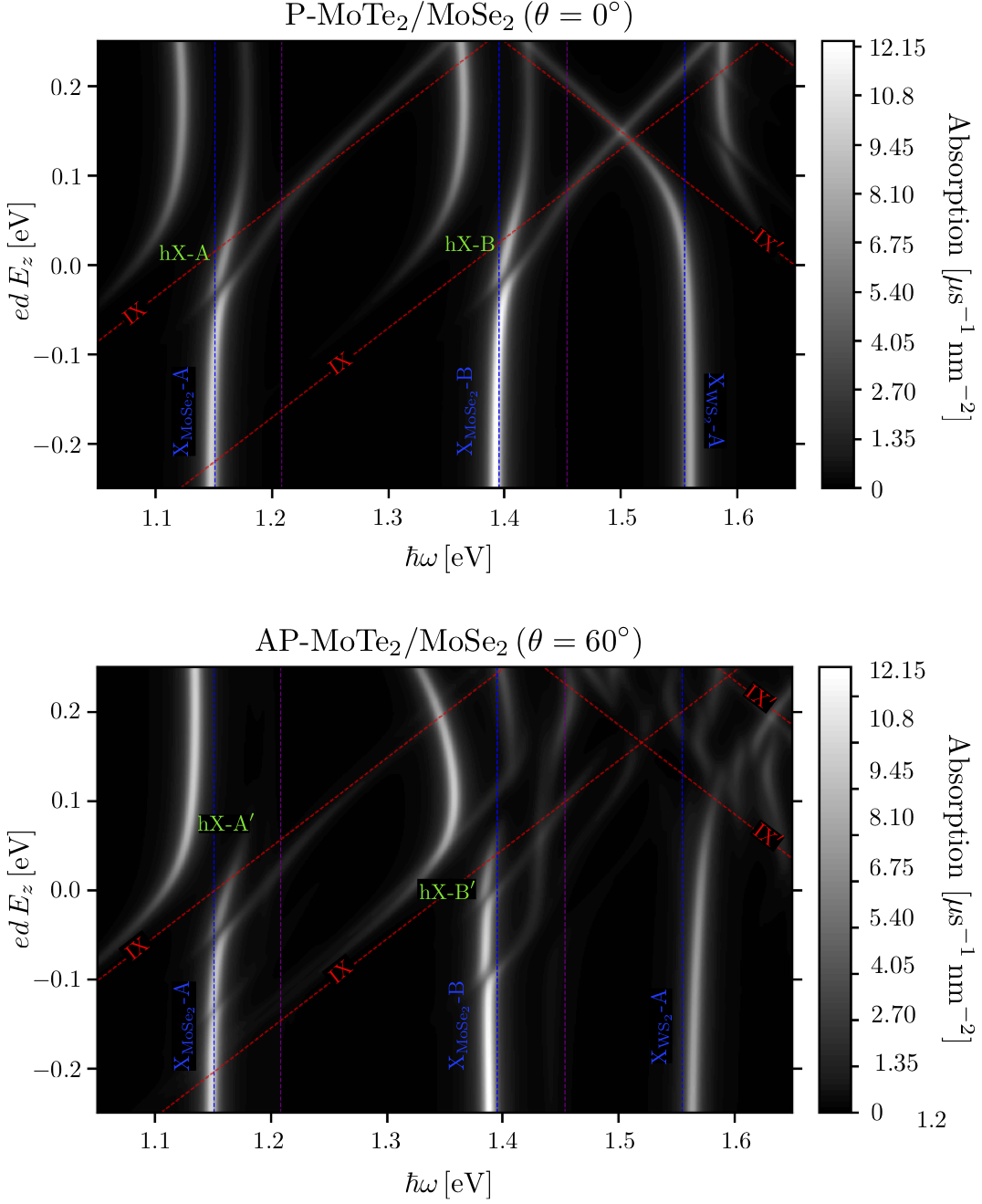}
\caption{Absorption spectra of parallel  (top) and anti-parallel (bottom) MoTe${}_2$/MoSe${}_2$, for varying out-of-plane electric field. The energies of various intralayer and interlayer exciton states, in the limit of $t_{\rm c}=t_{\rm v}=0$, are shown for reference. Line shapes are assumed of Lorentzian form, with broadening of $5\,{\rm meV}$.}
\label{fig:hXAbsEz_MoTe2_MoSe2}
\end{center}
\end{figure}

\begin{figure}[t!]
\begin{center}
\includegraphics[width=0.9\columnwidth]{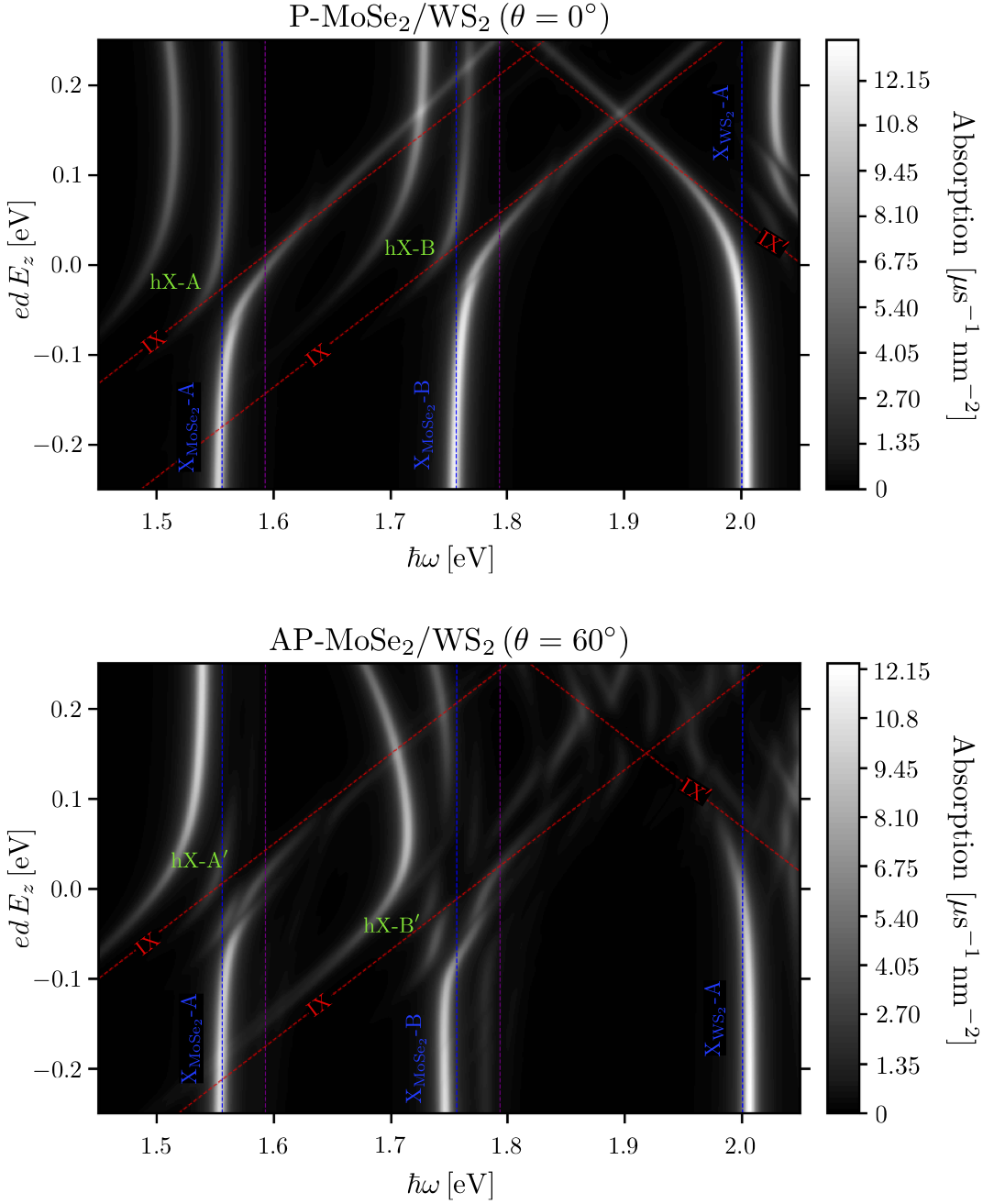}
\caption{Absorption spectra of parallel  (top) and anti-parallel (bottom) MoSe${}_2$/WS${}_2$, for varying out-of-plane electric field. The energies of various intralayer and interlayer exciton states, in the limit of $t_{\rm c}=t_{\rm v}=0$, are shown for reference. Line shapes are assumed of Lorentzian form, with broadening of $5\,{\rm meV}$.}
\label{fig:hXAbsEz_MoSe2_WS2}
\end{center}
\end{figure}

In addition to the sharp hX energy modulation shown in Figs.\ \ref{fig:Xdispersions}(c) and \ref{fig:Xdispersions}(d), the higher-energy features in the absorption spectra also reflect the formation of a secondary pair of hXs for P stacking, and two secondary pairs of hXs for AP stacking. In the former case, the pair of lines labeled hX-B originates as the ${\rm MX}_2$ intralayer B exciton (${\rm X}_{\uparrow\uparrow}^{++}$) hybridizes resonantly with ${\rm IX}_{\uparrow\uparrow}^{++}$, through interlayer electron tunneling. This type of mixing also occurs in the latter case of AP stacking, producing the pair of lines labeled hX-B' in Fig.\ \ref{fig:AbsVsTh}. Surprisingly, an additional near resonance between the ${\rm M'X_2'}$ A exciton (${\rm X}'{}_{\uparrow\uparrow}^{--}$) and the interlayer exciton ${\rm IX}_{\uparrow\uparrow}^{-+}$ appears for relatively large misalignment angles, $60^\circ - \theta \approx 8^\circ$, giving rise to a third pair of hXs, labeled hX-C. By contrast to all previously discussed cases, hX-C are produced by interlayer tunneling of holes, as sketched in the right panels of Fig.\ \ref{fig:AbsVsTh}. This type of mixing is possible only for AP stacking in both MoTe${}_2$/MoSe${}_2$ and MoSe${}_2$/WS${}_2$, where the top valence band of the ${\rm MX_2}$ layer and that of the ${\rm M'X_2'}$ have opposite spin quantum numbers. The smooth twist-angle crossover from hX-B to hX-C  produces a clear absorption line that shifts with increased misalignment by a remarkable $200\,{\rm meV}$. This is enabled by the large interlayer hole tunneling matrix element $|t_{\rm v}|>|t_{\rm c}|$, which gives strong mixing between the ${\rm M'X_2'}$ A exciton and ${\rm IX}_{\uparrow\uparrow}^{-+}$ even at $\theta \approx 60^\circ$, where the detuning between these two exciton states is relatively large. Figs.\ \ref{fig:hXAbsEz_MoTe2_MoSe2} and \ref{fig:hXAbsEz_MoSe2_WS2} show that sufficiently large, positive electric fields bring down the higher-energy ${\rm IX}'{}_{ss}^{+\pm}$ states, due to their positive electric dipole moments (see Fig.\ \ref{fig:XXpYZ}), allowing them to also hybridize with the ${\rm M'X_2'}$ B exciton to produce the complex absorption signatures appearing in the top-right corner of each panel. 

\subsection{hX fine structure due to mSL-induced umklapp electron-photon interaction}\label{sec:Afolding}
Perhaps, the most important features appearing in Fig.\ \ref{fig:AbsVsTh} are the additional absorption lines indicated by white arrows, which accompany the hX-A and hX-B signatures for $\theta\approx 0^\circ$ or $60^\circ$, especially pronounced for MoSe${}_2$/WS${}_2$. These lines originate from the minibands obtained by the first folding of the A or B intralayer exciton dispersion into the mBZ, to the new $\gamma$ point. These states become optically active due to umklapp photon absorption processes, in which a mSL Bragg vector is transferred to the crystal, thus making exciton states with finite momenta $\bb_{0,\pm1}$, $\bb_{\pm1,0}$ and $\bb_{\pm1,\pm1}$ bright. This is depicted in the right panels of Fig.\ \ref{fig:AbsVsTh}. The presence of these lines can provide direct evidence of moir\'e superlattice effects. We have verified that, for aligned MoTe${}_2$/MoSe${}_2$ and MoSe${}_2$/WS${}_2$ heterostructures, the three-peak spectra produced by the two main hX lines and the third umklapp line are robust to variation of the main theoretical parameters considered (the interlayer electron hopping energy $t_{\rm c}$, and the conduction-band masses of the two TMD layers), and should thus be visible in real samples with different preparation methods, and under different conditions. This is shown in Figures \ref{fig:hXvsMassMoTe2} and \ref{fig:hXvsMassMoSe2} of Appendix \ref{app:parametric}.

\begin{figure}[t!]
\begin{center}
\includegraphics[width=\columnwidth]{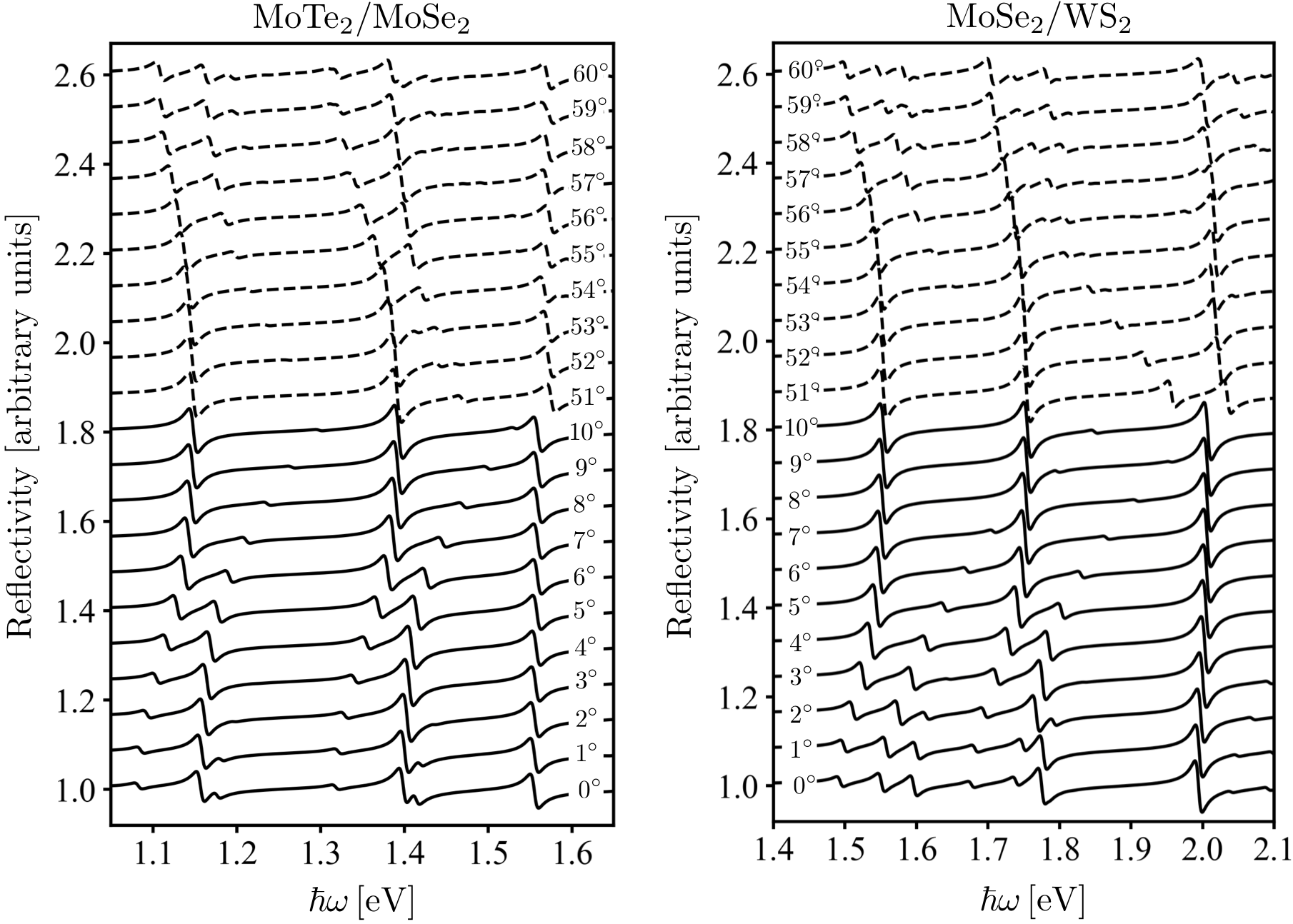}
\caption{Reflectivity spectra, typically measured in optics experiments, of MoTe${}_2$/MoSe${}_2$ and MoSe${}_2$/WS${}_2$ for various twist angles near perfect alignment and anti-alignment. The absorption peaks predicted in Fig.\ \ref{fig:AbsVsTh} translate into double features, where the exciton energy can be read as the point of maximum negative derivative between the peak and the trough.}
\label{fig:Reflectivity}
\end{center}
\end{figure}

Figs.\ \ref{fig:hXAbsEz_MoTe2_MoSe2} and \ref{fig:hXAbsEz_MoSe2_WS2} show that the umklapp photon absorption lines disappear below some negative electric field value, but remain bright for $E_z>0$, and can be identified with the purple reference lines corresponding to the energy of the first folded A or B $\gamma$-exciton, up to an overall red shift due to interaction with higher-energy IX minibands. The splitting of this triad formed by the two hX-A or -B lines and the first A or B umklapp photon absorption line, into a pair of lines that do not undergo a Stark shift, plus one that does, can serve to identify both the hybridized exciton physics and the moir\'e superlattice effects in an experimental setting. As a final remark, we note that the optical spectra of TMD heterostructures are normally measured through their reflectivity, rather than absorption properties. Figure \ref{fig:Reflectivity} shows our prediction for the reflectivity spectra of MoTe${}_2$/MoSe${}_2$ and MoSe${}_2$/WS${}_2$, obtained from our absorption calculations via a Kramers-Kronig relation \cite{Landau_ED}.

\section{Conclusions}\label{sec:conclusions}
We have studied the interplay between band alignment and the presence of emergent moir\'e superlattices (mSL) in twisted heterobilayers of transition-metal dichalcogenides (TMDs). Starting from a microscopic interlayer tunneling Hamiltonian, we have derived effective harmonic-potentials based on perturbation theory, to describe the effects of moir\'e patterns on the electron and hole bands of TMD heterostructures with large interlayer offsets between carrier band edges. We have shown that this approach fails in TMD homobilayers, and in heterostructures such as MoSe${}_2$/WS${}_2$ and MoTe${}_2$/MoSe${}_2$, where bands of the two constituent monolayers hybridize resonantly. Our results show that the influence of higher moir\'e superlattice minibands for the low-energy electron band structure in these heterobilayers becomes increasingly important as the interlayer band-edges offset is reduced; in other words, that resonant interlayer hybridization amplifies the moir\'e superlattice effects on the electronic structure. By treating hybridization effects exactly, we have predicted the appearance of van Hove singularities near the conduction miniband edges in these materials close to perfect alignment, of potential interest for the study of strongly correlated electron physics in TMDs.

We have also developed a general description of low-energy excitons in TMD heterobilayers, and found that the small interlayer conduction-band-edge detunings in MoSe${}_2$/WS${}_2$ and MoTe${}_2$/MoSe${}_2$ result in nearly degenerate intralayer and interlayer exciton states, with the resonant condition further enhanced by the difference in binding energies of these two exciton species. This gives rise to hybridized excitons (hXs), which inherit the brightness of intralayer excitons and the polar nature of interlayer excitons. Presently, our model neglects the effects of periodic strain that may develop in each layer of the heterostructure, and which can affect the energies of the band edges. Such effects may be most important for the best lattice-matched and highly aligned structures (\emph{e.g.}, WSe${}_2$/MoSe${}_2$ and WSe${}_2$/WS${}_2$), as well as homobilayers, and should be added on top of the hybridization effects studied here. Using experimental values for the exciton energies reported in the literature on TMD heterobilayers, we have evaluated the full optical spectra of MoSe${}_2$/WS${}_2$ and MoTe${}_2$/MoSe${}_2$ heterostructures, and made predictions for explicit signatures of strong intralayer-interlayer exciton hybridization, and of the presence of the moir\'e superlattice. Hence, we predict that mSL-modified hXs should be ubiquitous to TMD heterostructures, and dominate the low-energy spectrum of closely aligned TMD heterobilayers with near resonant band edges, in agreement with recent experimental developments \cite{twist_angle2018}.

\acknowledgements{The authors acknowledge funding by the EU Graphene Flagship Project, ERC Synergy Grant Hetero2D, EPSRC Grand Challenges Grant, Lloyd Register Foundation Nanotechnology Grant, European Quantum Technologies Flagship Project, and EPSRC grant EP/P026850/1. We would like to thank E.M.\ Alexeev, M.\ Danovich, F.H.L.\ Koppens, A.\ Kozikov, K.S.\ Novoselov, M.\ Potemski, A.I.\ Tartakovskii, J.R.\ Wallbank and V.\ Z\'olyomi for fruitful discussions at various stages of this work.}

\newpage

\bibliography{references}

\begin{thebibliography}{71}%
\makeatletter
\providecommand \@ifxundefined [1]{%
 \@ifx{#1\undefined}
}%
\providecommand \@ifnum [1]{%
 \ifnum #1\expandafter \@firstoftwo
 \else \expandafter \@secondoftwo
 \fi
}%
\providecommand \@ifx [1]{%
 \ifx #1\expandafter \@firstoftwo
 \else \expandafter \@secondoftwo
 \fi
}%
\providecommand \natexlab [1]{#1}%
\providecommand \enquote  [1]{``#1''}%
\providecommand \bibnamefont  [1]{#1}%
\providecommand \bibfnamefont [1]{#1}%
\providecommand \citenamefont [1]{#1}%
\providecommand \href@noop [0]{\@secondoftwo}%
\providecommand \href [0]{\begingroup \@sanitize@url \@href}%
\providecommand \@href[1]{\@@startlink{#1}\@@href}%
\providecommand \@@href[1]{\endgroup#1\@@endlink}%
\providecommand \@sanitize@url [0]{\catcode `\\12\catcode `\$12\catcode
  `\&12\catcode `\#12\catcode `\^12\catcode `\_12\catcode `\%12\relax}%
\providecommand \@@startlink[1]{}%
\providecommand \@@endlink[0]{}%
\providecommand \url  [0]{\begingroup\@sanitize@url \@url }%
\providecommand \@url [1]{\endgroup\@href {#1}{\urlprefix }}%
\providecommand \urlprefix  [0]{URL }%
\providecommand \Eprint [0]{\href }%
\providecommand \doibase [0]{http://dx.doi.org/}%
\providecommand \selectlanguage [0]{\@gobble}%
\providecommand \bibinfo  [0]{\@secondoftwo}%
\providecommand \bibfield  [0]{\@secondoftwo}%
\providecommand \translation [1]{[#1]}%
\providecommand \BibitemOpen [0]{}%
\providecommand \bibitemStop [0]{}%
\providecommand \bibitemNoStop [0]{.\EOS\space}%
\providecommand \EOS [0]{\spacefactor3000\relax}%
\providecommand \BibitemShut  [1]{\csname bibitem#1\endcsname}%
\let\auto@bib@innerbib\@empty
\bibitem [{\citenamefont {Geim}\ and\ \citenamefont
  {Grigorieva}(2013)}]{GeimNature2013}%
  \BibitemOpen
  \bibfield  {author} {\bibinfo {author} {\bibfnamefont {A.~K.}\ \bibnamefont
  {Geim}}\ and\ \bibinfo {author} {\bibfnamefont {I.~V.}\ \bibnamefont
  {Grigorieva}},\ }\href@noop {} {\bibfield  {journal} {\bibinfo  {journal}
  {Nature}\ }\textbf {\bibinfo {volume} {499}},\ \bibinfo {pages} {419}
  (\bibinfo {year} {2013})}\BibitemShut {NoStop}%
\bibitem [{\citenamefont {Novoselov}\ \emph {et~al.}(2016)\citenamefont
  {Novoselov}, \citenamefont {Mishchenko}, \citenamefont {Carvalho},\ and\
  \citenamefont {Castro~Neto}}]{vdW_review}%
  \BibitemOpen
  \bibfield  {author} {\bibinfo {author} {\bibfnamefont {K.~S.}\ \bibnamefont
  {Novoselov}}, \bibinfo {author} {\bibfnamefont {A.}~\bibnamefont
  {Mishchenko}}, \bibinfo {author} {\bibfnamefont {A.}~\bibnamefont
  {Carvalho}}, \ and\ \bibinfo {author} {\bibfnamefont {A.~H.}\ \bibnamefont
  {Castro~Neto}},\ }\href {\doibase 10.1126/science.aac9439} {\bibfield
  {journal} {\bibinfo  {journal} {Science}\ }\textbf {\bibinfo {volume} {353}}
  (\bibinfo {year} {2016}),\ 10.1126/science.aac9439}\BibitemShut {NoStop}%
\bibitem [{\citenamefont {Ponomarenko}\ \emph {et~al.}(2013)\citenamefont
  {Ponomarenko}, \citenamefont {Gorbachev}, \citenamefont {Yu}, \citenamefont
  {Elias}, \citenamefont {Jalil}, \citenamefont {Patel}, \citenamefont
  {Mishchenko}, \citenamefont {Mayorov}, \citenamefont {Woods}, \citenamefont
  {Wallbank}, \citenamefont {Mucha-Kruczynski}, \citenamefont {Piot},
  \citenamefont {Potemski}, \citenamefont {Grigorieva}, \citenamefont
  {Novoselov}, \citenamefont {Guinea}, \citenamefont {Fal'ko},\ and\
  \citenamefont {Geim}}]{ponomarenko_GhBN_2013}%
  \BibitemOpen
  \bibfield  {author} {\bibinfo {author} {\bibfnamefont {L.~A.}\ \bibnamefont
  {Ponomarenko}}, \bibinfo {author} {\bibfnamefont {R.~V.}\ \bibnamefont
  {Gorbachev}}, \bibinfo {author} {\bibfnamefont {G.~L.}\ \bibnamefont {Yu}},
  \bibinfo {author} {\bibfnamefont {D.~C.}\ \bibnamefont {Elias}}, \bibinfo
  {author} {\bibfnamefont {R.}~\bibnamefont {Jalil}}, \bibinfo {author}
  {\bibfnamefont {A.~A.}\ \bibnamefont {Patel}}, \bibinfo {author}
  {\bibfnamefont {A.}~\bibnamefont {Mishchenko}}, \bibinfo {author}
  {\bibfnamefont {A.~S.}\ \bibnamefont {Mayorov}}, \bibinfo {author}
  {\bibfnamefont {C.~R.}\ \bibnamefont {Woods}}, \bibinfo {author}
  {\bibfnamefont {J.~R.}\ \bibnamefont {Wallbank}}, \bibinfo {author}
  {\bibfnamefont {M.}~\bibnamefont {Mucha-Kruczynski}}, \bibinfo {author}
  {\bibfnamefont {B.~A.}\ \bibnamefont {Piot}}, \bibinfo {author}
  {\bibfnamefont {M.}~\bibnamefont {Potemski}}, \bibinfo {author}
  {\bibfnamefont {I.~V.}\ \bibnamefont {Grigorieva}}, \bibinfo {author}
  {\bibfnamefont {K.~S.}\ \bibnamefont {Novoselov}}, \bibinfo {author}
  {\bibfnamefont {F.}~\bibnamefont {Guinea}}, \bibinfo {author} {\bibfnamefont
  {V.~I.}\ \bibnamefont {Fal'ko}}, \ and\ \bibinfo {author} {\bibfnamefont
  {A.~K.}\ \bibnamefont {Geim}},\ }\href
  {http://dx.doi.org/10.1038/nature12187} {\bibfield  {journal} {\bibinfo
  {journal} {Nature}\ }\textbf {\bibinfo {volume} {497}},\ \bibinfo {pages}
  {594} (\bibinfo {year} {2013})}\BibitemShut {NoStop}%
\bibitem [{\citenamefont {Dean}\ \emph {et~al.}(2013)\citenamefont {Dean},
  \citenamefont {Wang}, \citenamefont {Maher}, \citenamefont {Forsythe},
  \citenamefont {Ghahari}, \citenamefont {Gao}, \citenamefont {Katoch},
  \citenamefont {Ishigami}, \citenamefont {Moon}, \citenamefont {Koshino},
  \citenamefont {Taniguchi}, \citenamefont {Watanabe}, \citenamefont {Shepard},
  \citenamefont {Hone},\ and\ \citenamefont {Kim}}]{dean_GhBN_2013}%
  \BibitemOpen
  \bibfield  {author} {\bibinfo {author} {\bibfnamefont {C.~R.}\ \bibnamefont
  {Dean}}, \bibinfo {author} {\bibfnamefont {L.}~\bibnamefont {Wang}}, \bibinfo
  {author} {\bibfnamefont {P.}~\bibnamefont {Maher}}, \bibinfo {author}
  {\bibfnamefont {C.}~\bibnamefont {Forsythe}}, \bibinfo {author}
  {\bibfnamefont {F.}~\bibnamefont {Ghahari}}, \bibinfo {author} {\bibfnamefont
  {Y.}~\bibnamefont {Gao}}, \bibinfo {author} {\bibfnamefont {J.}~\bibnamefont
  {Katoch}}, \bibinfo {author} {\bibfnamefont {M.}~\bibnamefont {Ishigami}},
  \bibinfo {author} {\bibfnamefont {P.}~\bibnamefont {Moon}}, \bibinfo {author}
  {\bibfnamefont {M.}~\bibnamefont {Koshino}}, \bibinfo {author} {\bibfnamefont
  {T.}~\bibnamefont {Taniguchi}}, \bibinfo {author} {\bibfnamefont
  {K.}~\bibnamefont {Watanabe}}, \bibinfo {author} {\bibfnamefont {K.~L.}\
  \bibnamefont {Shepard}}, \bibinfo {author} {\bibfnamefont {J.}~\bibnamefont
  {Hone}}, \ and\ \bibinfo {author} {\bibfnamefont {P.}~\bibnamefont {Kim}},\
  }\href {http://dx.doi.org/10.1038/nature12186} {\bibfield  {journal}
  {\bibinfo  {journal} {Nature}\ }\textbf {\bibinfo {volume} {497}},\ \bibinfo
  {pages} {598} (\bibinfo {year} {2013})}\BibitemShut {NoStop}%
\bibitem [{\citenamefont {Hunt}\ \emph {et~al.}(2013)\citenamefont {Hunt},
  \citenamefont {Sanchez-Yamagishi}, \citenamefont {Young}, \citenamefont
  {Yankowitz}, \citenamefont {LeRoy}, \citenamefont {Watanabe}, \citenamefont
  {Taniguchi}, \citenamefont {Moon}, \citenamefont {Koshino}, \citenamefont
  {Jarillo-Herrero},\ and\ \citenamefont {Ashoori}}]{hunt_GhBN_2013}%
  \BibitemOpen
  \bibfield  {author} {\bibinfo {author} {\bibfnamefont {B.}~\bibnamefont
  {Hunt}}, \bibinfo {author} {\bibfnamefont {J.~D.}\ \bibnamefont
  {Sanchez-Yamagishi}}, \bibinfo {author} {\bibfnamefont {A.~F.}\ \bibnamefont
  {Young}}, \bibinfo {author} {\bibfnamefont {M.}~\bibnamefont {Yankowitz}},
  \bibinfo {author} {\bibfnamefont {B.~J.}\ \bibnamefont {LeRoy}}, \bibinfo
  {author} {\bibfnamefont {K.}~\bibnamefont {Watanabe}}, \bibinfo {author}
  {\bibfnamefont {T.}~\bibnamefont {Taniguchi}}, \bibinfo {author}
  {\bibfnamefont {P.}~\bibnamefont {Moon}}, \bibinfo {author} {\bibfnamefont
  {M.}~\bibnamefont {Koshino}}, \bibinfo {author} {\bibfnamefont
  {P.}~\bibnamefont {Jarillo-Herrero}}, \ and\ \bibinfo {author} {\bibfnamefont
  {R.~C.}\ \bibnamefont {Ashoori}},\ }\href {\doibase 10.1126/science.1237240}
  {\bibfield  {journal} {\bibinfo  {journal} {Science}\ }\textbf {\bibinfo
  {volume} {340}},\ \bibinfo {pages} {1427} (\bibinfo {year}
  {2013})}\BibitemShut {NoStop}%
\bibitem [{\citenamefont {Li}\ \emph {et~al.}(2010)\citenamefont {Li},
  \citenamefont {Luican}, \citenamefont {{Lopes Dos Santos}}, \citenamefont
  {{Castro Neto}}, \citenamefont {Reina}, \citenamefont {Kong},\ and\
  \citenamefont {Andrei}}]{Li2010}%
  \BibitemOpen
  \bibfield  {author} {\bibinfo {author} {\bibfnamefont {G.}~\bibnamefont
  {Li}}, \bibinfo {author} {\bibfnamefont {A.}~\bibnamefont {Luican}}, \bibinfo
  {author} {\bibfnamefont {J.~M.~B.}\ \bibnamefont {{Lopes Dos Santos}}},
  \bibinfo {author} {\bibfnamefont {A.~H.}\ \bibnamefont {{Castro Neto}}},
  \bibinfo {author} {\bibfnamefont {A.}~\bibnamefont {Reina}}, \bibinfo
  {author} {\bibfnamefont {J.}~\bibnamefont {Kong}}, \ and\ \bibinfo {author}
  {\bibfnamefont {E.~Y.}\ \bibnamefont {Andrei}},\ }\href@noop {} {\bibfield
  {journal} {\bibinfo  {journal} {Nat. Phys.}\ }\textbf {\bibinfo {volume}
  {6}},\ \bibinfo {pages} {109} (\bibinfo {year} {2010})}\BibitemShut {NoStop}%
\bibitem [{\citenamefont {Yankowitz}\ \emph {et~al.}(2012)\citenamefont
  {Yankowitz}, \citenamefont {Xue}, \citenamefont {Cormode}, \citenamefont
  {Sanchez-Yamagishi}, \citenamefont {Watanabe}, \citenamefont {Taniguchi},
  \citenamefont {Jarillo-Herrero}, \citenamefont {Jacquod},\ and\ \citenamefont
  {LeRoy}}]{Yankowitz2012}%
  \BibitemOpen
  \bibfield  {author} {\bibinfo {author} {\bibfnamefont {M.}~\bibnamefont
  {Yankowitz}}, \bibinfo {author} {\bibfnamefont {J.}~\bibnamefont {Xue}},
  \bibinfo {author} {\bibfnamefont {D.}~\bibnamefont {Cormode}}, \bibinfo
  {author} {\bibfnamefont {J.~D.}\ \bibnamefont {Sanchez-Yamagishi}}, \bibinfo
  {author} {\bibfnamefont {K.}~\bibnamefont {Watanabe}}, \bibinfo {author}
  {\bibfnamefont {T.}~\bibnamefont {Taniguchi}}, \bibinfo {author}
  {\bibfnamefont {P.}~\bibnamefont {Jarillo-Herrero}}, \bibinfo {author}
  {\bibfnamefont {P.}~\bibnamefont {Jacquod}}, \ and\ \bibinfo {author}
  {\bibfnamefont {B.~J.}\ \bibnamefont {LeRoy}},\ }\href {\doibase
  10.1038/nphys2272} {\bibfield  {journal} {\bibinfo  {journal} {Nat. Phys.}\
  }\textbf {\bibinfo {volume} {8}},\ \bibinfo {pages} {382} (\bibinfo {year}
  {2012})}\BibitemShut {NoStop}%
\bibitem [{\citenamefont {Krishna~Kumar}\ \emph {et~al.}(2018)\citenamefont
  {Krishna~Kumar}, \citenamefont {Mishchenko}, \citenamefont {Chen},
  \citenamefont {Pezzini}, \citenamefont {Auton}, \citenamefont {Ponomarenko},
  \citenamefont {Zeitler}, \citenamefont {Eaves}, \citenamefont
  {Fal{\textquoteright}ko},\ and\ \citenamefont {Geim}}]{roshan_2018}%
  \BibitemOpen
  \bibfield  {author} {\bibinfo {author} {\bibfnamefont {R.}~\bibnamefont
  {Krishna~Kumar}}, \bibinfo {author} {\bibfnamefont {A.}~\bibnamefont
  {Mishchenko}}, \bibinfo {author} {\bibfnamefont {X.}~\bibnamefont {Chen}},
  \bibinfo {author} {\bibfnamefont {S.}~\bibnamefont {Pezzini}}, \bibinfo
  {author} {\bibfnamefont {G.~H.}\ \bibnamefont {Auton}}, \bibinfo {author}
  {\bibfnamefont {L.~A.}\ \bibnamefont {Ponomarenko}}, \bibinfo {author}
  {\bibfnamefont {U.}~\bibnamefont {Zeitler}}, \bibinfo {author} {\bibfnamefont
  {L.}~\bibnamefont {Eaves}}, \bibinfo {author} {\bibfnamefont {V.~I.}\
  \bibnamefont {Fal{\textquoteright}ko}}, \ and\ \bibinfo {author}
  {\bibfnamefont {A.~K.}\ \bibnamefont {Geim}},\ }\href {\doibase
  10.1073/pnas.1804572115} {\bibfield  {journal} {\bibinfo  {journal} {PNAS}\
  }\textbf {\bibinfo {volume} {115}},\ \bibinfo {pages} {5135} (\bibinfo {year}
  {2018})}\BibitemShut {NoStop}%
\bibitem [{\citenamefont {Yu}\ \emph {et~al.}(2014)\citenamefont {Yu},
  \citenamefont {Gorbachev}, \citenamefont {Tu}, \citenamefont {Kretinin},
  \citenamefont {Cao}, \citenamefont {Jalil}, \citenamefont {Withers},
  \citenamefont {Ponomarenko}, \citenamefont {Piot}, \citenamefont {Potemski},
  \citenamefont {Elias}, \citenamefont {Chen}, \citenamefont {Watanabe},
  \citenamefont {Taniguchi}, \citenamefont {Grigorieva}, \citenamefont
  {Novoselov}, \citenamefont {Fal'ko}, \citenamefont {Geim},\ and\
  \citenamefont {Mishchenko}}]{capacitance_2014}%
  \BibitemOpen
  \bibfield  {author} {\bibinfo {author} {\bibfnamefont {G.~L.}\ \bibnamefont
  {Yu}}, \bibinfo {author} {\bibfnamefont {R.~V.}\ \bibnamefont {Gorbachev}},
  \bibinfo {author} {\bibfnamefont {J.~S.}\ \bibnamefont {Tu}}, \bibinfo
  {author} {\bibfnamefont {A.~V.}\ \bibnamefont {Kretinin}}, \bibinfo {author}
  {\bibfnamefont {Y.}~\bibnamefont {Cao}}, \bibinfo {author} {\bibfnamefont
  {R.}~\bibnamefont {Jalil}}, \bibinfo {author} {\bibfnamefont
  {F.}~\bibnamefont {Withers}}, \bibinfo {author} {\bibfnamefont {L.~A.}\
  \bibnamefont {Ponomarenko}}, \bibinfo {author} {\bibfnamefont {B.~A.}\
  \bibnamefont {Piot}}, \bibinfo {author} {\bibfnamefont {M.}~\bibnamefont
  {Potemski}}, \bibinfo {author} {\bibfnamefont {D.~C.}\ \bibnamefont {Elias}},
  \bibinfo {author} {\bibfnamefont {X.}~\bibnamefont {Chen}}, \bibinfo {author}
  {\bibfnamefont {K.}~\bibnamefont {Watanabe}}, \bibinfo {author}
  {\bibfnamefont {T.}~\bibnamefont {Taniguchi}}, \bibinfo {author}
  {\bibfnamefont {I.~V.}\ \bibnamefont {Grigorieva}}, \bibinfo {author}
  {\bibfnamefont {K.~S.}\ \bibnamefont {Novoselov}}, \bibinfo {author}
  {\bibfnamefont {V.~I.}\ \bibnamefont {Fal'ko}}, \bibinfo {author}
  {\bibfnamefont {A.~K.}\ \bibnamefont {Geim}}, \ and\ \bibinfo {author}
  {\bibfnamefont {A.}~\bibnamefont {Mishchenko}},\ }\href
  {http://dx.doi.org/10.1038/nphys2979} {\bibfield  {journal} {\bibinfo
  {journal} {Nat. Phys.}\ }\textbf {\bibinfo {volume} {10}},\ \bibinfo {pages}
  {525} (\bibinfo {year} {2014})}\BibitemShut {NoStop}%
\bibitem [{\citenamefont {Ni}\ \emph {et~al.}(2015)\citenamefont {Ni},
  \citenamefont {Wang}, \citenamefont {Wu}, \citenamefont {Fei}, \citenamefont
  {Goldflam}, \citenamefont {Keilmann}, \citenamefont {{\"O}zyilmaz},
  \citenamefont {Castro~Neto}, \citenamefont {Xie}, \citenamefont {Fogler},\
  and\ \citenamefont {Basov}}]{IR_2015}%
  \BibitemOpen
  \bibfield  {author} {\bibinfo {author} {\bibfnamefont {G.~X.}\ \bibnamefont
  {Ni}}, \bibinfo {author} {\bibfnamefont {H.}~\bibnamefont {Wang}}, \bibinfo
  {author} {\bibfnamefont {J.~S.}\ \bibnamefont {Wu}}, \bibinfo {author}
  {\bibfnamefont {Z.}~\bibnamefont {Fei}}, \bibinfo {author} {\bibfnamefont
  {M.~D.}\ \bibnamefont {Goldflam}}, \bibinfo {author} {\bibfnamefont
  {F.}~\bibnamefont {Keilmann}}, \bibinfo {author} {\bibfnamefont
  {B.}~\bibnamefont {{\"O}zyilmaz}}, \bibinfo {author} {\bibfnamefont {A.~H.}\
  \bibnamefont {Castro~Neto}}, \bibinfo {author} {\bibfnamefont {X.~M.}\
  \bibnamefont {Xie}}, \bibinfo {author} {\bibfnamefont {M.~M.}\ \bibnamefont
  {Fogler}}, \ and\ \bibinfo {author} {\bibfnamefont {D.~N.}\ \bibnamefont
  {Basov}},\ }\href {http://dx.doi.org/10.1038/nmat4425} {\bibfield  {journal}
  {\bibinfo  {journal} {Nat. Mat.}\ }\textbf {\bibinfo {volume} {14}},\
  \bibinfo {pages} {1217} (\bibinfo {year} {2015})}\BibitemShut {NoStop}%
\bibitem [{\citenamefont {Hsu}\ \emph {et~al.}(2014)\citenamefont {Hsu},
  \citenamefont {Zhao}, \citenamefont {Li}, \citenamefont {Chen}, \citenamefont
  {Chiu}, \citenamefont {Chang}, \citenamefont {Chou},\ and\ \citenamefont
  {Chang}}]{ting_acsnano_2014}%
  \BibitemOpen
  \bibfield  {author} {\bibinfo {author} {\bibfnamefont {W.-T.}\ \bibnamefont
  {Hsu}}, \bibinfo {author} {\bibfnamefont {Z.-A.}\ \bibnamefont {Zhao}},
  \bibinfo {author} {\bibfnamefont {L.-J.}\ \bibnamefont {Li}}, \bibinfo
  {author} {\bibfnamefont {C.-H.}\ \bibnamefont {Chen}}, \bibinfo {author}
  {\bibfnamefont {M.-H.}\ \bibnamefont {Chiu}}, \bibinfo {author}
  {\bibfnamefont {P.-S.}\ \bibnamefont {Chang}}, \bibinfo {author}
  {\bibfnamefont {Y.-C.}\ \bibnamefont {Chou}}, \ and\ \bibinfo {author}
  {\bibfnamefont {W.-H.}\ \bibnamefont {Chang}},\ }\href {\doibase
  10.1021/nn500228r} {\bibfield  {journal} {\bibinfo  {journal} {ACS Nano}\
  }\textbf {\bibinfo {volume} {8}},\ \bibinfo {pages} {2951} (\bibinfo {year}
  {2014})}\BibitemShut {NoStop}%
\bibitem [{\citenamefont {Rigosi}\ \emph {et~al.}(2015)\citenamefont {Rigosi},
  \citenamefont {Hill}, \citenamefont {Li}, \citenamefont {Chernikov},\ and\
  \citenamefont {Heinz}}]{rigosi_nanolett_2015}%
  \BibitemOpen
  \bibfield  {author} {\bibinfo {author} {\bibfnamefont {A.~F.}\ \bibnamefont
  {Rigosi}}, \bibinfo {author} {\bibfnamefont {H.~M.}\ \bibnamefont {Hill}},
  \bibinfo {author} {\bibfnamefont {Y.}~\bibnamefont {Li}}, \bibinfo {author}
  {\bibfnamefont {A.}~\bibnamefont {Chernikov}}, \ and\ \bibinfo {author}
  {\bibfnamefont {T.~F.}\ \bibnamefont {Heinz}},\ }\href {\doibase
  10.1021/acs.nanolett.5b01055} {\bibfield  {journal} {\bibinfo  {journal}
  {Nano Lett.}\ }\textbf {\bibinfo {volume} {15}},\ \bibinfo {pages} {5033}
  (\bibinfo {year} {2015})}\BibitemShut {NoStop}%
\bibitem [{\citenamefont {Hill}\ \emph {et~al.}(2016)\citenamefont {Hill},
  \citenamefont {Rigosi}, \citenamefont {Rim}, \citenamefont {Flynn},\ and\
  \citenamefont {Heinz}}]{hill_nanolett_2016}%
  \BibitemOpen
  \bibfield  {author} {\bibinfo {author} {\bibfnamefont {H.~M.}\ \bibnamefont
  {Hill}}, \bibinfo {author} {\bibfnamefont {A.~F.}\ \bibnamefont {Rigosi}},
  \bibinfo {author} {\bibfnamefont {K.~T.}\ \bibnamefont {Rim}}, \bibinfo
  {author} {\bibfnamefont {G.~W.}\ \bibnamefont {Flynn}}, \ and\ \bibinfo
  {author} {\bibfnamefont {T.~F.}\ \bibnamefont {Heinz}},\ }\href {\doibase
  10.1021/acs.nanolett.6b01007} {\bibfield  {journal} {\bibinfo  {journal}
  {Nano Lett.}\ }\textbf {\bibinfo {volume} {16}},\ \bibinfo {pages} {4831}
  (\bibinfo {year} {2016})}\BibitemShut {NoStop}%
\bibitem [{\citenamefont {Nayak}\ \emph {et~al.}(2017)\citenamefont {Nayak},
  \citenamefont {Horbatenko}, \citenamefont {Ahn}, \citenamefont {Kim},
  \citenamefont {Lee}, \citenamefont {Ma}, \citenamefont {Jang}, \citenamefont
  {Lim}, \citenamefont {Kim}, \citenamefont {Ryu}, \citenamefont {Cheong},
  \citenamefont {Park},\ and\ \citenamefont {Shin}}]{nayak_acsnano_2017}%
  \BibitemOpen
  \bibfield  {author} {\bibinfo {author} {\bibfnamefont {P.~K.}\ \bibnamefont
  {Nayak}}, \bibinfo {author} {\bibfnamefont {Y.}~\bibnamefont {Horbatenko}},
  \bibinfo {author} {\bibfnamefont {S.}~\bibnamefont {Ahn}}, \bibinfo {author}
  {\bibfnamefont {G.}~\bibnamefont {Kim}}, \bibinfo {author} {\bibfnamefont
  {J.-U.}\ \bibnamefont {Lee}}, \bibinfo {author} {\bibfnamefont {K.~Y.}\
  \bibnamefont {Ma}}, \bibinfo {author} {\bibfnamefont {A.-R.}\ \bibnamefont
  {Jang}}, \bibinfo {author} {\bibfnamefont {H.}~\bibnamefont {Lim}}, \bibinfo
  {author} {\bibfnamefont {D.}~\bibnamefont {Kim}}, \bibinfo {author}
  {\bibfnamefont {S.}~\bibnamefont {Ryu}}, \bibinfo {author} {\bibfnamefont
  {H.}~\bibnamefont {Cheong}}, \bibinfo {author} {\bibfnamefont
  {N.}~\bibnamefont {Park}}, \ and\ \bibinfo {author} {\bibfnamefont {H.~S.}\
  \bibnamefont {Shin}},\ }\href {\doibase 10.1021/acsnano.7b00640} {\bibfield
  {journal} {\bibinfo  {journal} {ACS Nano}\ }\textbf {\bibinfo {volume}
  {11}},\ \bibinfo {pages} {4041} (\bibinfo {year} {2017})}\BibitemShut
  {NoStop}%
\bibitem [{\citenamefont {Alexeev}\ \emph {et~al.}(2017)\citenamefont
  {Alexeev}, \citenamefont {Catanzaro}, \citenamefont {Skrypka}, \citenamefont
  {Nayak}, \citenamefont {Ahn}, \citenamefont {Pak}, \citenamefont {Lee},
  \citenamefont {Sohn}, \citenamefont {Novoselov}, \citenamefont {Shin},\ and\
  \citenamefont {Tartakovskii}}]{alexeev_nanolett_2017}%
  \BibitemOpen
  \bibfield  {author} {\bibinfo {author} {\bibfnamefont {E.~M.}\ \bibnamefont
  {Alexeev}}, \bibinfo {author} {\bibfnamefont {A.}~\bibnamefont {Catanzaro}},
  \bibinfo {author} {\bibfnamefont {O.~V.}\ \bibnamefont {Skrypka}}, \bibinfo
  {author} {\bibfnamefont {P.~K.}\ \bibnamefont {Nayak}}, \bibinfo {author}
  {\bibfnamefont {S.}~\bibnamefont {Ahn}}, \bibinfo {author} {\bibfnamefont
  {S.}~\bibnamefont {Pak}}, \bibinfo {author} {\bibfnamefont {J.}~\bibnamefont
  {Lee}}, \bibinfo {author} {\bibfnamefont {J.~I.}\ \bibnamefont {Sohn}},
  \bibinfo {author} {\bibfnamefont {K.~S.}\ \bibnamefont {Novoselov}}, \bibinfo
  {author} {\bibfnamefont {H.~S.}\ \bibnamefont {Shin}}, \ and\ \bibinfo
  {author} {\bibfnamefont {A.~I.}\ \bibnamefont {Tartakovskii}},\ }\href
  {\doibase 10.1021/acs.nanolett.7b01763} {\bibfield  {journal} {\bibinfo
  {journal} {Nano Lett.}\ }\textbf {\bibinfo {volume} {17}},\ \bibinfo {pages}
  {5342} (\bibinfo {year} {2017})}\BibitemShut {NoStop}%
\bibitem [{\citenamefont {Zhang}\ \emph {et~al.}(2017)\citenamefont {Zhang},
  \citenamefont {Chuu}, \citenamefont {Ren}, \citenamefont {Li}, \citenamefont
  {Li}, \citenamefont {Jin}, \citenamefont {Chou},\ and\ \citenamefont
  {Shih}}]{Zhange1601459}%
  \BibitemOpen
  \bibfield  {author} {\bibinfo {author} {\bibfnamefont {C.}~\bibnamefont
  {Zhang}}, \bibinfo {author} {\bibfnamefont {C.-P.}\ \bibnamefont {Chuu}},
  \bibinfo {author} {\bibfnamefont {X.}~\bibnamefont {Ren}}, \bibinfo {author}
  {\bibfnamefont {M.-Y.}\ \bibnamefont {Li}}, \bibinfo {author} {\bibfnamefont
  {L.-J.}\ \bibnamefont {Li}}, \bibinfo {author} {\bibfnamefont
  {C.}~\bibnamefont {Jin}}, \bibinfo {author} {\bibfnamefont {M.-Y.}\
  \bibnamefont {Chou}}, \ and\ \bibinfo {author} {\bibfnamefont {C.-K.}\
  \bibnamefont {Shih}},\ }\href {\doibase 10.1126/sciadv.1601459} {\bibfield
  {journal} {\bibinfo  {journal} {Sci. Adv.}\ }\textbf {\bibinfo {volume} {3}}
  (\bibinfo {year} {2017}),\ 10.1126/sciadv.1601459}\BibitemShut {NoStop}%
\bibitem [{\citenamefont {Mak}\ \emph {et~al.}(2010)\citenamefont {Mak},
  \citenamefont {Lee}, \citenamefont {Hone}, \citenamefont {Shan},\ and\
  \citenamefont {Heinz}}]{mak_MoS2_2010}%
  \BibitemOpen
  \bibfield  {author} {\bibinfo {author} {\bibfnamefont {K.~F.}\ \bibnamefont
  {Mak}}, \bibinfo {author} {\bibfnamefont {C.}~\bibnamefont {Lee}}, \bibinfo
  {author} {\bibfnamefont {J.}~\bibnamefont {Hone}}, \bibinfo {author}
  {\bibfnamefont {J.}~\bibnamefont {Shan}}, \ and\ \bibinfo {author}
  {\bibfnamefont {T.~F.}\ \bibnamefont {Heinz}},\ }\href {\doibase
  10.1103/PhysRevLett.105.136805} {\bibfield  {journal} {\bibinfo  {journal}
  {Phys. Rev. Lett.}\ }\textbf {\bibinfo {volume} {105}},\ \bibinfo {pages}
  {136805} (\bibinfo {year} {2010})}\BibitemShut {NoStop}%
\bibitem [{\citenamefont {Splendiani}\ \emph {et~al.}(2010)\citenamefont
  {Splendiani}, \citenamefont {Sun}, \citenamefont {Zhang}, \citenamefont {Li},
  \citenamefont {Kim}, \citenamefont {Chim}, \citenamefont {Galli},\ and\
  \citenamefont {Wang}}]{splendiani_MoS2_2010}%
  \BibitemOpen
  \bibfield  {author} {\bibinfo {author} {\bibfnamefont {A.}~\bibnamefont
  {Splendiani}}, \bibinfo {author} {\bibfnamefont {L.}~\bibnamefont {Sun}},
  \bibinfo {author} {\bibfnamefont {Y.}~\bibnamefont {Zhang}}, \bibinfo
  {author} {\bibfnamefont {T.}~\bibnamefont {Li}}, \bibinfo {author}
  {\bibfnamefont {J.}~\bibnamefont {Kim}}, \bibinfo {author} {\bibfnamefont
  {C.-Y.}\ \bibnamefont {Chim}}, \bibinfo {author} {\bibfnamefont
  {G.}~\bibnamefont {Galli}}, \ and\ \bibinfo {author} {\bibfnamefont
  {F.}~\bibnamefont {Wang}},\ }\href {\doibase 10.1021/nl903868w} {\bibfield
  {journal} {\bibinfo  {journal} {Nano Lett.}\ }\textbf {\bibinfo {volume}
  {10}},\ \bibinfo {pages} {1271} (\bibinfo {year} {2010})}\BibitemShut
  {NoStop}%
\bibitem [{\citenamefont {Wang}\ \emph {et~al.}(2018)\citenamefont {Wang},
  \citenamefont {Chernikov}, \citenamefont {Glazov}, \citenamefont {Heinz},
  \citenamefont {Marie}, \citenamefont {Amand},\ and\ \citenamefont
  {Urbaszek}}]{review_2018}%
  \BibitemOpen
  \bibfield  {author} {\bibinfo {author} {\bibfnamefont {G.}~\bibnamefont
  {Wang}}, \bibinfo {author} {\bibfnamefont {A.}~\bibnamefont {Chernikov}},
  \bibinfo {author} {\bibfnamefont {M.~M.}\ \bibnamefont {Glazov}}, \bibinfo
  {author} {\bibfnamefont {T.~F.}\ \bibnamefont {Heinz}}, \bibinfo {author}
  {\bibfnamefont {X.}~\bibnamefont {Marie}}, \bibinfo {author} {\bibfnamefont
  {T.}~\bibnamefont {Amand}}, \ and\ \bibinfo {author} {\bibfnamefont
  {B.}~\bibnamefont {Urbaszek}},\ }\href {\doibase
  10.1103/RevModPhys.90.021001} {\bibfield  {journal} {\bibinfo  {journal}
  {Rev. Mod. Phys.}\ }\textbf {\bibinfo {volume} {90}},\ \bibinfo {pages}
  {021001} (\bibinfo {year} {2018})}\BibitemShut {NoStop}%
\bibitem [{\citenamefont {Yao}\ \emph {et~al.}(2008)\citenamefont {Yao},
  \citenamefont {Xiao},\ and\ \citenamefont {Niu}}]{wangyao_prb_2008}%
  \BibitemOpen
  \bibfield  {author} {\bibinfo {author} {\bibfnamefont {W.}~\bibnamefont
  {Yao}}, \bibinfo {author} {\bibfnamefont {D.}~\bibnamefont {Xiao}}, \ and\
  \bibinfo {author} {\bibfnamefont {Q.}~\bibnamefont {Niu}},\ }\href {\doibase
  10.1103/PhysRevB.77.235406} {\bibfield  {journal} {\bibinfo  {journal} {Phys.
  Rev. B}\ }\textbf {\bibinfo {volume} {77}},\ \bibinfo {pages} {235406}
  (\bibinfo {year} {2008})}\BibitemShut {NoStop}%
\bibitem [{\citenamefont {Zeng}\ \emph {et~al.}(2012)\citenamefont {Zeng},
  \citenamefont {Dai}, \citenamefont {Yao}, \citenamefont {Xiao},\ and\
  \citenamefont {Cui}}]{zeng_natnano_2012}%
  \BibitemOpen
  \bibfield  {author} {\bibinfo {author} {\bibfnamefont {H.}~\bibnamefont
  {Zeng}}, \bibinfo {author} {\bibfnamefont {J.}~\bibnamefont {Dai}}, \bibinfo
  {author} {\bibfnamefont {W.}~\bibnamefont {Yao}}, \bibinfo {author}
  {\bibfnamefont {D.}~\bibnamefont {Xiao}}, \ and\ \bibinfo {author}
  {\bibfnamefont {X.}~\bibnamefont {Cui}},\ }\href
  {http://dx.doi.org/10.1038/nnano.2012.95} {\bibfield  {journal} {\bibinfo
  {journal} {Nat. Nanotechnol.}\ }\textbf {\bibinfo {volume} {7}},\ \bibinfo
  {pages} {490} (\bibinfo {year} {2012})}\BibitemShut {NoStop}%
\bibitem [{\citenamefont {Mak}\ \emph {et~al.}(2012)\citenamefont {Mak},
  \citenamefont {He}, \citenamefont {Shan},\ and\ \citenamefont
  {Heinz}}]{mak_natnano_2012}%
  \BibitemOpen
  \bibfield  {author} {\bibinfo {author} {\bibfnamefont {K.~F.}\ \bibnamefont
  {Mak}}, \bibinfo {author} {\bibfnamefont {K.}~\bibnamefont {He}}, \bibinfo
  {author} {\bibfnamefont {J.}~\bibnamefont {Shan}}, \ and\ \bibinfo {author}
  {\bibfnamefont {T.~F.}\ \bibnamefont {Heinz}},\ }\href
  {http://dx.doi.org/10.1038/nnano.2012.96} {\bibfield  {journal} {\bibinfo
  {journal} {Nat. Nanotechnol.}\ }\textbf {\bibinfo {volume} {7}},\ \bibinfo
  {pages} {494} (\bibinfo {year} {2012})}\BibitemShut {NoStop}%
\bibitem [{\citenamefont {Kuwabara}\ \emph {et~al.}(1990)\citenamefont
  {Kuwabara}, \citenamefont {Clarke},\ and\ \citenamefont
  {Smith}}]{kuwabara_1990}%
  \BibitemOpen
  \bibfield  {author} {\bibinfo {author} {\bibfnamefont {M.}~\bibnamefont
  {Kuwabara}}, \bibinfo {author} {\bibfnamefont {D.~R.}\ \bibnamefont
  {Clarke}}, \ and\ \bibinfo {author} {\bibfnamefont {D.~A.}\ \bibnamefont
  {Smith}},\ }\href {\doibase 10.1063/1.102906} {\bibfield  {journal} {\bibinfo
   {journal} {Applied Physics Letters}\ }\textbf {\bibinfo {volume} {56}},\
  \bibinfo {pages} {2396} (\bibinfo {year} {1990})}\BibitemShut {NoStop}%
\bibitem [{\citenamefont {Wu}\ \emph {et~al.}(2018{\natexlab{a}})\citenamefont
  {Wu}, \citenamefont {Lovorn}, \citenamefont {Tutuc},\ and\ \citenamefont
  {MacDonald}}]{macdonald_hubbard}%
  \BibitemOpen
  \bibfield  {author} {\bibinfo {author} {\bibfnamefont {F.}~\bibnamefont
  {Wu}}, \bibinfo {author} {\bibfnamefont {T.}~\bibnamefont {Lovorn}}, \bibinfo
  {author} {\bibfnamefont {E.}~\bibnamefont {Tutuc}}, \ and\ \bibinfo {author}
  {\bibfnamefont {A.~H.}\ \bibnamefont {MacDonald}},\ }\href {\doibase
  10.1103/PhysRevLett.121.026402} {\bibfield  {journal} {\bibinfo  {journal}
  {Phys. Rev. Lett.}\ }\textbf {\bibinfo {volume} {121}},\ \bibinfo {pages}
  {026402} (\bibinfo {year} {2018}{\natexlab{a}})}\BibitemShut {NoStop}%
\bibitem [{\citenamefont {Cao}\ \emph {et~al.}(2018)\citenamefont {Cao},
  \citenamefont {Fatemi}, \citenamefont {Fang}, \citenamefont {Watanabe},
  \citenamefont {Taniguchi}, \citenamefont {Kaxiras},\ and\ \citenamefont
  {Jarillo-Herrero}}]{Cao2018}%
  \BibitemOpen
  \bibfield  {author} {\bibinfo {author} {\bibfnamefont {Y.}~\bibnamefont
  {Cao}}, \bibinfo {author} {\bibfnamefont {V.}~\bibnamefont {Fatemi}},
  \bibinfo {author} {\bibfnamefont {S.}~\bibnamefont {Fang}}, \bibinfo {author}
  {\bibfnamefont {K.}~\bibnamefont {Watanabe}}, \bibinfo {author}
  {\bibfnamefont {T.}~\bibnamefont {Taniguchi}}, \bibinfo {author}
  {\bibfnamefont {E.}~\bibnamefont {Kaxiras}}, \ and\ \bibinfo {author}
  {\bibfnamefont {P.}~\bibnamefont {Jarillo-Herrero}},\ }\href@noop {}
  {\bibfield  {journal} {\bibinfo  {journal} {Nature}\ }\textbf {\bibinfo
  {volume} {556}},\ \bibinfo {pages} {43} (\bibinfo {year} {2018})}\BibitemShut
  {NoStop}%
\bibitem [{\citenamefont {Gong}\ \emph {et~al.}(2013)\citenamefont {Gong},
  \citenamefont {Zhang}, \citenamefont {Wang}, \citenamefont {Colombo},
  \citenamefont {Wallace},\ and\ \citenamefont {Cho}}]{band_alignment}%
  \BibitemOpen
  \bibfield  {author} {\bibinfo {author} {\bibfnamefont {C.}~\bibnamefont
  {Gong}}, \bibinfo {author} {\bibfnamefont {H.}~\bibnamefont {Zhang}},
  \bibinfo {author} {\bibfnamefont {W.}~\bibnamefont {Wang}}, \bibinfo {author}
  {\bibfnamefont {L.}~\bibnamefont {Colombo}}, \bibinfo {author} {\bibfnamefont
  {R.~M.}\ \bibnamefont {Wallace}}, \ and\ \bibinfo {author} {\bibfnamefont
  {K.}~\bibnamefont {Cho}},\ }\href {\doibase 10.1063/1.4817409} {\bibfield
  {journal} {\bibinfo  {journal} {Appl. Phys. Lett.}\ }\textbf {\bibinfo
  {volume} {103}},\ \bibinfo {pages} {053513} (\bibinfo {year}
  {2013})}\BibitemShut {NoStop}%
\bibitem [{\citenamefont {Xu}\ \emph {et~al.}(2018)\citenamefont {Xu},
  \citenamefont {Xu}, \citenamefont {Zhang}, \citenamefont {Peng},
  \citenamefont {Shao}, \citenamefont {Ni}, \citenamefont {Li}, \citenamefont
  {Yao}, \citenamefont {Lu}, \citenamefont {Zhu},\ and\ \citenamefont
  {Soukoulis}}]{first_principles_2018}%
  \BibitemOpen
  \bibfield  {author} {\bibinfo {author} {\bibfnamefont {K.}~\bibnamefont
  {Xu}}, \bibinfo {author} {\bibfnamefont {Y.}~\bibnamefont {Xu}}, \bibinfo
  {author} {\bibfnamefont {H.}~\bibnamefont {Zhang}}, \bibinfo {author}
  {\bibfnamefont {B.}~\bibnamefont {Peng}}, \bibinfo {author} {\bibfnamefont
  {H.}~\bibnamefont {Shao}}, \bibinfo {author} {\bibfnamefont {G.}~\bibnamefont
  {Ni}}, \bibinfo {author} {\bibfnamefont {J.}~\bibnamefont {Li}}, \bibinfo
  {author} {\bibfnamefont {M.}~\bibnamefont {Yao}}, \bibinfo {author}
  {\bibfnamefont {H.}~\bibnamefont {Lu}}, \bibinfo {author} {\bibfnamefont
  {H.}~\bibnamefont {Zhu}}, \ and\ \bibinfo {author} {\bibfnamefont {C.~M.}\
  \bibnamefont {Soukoulis}},\ }\href {\doibase 10.1039/C8CP05522J} {\bibfield
  {journal} {\bibinfo  {journal} {Phys. Chem. Chem. Phys.}\ }\textbf {\bibinfo
  {volume} {20}},\ \bibinfo {pages} {30351} (\bibinfo {year}
  {2018})}\BibitemShut {NoStop}%
\bibitem [{\citenamefont {Kozawa}\ \emph {et~al.}(2016)\citenamefont {Kozawa},
  \citenamefont {Carvalho}, \citenamefont {Verzhbitskiy}, \citenamefont
  {Giustiniano}, \citenamefont {Miyauchi}, \citenamefont {Mouri}, \citenamefont
  {Neto}, \citenamefont {Matsuda},\ and\ \citenamefont {Eda}}]{Kozawa2015}%
  \BibitemOpen
  \bibfield  {author} {\bibinfo {author} {\bibfnamefont {D.}~\bibnamefont
  {Kozawa}}, \bibinfo {author} {\bibfnamefont {A.}~\bibnamefont {Carvalho}},
  \bibinfo {author} {\bibfnamefont {I.}~\bibnamefont {Verzhbitskiy}}, \bibinfo
  {author} {\bibfnamefont {F.}~\bibnamefont {Giustiniano}}, \bibinfo {author}
  {\bibfnamefont {Y.}~\bibnamefont {Miyauchi}}, \bibinfo {author}
  {\bibfnamefont {S.}~\bibnamefont {Mouri}}, \bibinfo {author} {\bibfnamefont
  {A.~H.~C.}\ \bibnamefont {Neto}}, \bibinfo {author} {\bibfnamefont
  {K.}~\bibnamefont {Matsuda}}, \ and\ \bibinfo {author} {\bibfnamefont
  {G.}~\bibnamefont {Eda}},\ }\href@noop {} {\bibfield  {journal} {\bibinfo
  {journal} {Nano Lett.}\ }\textbf {\bibinfo {volume} {16}},\ \bibinfo {pages}
  {4087} (\bibinfo {year} {2016})}\BibitemShut {NoStop}%
\bibitem [{\citenamefont {Bistritzer}\ and\ \citenamefont
  {MacDonald}(2011)}]{macdonald_pnas}%
  \BibitemOpen
  \bibfield  {author} {\bibinfo {author} {\bibfnamefont {R.}~\bibnamefont
  {Bistritzer}}\ and\ \bibinfo {author} {\bibfnamefont {A.~H.}\ \bibnamefont
  {MacDonald}},\ }\href@noop {} {\bibfield  {journal} {\bibinfo  {journal}
  {PNAS}\ }\textbf {\bibinfo {volume} {108}},\ \bibinfo {pages} {12233}
  (\bibinfo {year} {2011})}\BibitemShut {NoStop}%
\bibitem [{\citenamefont {Wu}\ \emph {et~al.}(2017)\citenamefont {Wu},
  \citenamefont {Lovorn},\ and\ \citenamefont {MacDonald}}]{macdonald_intra}%
  \BibitemOpen
  \bibfield  {author} {\bibinfo {author} {\bibfnamefont {F.}~\bibnamefont
  {Wu}}, \bibinfo {author} {\bibfnamefont {T.}~\bibnamefont {Lovorn}}, \ and\
  \bibinfo {author} {\bibfnamefont {A.~H.}\ \bibnamefont {MacDonald}},\ }\href
  {\doibase 10.1103/PhysRevLett.118.147401} {\bibfield  {journal} {\bibinfo
  {journal} {Phys. Rev. Lett.}\ }\textbf {\bibinfo {volume} {118}},\ \bibinfo
  {pages} {147401} (\bibinfo {year} {2017})}\BibitemShut {NoStop}%
\bibitem [{\citenamefont {Yu}\ \emph {et~al.}(2017)\citenamefont {Yu},
  \citenamefont {Liu}, \citenamefont {Tang}, \citenamefont {Xu},\ and\
  \citenamefont {Yao}}]{hongyi_moire}%
  \BibitemOpen
  \bibfield  {author} {\bibinfo {author} {\bibfnamefont {H.}~\bibnamefont
  {Yu}}, \bibinfo {author} {\bibfnamefont {G.-B.}\ \bibnamefont {Liu}},
  \bibinfo {author} {\bibfnamefont {J.}~\bibnamefont {Tang}}, \bibinfo {author}
  {\bibfnamefont {X.}~\bibnamefont {Xu}}, \ and\ \bibinfo {author}
  {\bibfnamefont {W.}~\bibnamefont {Yao}},\ }\href {\doibase
  10.1126/sciadv.1701696} {\bibfield  {journal} {\bibinfo  {journal} {Sci.
  Adv.}\ }\textbf {\bibinfo {volume} {3}},\ \bibinfo {pages} {e1701696}
  (\bibinfo {year} {2017})}\BibitemShut {NoStop}%
\bibitem [{\citenamefont {Wu}\ \emph {et~al.}(2018{\natexlab{b}})\citenamefont
  {Wu}, \citenamefont {Lovorn},\ and\ \citenamefont {MacDonald}}]{Wu2017}%
  \BibitemOpen
  \bibfield  {author} {\bibinfo {author} {\bibfnamefont {F.}~\bibnamefont
  {Wu}}, \bibinfo {author} {\bibfnamefont {T.}~\bibnamefont {Lovorn}}, \ and\
  \bibinfo {author} {\bibfnamefont {A.~H.}\ \bibnamefont {MacDonald}},\ }\href
  {\doibase 10.1103/PhysRevB.97.035306} {\bibfield  {journal} {\bibinfo
  {journal} {Phys. Rev. B}\ }\textbf {\bibinfo {volume} {97}},\ \bibinfo
  {pages} {035306} (\bibinfo {year} {2018}{\natexlab{b}})}\BibitemShut
  {NoStop}%
\bibitem [{\citenamefont {Tong}\ \emph {et~al.}(2016)\citenamefont {Tong},
  \citenamefont {Yu}, \citenamefont {Zhu}, \citenamefont {Wang}, \citenamefont
  {Xu},\ and\ \citenamefont {Yao}}]{wang_yao_tvvtcc}%
  \BibitemOpen
  \bibfield  {author} {\bibinfo {author} {\bibfnamefont {Q.}~\bibnamefont
  {Tong}}, \bibinfo {author} {\bibfnamefont {H.}~\bibnamefont {Yu}}, \bibinfo
  {author} {\bibfnamefont {Q.}~\bibnamefont {Zhu}}, \bibinfo {author}
  {\bibfnamefont {Y.}~\bibnamefont {Wang}}, \bibinfo {author} {\bibfnamefont
  {X.}~\bibnamefont {Xu}}, \ and\ \bibinfo {author} {\bibfnamefont
  {W.}~\bibnamefont {Yao}},\ }\href {http://dx.doi.org/10.1038/nphys3968}
  {\bibfield  {journal} {\bibinfo  {journal} {Nat. Phys.}\ }\textbf {\bibinfo
  {volume} {13}},\ \bibinfo {pages} {356} (\bibinfo {year} {2016})}\BibitemShut
  {NoStop}%
\bibitem [{Note1()}]{Note1}%
  \BibitemOpen
  \bibinfo {note} {An alternative nomenclature is used, \protect \emph {e.g.},
  in Refs.\ \protect \rev@citealpnum {wang_yao_tvvtcc,wangyao_coupling}, where
  P and AP stacking configurations are referred to as R and H stacking,
  respectively. We choose the former convention to avoid confusion with
  standard nomenclature for commensurate stacking.}\BibitemShut {Stop}%
\bibitem [{\citenamefont {Constantinescu}\ \emph {et~al.}(2013)\citenamefont
  {Constantinescu}, \citenamefont {Kuc},\ and\ \citenamefont
  {Heine}}]{tmd_stacking_2013}%
  \BibitemOpen
  \bibfield  {author} {\bibinfo {author} {\bibfnamefont {G.}~\bibnamefont
  {Constantinescu}}, \bibinfo {author} {\bibfnamefont {A.}~\bibnamefont {Kuc}},
  \ and\ \bibinfo {author} {\bibfnamefont {T.}~\bibnamefont {Heine}},\ }\href
  {\doibase 10.1103/PhysRevLett.111.036104} {\bibfield  {journal} {\bibinfo
  {journal} {Phys. Rev. Lett.}\ }\textbf {\bibinfo {volume} {111}},\ \bibinfo
  {pages} {036104} (\bibinfo {year} {2013})}\BibitemShut {NoStop}%
\bibitem [{\citenamefont {He}\ \emph {et~al.}(2014)\citenamefont {He},
  \citenamefont {Hummer},\ and\ \citenamefont {Franchini}}]{tmd_stacking_2014}%
  \BibitemOpen
  \bibfield  {author} {\bibinfo {author} {\bibfnamefont {J.}~\bibnamefont
  {He}}, \bibinfo {author} {\bibfnamefont {K.}~\bibnamefont {Hummer}}, \ and\
  \bibinfo {author} {\bibfnamefont {C.}~\bibnamefont {Franchini}},\ }\href
  {\doibase 10.1103/PhysRevB.89.075409} {\bibfield  {journal} {\bibinfo
  {journal} {Phys. Rev. B}\ }\textbf {\bibinfo {volume} {89}},\ \bibinfo
  {pages} {075409} (\bibinfo {year} {2014})}\BibitemShut {NoStop}%
\bibitem [{\citenamefont {Korm{\'a}nyos}\ \emph {et~al.}(2015)\citenamefont
  {Korm{\'a}nyos}, \citenamefont {Burkard}, \citenamefont {Gmitra},
  \citenamefont {Fabian}, \citenamefont {Z{\'o}lyomi}, \citenamefont
  {Drummond},\ and\ \citenamefont {Fal'ko}}]{kdotp}%
  \BibitemOpen
  \bibfield  {author} {\bibinfo {author} {\bibfnamefont {A.}~\bibnamefont
  {Korm{\'a}nyos}}, \bibinfo {author} {\bibfnamefont {G.}~\bibnamefont
  {Burkard}}, \bibinfo {author} {\bibfnamefont {M.}~\bibnamefont {Gmitra}},
  \bibinfo {author} {\bibfnamefont {J.}~\bibnamefont {Fabian}}, \bibinfo
  {author} {\bibfnamefont {V.}~\bibnamefont {Z{\'o}lyomi}}, \bibinfo {author}
  {\bibfnamefont {N.~D.}\ \bibnamefont {Drummond}}, \ and\ \bibinfo {author}
  {\bibfnamefont {V.}~\bibnamefont {Fal'ko}},\ }\href
  {http://stacks.iop.org/2053-1583/2/i=2/a=022001} {\bibfield  {journal}
  {\bibinfo  {journal} {2D Materials}\ }\textbf {\bibinfo {volume} {2}},\
  \bibinfo {pages} {022001} (\bibinfo {year} {2015})}\BibitemShut {NoStop}%
\bibitem [{\citenamefont {Kyl\"anp\"a\"a}\ and\ \citenamefont
  {Komsa}(2015)}]{komsa_2015}%
  \BibitemOpen
  \bibfield  {author} {\bibinfo {author} {\bibfnamefont {I.}~\bibnamefont
  {Kyl\"anp\"a\"a}}\ and\ \bibinfo {author} {\bibfnamefont {H.-P.}\
  \bibnamefont {Komsa}},\ }\href {\doibase 10.1103/PhysRevB.92.205418}
  {\bibfield  {journal} {\bibinfo  {journal} {Phys. Rev. B}\ }\textbf {\bibinfo
  {volume} {92}},\ \bibinfo {pages} {205418} (\bibinfo {year}
  {2015})}\BibitemShut {NoStop}%
\bibitem [{\citenamefont {Cheiwchanchamnangij}\ and\ \citenamefont
  {Lambrecht}(2012)}]{Cheiwchanchamnangij2012}%
  \BibitemOpen
  \bibfield  {author} {\bibinfo {author} {\bibfnamefont {T.}~\bibnamefont
  {Cheiwchanchamnangij}}\ and\ \bibinfo {author} {\bibfnamefont {W.~R.~L.}\
  \bibnamefont {Lambrecht}},\ }\href {\doibase 10.1103/PhysRevB.85.205302}
  {\bibfield  {journal} {\bibinfo  {journal} {Phys. Rev. B}\ }\textbf {\bibinfo
  {volume} {85}},\ \bibinfo {pages} {205302} (\bibinfo {year}
  {2012})}\BibitemShut {NoStop}%
\bibitem [{\citenamefont {Wang}\ \emph
  {et~al.}(2017{\natexlab{a}})\citenamefont {Wang}, \citenamefont {Wang},
  \citenamefont {Yao}, \citenamefont {Liu},\ and\ \citenamefont
  {Yu}}]{wangyao_coupling}%
  \BibitemOpen
  \bibfield  {author} {\bibinfo {author} {\bibfnamefont {Y.}~\bibnamefont
  {Wang}}, \bibinfo {author} {\bibfnamefont {Z.}~\bibnamefont {Wang}}, \bibinfo
  {author} {\bibfnamefont {W.}~\bibnamefont {Yao}}, \bibinfo {author}
  {\bibfnamefont {G.-B.}\ \bibnamefont {Liu}}, \ and\ \bibinfo {author}
  {\bibfnamefont {H.}~\bibnamefont {Yu}},\ }\href {\doibase
  10.1103/PhysRevB.95.115429} {\bibfield  {journal} {\bibinfo  {journal} {Phys.
  Rev. B}\ }\textbf {\bibinfo {volume} {95}},\ \bibinfo {pages} {115429}
  (\bibinfo {year} {2017}{\natexlab{a}})}\BibitemShut {NoStop}%
\bibitem [{\citenamefont {Korm\'anyos}\ \emph {et~al.}(2018)\citenamefont
  {Korm\'anyos}, \citenamefont {Z\'olyomi}, \citenamefont {Fal'ko},\ and\
  \citenamefont {Burkard}}]{kormanyos_kdotp_2018}%
  \BibitemOpen
  \bibfield  {author} {\bibinfo {author} {\bibfnamefont {A.}~\bibnamefont
  {Korm\'anyos}}, \bibinfo {author} {\bibfnamefont {V.}~\bibnamefont
  {Z\'olyomi}}, \bibinfo {author} {\bibfnamefont {V.~I.}\ \bibnamefont
  {Fal'ko}}, \ and\ \bibinfo {author} {\bibfnamefont {G.}~\bibnamefont
  {Burkard}},\ }\href {\doibase 10.1103/PhysRevB.98.035408} {\bibfield
  {journal} {\bibinfo  {journal} {Phys. Rev. B}\ }\textbf {\bibinfo {volume}
  {98}},\ \bibinfo {pages} {035408} (\bibinfo {year} {2018})}\BibitemShut
  {NoStop}%
\bibitem [{\citenamefont {Alexeev}\ \emph {et~al.}(2019)\citenamefont
  {Alexeev}, \citenamefont {Ruiz-Tijerina}, \citenamefont {Danovich},
  \citenamefont {Hamer}, \citenamefont {Terry}, \citenamefont {Nayak},
  \citenamefont {Ahn}, \citenamefont {Pak}, \citenamefont {Lee}, \citenamefont
  {Sohn}, \citenamefont {Molas}, \citenamefont {Koperski}, \citenamefont
  {Watanabe}, \citenamefont {Taniguchi}, \citenamefont {Novoselov},
  \citenamefont {Gorbachev}, \citenamefont {Shin}, \citenamefont {Fal'ko},\
  and\ \citenamefont {Tartakovskii}}]{twist_angle2018}%
  \BibitemOpen
  \bibfield  {author} {\bibinfo {author} {\bibfnamefont {E.~M.}\ \bibnamefont
  {Alexeev}}, \bibinfo {author} {\bibfnamefont {D.~A.}\ \bibnamefont
  {Ruiz-Tijerina}}, \bibinfo {author} {\bibfnamefont {M.}~\bibnamefont
  {Danovich}}, \bibinfo {author} {\bibfnamefont {M.~J.}\ \bibnamefont {Hamer}},
  \bibinfo {author} {\bibfnamefont {D.~J.}\ \bibnamefont {Terry}}, \bibinfo
  {author} {\bibfnamefont {P.~K.}\ \bibnamefont {Nayak}}, \bibinfo {author}
  {\bibfnamefont {S.}~\bibnamefont {Ahn}}, \bibinfo {author} {\bibfnamefont
  {S.}~\bibnamefont {Pak}}, \bibinfo {author} {\bibfnamefont {J.}~\bibnamefont
  {Lee}}, \bibinfo {author} {\bibfnamefont {J.~I.}\ \bibnamefont {Sohn}},
  \bibinfo {author} {\bibfnamefont {M.~R.}\ \bibnamefont {Molas}}, \bibinfo
  {author} {\bibfnamefont {M.}~\bibnamefont {Koperski}}, \bibinfo {author}
  {\bibfnamefont {K.}~\bibnamefont {Watanabe}}, \bibinfo {author}
  {\bibfnamefont {T.}~\bibnamefont {Taniguchi}}, \bibinfo {author}
  {\bibfnamefont {K.~S.}\ \bibnamefont {Novoselov}}, \bibinfo {author}
  {\bibfnamefont {R.~V.}\ \bibnamefont {Gorbachev}}, \bibinfo {author}
  {\bibfnamefont {H.~S.}\ \bibnamefont {Shin}}, \bibinfo {author}
  {\bibfnamefont {V.~I.}\ \bibnamefont {Fal'ko}}, \ and\ \bibinfo {author}
  {\bibfnamefont {A.~I.}\ \bibnamefont {Tartakovskii}},\ }\href {\doibase
  10.1038/s41586-019-0986-9} {\bibfield  {journal} {\bibinfo  {journal}
  {Nature}\ }\textbf {\bibinfo {volume} {567}},\ \bibinfo {pages} {81}
  (\bibinfo {year} {2019})}\BibitemShut {NoStop}%
\bibitem [{\citenamefont {Mostaani}\ \emph {et~al.}(2017)\citenamefont
  {Mostaani}, \citenamefont {Szyniszewski}, \citenamefont {Price},
  \citenamefont {Maezono}, \citenamefont {Danovich}, \citenamefont {Hunt},
  \citenamefont {Drummond},\ and\ \citenamefont
  {Fal'ko}}]{mostaani_excitonic_prb_2017}%
  \BibitemOpen
  \bibfield  {author} {\bibinfo {author} {\bibfnamefont {E.}~\bibnamefont
  {Mostaani}}, \bibinfo {author} {\bibfnamefont {M.}~\bibnamefont
  {Szyniszewski}}, \bibinfo {author} {\bibfnamefont {C.~H.}\ \bibnamefont
  {Price}}, \bibinfo {author} {\bibfnamefont {R.}~\bibnamefont {Maezono}},
  \bibinfo {author} {\bibfnamefont {M.}~\bibnamefont {Danovich}}, \bibinfo
  {author} {\bibfnamefont {R.~J.}\ \bibnamefont {Hunt}}, \bibinfo {author}
  {\bibfnamefont {N.~D.}\ \bibnamefont {Drummond}}, \ and\ \bibinfo {author}
  {\bibfnamefont {V.~I.}\ \bibnamefont {Fal'ko}},\ }\href {\doibase
  10.1103/PhysRevB.96.075431} {\bibfield  {journal} {\bibinfo  {journal} {Phys.
  Rev. B}\ }\textbf {\bibinfo {volume} {96}},\ \bibinfo {pages} {075431}
  (\bibinfo {year} {2017})}\BibitemShut {NoStop}%
\bibitem [{\citenamefont {Al-Hilli}\ and\ \citenamefont
  {Evans}(1972)}]{mose2_d}%
  \BibitemOpen
  \bibfield  {author} {\bibinfo {author} {\bibfnamefont {A.~A.}\ \bibnamefont
  {Al-Hilli}}\ and\ \bibinfo {author} {\bibfnamefont {B.~L.}\ \bibnamefont
  {Evans}},\ }\href {\doibase 10.1016/0022-0248(72)90129-7} {\bibfield
  {journal} {\bibinfo  {journal} {Journal of Crystal Growth}\ }\textbf
  {\bibinfo {volume} {15}},\ \bibinfo {pages} {93} (\bibinfo {year}
  {1972})}\BibitemShut {NoStop}%
\bibitem [{\citenamefont {Kindermann}\ \emph {et~al.}(2012)\citenamefont
  {Kindermann}, \citenamefont {Uchoa},\ and\ \citenamefont {Miller}}]{uchoa}%
  \BibitemOpen
  \bibfield  {author} {\bibinfo {author} {\bibfnamefont {M.}~\bibnamefont
  {Kindermann}}, \bibinfo {author} {\bibfnamefont {B.}~\bibnamefont {Uchoa}}, \
  and\ \bibinfo {author} {\bibfnamefont {D.~L.}\ \bibnamefont {Miller}},\
  }\href {\doibase 10.1103/PhysRevB.86.115415} {\bibfield  {journal} {\bibinfo
  {journal} {Phys. Rev. B}\ }\textbf {\bibinfo {volume} {86}},\ \bibinfo
  {pages} {115415} (\bibinfo {year} {2012})}\BibitemShut {NoStop}%
\bibitem [{\citenamefont {Wallbank}\ \emph {et~al.}(2013)\citenamefont
  {Wallbank}, \citenamefont {Patel}, \citenamefont
  {Mucha-Kruczy\ifmmode~\acute{n}\else \'{n}\fi{}ski}, \citenamefont {Geim},\
  and\ \citenamefont {Fal'ko}}]{wallbank}%
  \BibitemOpen
  \bibfield  {author} {\bibinfo {author} {\bibfnamefont {J.~R.}\ \bibnamefont
  {Wallbank}}, \bibinfo {author} {\bibfnamefont {A.~A.}\ \bibnamefont {Patel}},
  \bibinfo {author} {\bibfnamefont {M.}~\bibnamefont
  {Mucha-Kruczy\ifmmode~\acute{n}\else \'{n}\fi{}ski}}, \bibinfo {author}
  {\bibfnamefont {A.~K.}\ \bibnamefont {Geim}}, \ and\ \bibinfo {author}
  {\bibfnamefont {V.~I.}\ \bibnamefont {Fal'ko}},\ }\href {\doibase
  10.1103/PhysRevB.87.245408} {\bibfield  {journal} {\bibinfo  {journal} {Phys.
  Rev. B}\ }\textbf {\bibinfo {volume} {87}},\ \bibinfo {pages} {245408}
  (\bibinfo {year} {2013})}\BibitemShut {NoStop}%
\bibitem [{\citenamefont {Schrieffer}\ and\ \citenamefont
  {Wolff}(1966)}]{schrieffer_wolff}%
  \BibitemOpen
  \bibfield  {author} {\bibinfo {author} {\bibfnamefont {J.~R.}\ \bibnamefont
  {Schrieffer}}\ and\ \bibinfo {author} {\bibfnamefont {P.~A.}\ \bibnamefont
  {Wolff}},\ }\href {\doibase 10.1103/PhysRev.149.491} {\bibfield  {journal}
  {\bibinfo  {journal} {Phys. Rev.}\ }\textbf {\bibinfo {volume} {149}},\
  \bibinfo {pages} {491} (\bibinfo {year} {1966})}\BibitemShut {NoStop}%
\bibitem [{\citenamefont {Wallbank}\ \emph {et~al.}(2015)\citenamefont
  {Wallbank}, \citenamefont {Mucha-Kruczyński}, \citenamefont {Chen},\ and\
  \citenamefont {Fal'ko}}]{wallbank_annalen_2015}%
  \BibitemOpen
  \bibfield  {author} {\bibinfo {author} {\bibfnamefont {J.~R.}\ \bibnamefont
  {Wallbank}}, \bibinfo {author} {\bibfnamefont {M.}~\bibnamefont
  {Mucha-Kruczyński}}, \bibinfo {author} {\bibfnamefont {X.}~\bibnamefont
  {Chen}}, \ and\ \bibinfo {author} {\bibfnamefont {V.~I.}\ \bibnamefont
  {Fal'ko}},\ }\href {\doibase 10.1002/andp.201400204} {\bibfield  {journal}
  {\bibinfo  {journal} {Annalen der Physik}\ }\textbf {\bibinfo {volume}
  {527}},\ \bibinfo {pages} {359} (\bibinfo {year} {2015})}\BibitemShut
  {NoStop}%
\bibitem [{\citenamefont {Koshino}\ and\ \citenamefont
  {Moon}(2015)}]{koshino_incommensurate}%
  \BibitemOpen
  \bibfield  {author} {\bibinfo {author} {\bibfnamefont {M.}~\bibnamefont
  {Koshino}}\ and\ \bibinfo {author} {\bibfnamefont {P.}~\bibnamefont {Moon}},\
  }\href {\doibase 10.7566/JPSJ.84.121001} {\bibfield  {journal} {\bibinfo
  {journal} {Journal of the Physical Society of Japan}\ }\textbf {\bibinfo
  {volume} {84}},\ \bibinfo {pages} {121001} (\bibinfo {year}
  {2015})}\BibitemShut {NoStop}%
\bibitem [{\citenamefont {Xiao}\ \emph {et~al.}(2012)\citenamefont {Xiao},
  \citenamefont {Liu}, \citenamefont {Feng}, \citenamefont {Xu},\ and\
  \citenamefont {Yao}}]{wangyao_spin_valley}%
  \BibitemOpen
  \bibfield  {author} {\bibinfo {author} {\bibfnamefont {D.}~\bibnamefont
  {Xiao}}, \bibinfo {author} {\bibfnamefont {G.-B.}\ \bibnamefont {Liu}},
  \bibinfo {author} {\bibfnamefont {W.}~\bibnamefont {Feng}}, \bibinfo {author}
  {\bibfnamefont {X.}~\bibnamefont {Xu}}, \ and\ \bibinfo {author}
  {\bibfnamefont {W.}~\bibnamefont {Yao}},\ }\href {\doibase
  10.1103/PhysRevLett.108.196802} {\bibfield  {journal} {\bibinfo  {journal}
  {Phys. Rev. Lett.}\ }\textbf {\bibinfo {volume} {108}},\ \bibinfo {pages}
  {196802} (\bibinfo {year} {2012})}\BibitemShut {NoStop}%
\bibitem [{\citenamefont {Rivera}\ \emph {et~al.}(2015)\citenamefont {Rivera},
  \citenamefont {Schaibley}, \citenamefont {Jones}, \citenamefont {Ross},
  \citenamefont {Wu}, \citenamefont {Aivazian}, \citenamefont {Klement},
  \citenamefont {Seyler}, \citenamefont {Clark}, \citenamefont {Ghimire},
  \citenamefont {Yan}, \citenamefont {Mandrus}, \citenamefont {Yao},\ and\
  \citenamefont {Xu}}]{MoSe2_WSe2_2015}%
  \BibitemOpen
  \bibfield  {author} {\bibinfo {author} {\bibfnamefont {P.}~\bibnamefont
  {Rivera}}, \bibinfo {author} {\bibfnamefont {J.~R.}\ \bibnamefont
  {Schaibley}}, \bibinfo {author} {\bibfnamefont {A.~M.}\ \bibnamefont
  {Jones}}, \bibinfo {author} {\bibfnamefont {J.~S.}\ \bibnamefont {Ross}},
  \bibinfo {author} {\bibfnamefont {S.}~\bibnamefont {Wu}}, \bibinfo {author}
  {\bibfnamefont {G.}~\bibnamefont {Aivazian}}, \bibinfo {author}
  {\bibfnamefont {P.}~\bibnamefont {Klement}}, \bibinfo {author} {\bibfnamefont
  {K.}~\bibnamefont {Seyler}}, \bibinfo {author} {\bibfnamefont
  {G.}~\bibnamefont {Clark}}, \bibinfo {author} {\bibfnamefont {N.~J.}\
  \bibnamefont {Ghimire}}, \bibinfo {author} {\bibfnamefont {J.}~\bibnamefont
  {Yan}}, \bibinfo {author} {\bibfnamefont {D.~G.}\ \bibnamefont {Mandrus}},
  \bibinfo {author} {\bibfnamefont {W.}~\bibnamefont {Yao}}, \ and\ \bibinfo
  {author} {\bibfnamefont {X.}~\bibnamefont {Xu}},\ }\href
  {http://dx.doi.org/10.1038/ncomms7242} {\bibfield  {journal} {\bibinfo
  {journal} {Nat. Commun.}\ }\textbf {\bibinfo {volume} {6}},\ \bibinfo {pages}
  {6242} (\bibinfo {year} {2015})}\BibitemShut {NoStop}%
\bibitem [{\citenamefont {Klein}\ \emph {et~al.}(2016)\citenamefont {Klein},
  \citenamefont {Wierzbowski}, \citenamefont {Regler}, \citenamefont {Becker},
  \citenamefont {Heimbach}, \citenamefont {Müller}, \citenamefont {Kaniber},\
  and\ \citenamefont {Finley}}]{klein_backgate_2016}%
  \BibitemOpen
  \bibfield  {author} {\bibinfo {author} {\bibfnamefont {J.}~\bibnamefont
  {Klein}}, \bibinfo {author} {\bibfnamefont {J.}~\bibnamefont {Wierzbowski}},
  \bibinfo {author} {\bibfnamefont {A.}~\bibnamefont {Regler}}, \bibinfo
  {author} {\bibfnamefont {J.}~\bibnamefont {Becker}}, \bibinfo {author}
  {\bibfnamefont {F.}~\bibnamefont {Heimbach}}, \bibinfo {author}
  {\bibfnamefont {K.}~\bibnamefont {Müller}}, \bibinfo {author} {\bibfnamefont
  {M.}~\bibnamefont {Kaniber}}, \ and\ \bibinfo {author} {\bibfnamefont
  {J.~J.}\ \bibnamefont {Finley}},\ }\href {\doibase
  10.1021/acs.nanolett.5b03954} {\bibfield  {journal} {\bibinfo  {journal}
  {Nano Lett.}\ }\textbf {\bibinfo {volume} {16}},\ \bibinfo {pages} {1554}
  (\bibinfo {year} {2016})}\BibitemShut {NoStop}%
\bibitem [{\citenamefont {Wang}\ \emph
  {et~al.}(2017{\natexlab{b}})\citenamefont {Wang}, \citenamefont {Stanev},
  \citenamefont {Valencia}, \citenamefont {Charles}, \citenamefont {Henning},
  \citenamefont {Sangwan}, \citenamefont {Lahiri}, \citenamefont {Mejia},
  \citenamefont {Sarangapani}, \citenamefont {Povolotskyi}, \citenamefont
  {Afzalian}, \citenamefont {Maassen}, \citenamefont {Klimeck}, \citenamefont
  {Hersam}, \citenamefont {Lauhon}, \citenamefont {Stern},\ and\ \citenamefont
  {Kubis}}]{wang_backgate_2017}%
  \BibitemOpen
  \bibfield  {author} {\bibinfo {author} {\bibfnamefont {K.-C.}\ \bibnamefont
  {Wang}}, \bibinfo {author} {\bibfnamefont {T.~K.}\ \bibnamefont {Stanev}},
  \bibinfo {author} {\bibfnamefont {D.}~\bibnamefont {Valencia}}, \bibinfo
  {author} {\bibfnamefont {J.}~\bibnamefont {Charles}}, \bibinfo {author}
  {\bibfnamefont {A.}~\bibnamefont {Henning}}, \bibinfo {author} {\bibfnamefont
  {V.~K.}\ \bibnamefont {Sangwan}}, \bibinfo {author} {\bibfnamefont
  {A.}~\bibnamefont {Lahiri}}, \bibinfo {author} {\bibfnamefont
  {D.}~\bibnamefont {Mejia}}, \bibinfo {author} {\bibfnamefont
  {P.}~\bibnamefont {Sarangapani}}, \bibinfo {author} {\bibfnamefont
  {M.}~\bibnamefont {Povolotskyi}}, \bibinfo {author} {\bibfnamefont
  {A.}~\bibnamefont {Afzalian}}, \bibinfo {author} {\bibfnamefont
  {J.}~\bibnamefont {Maassen}}, \bibinfo {author} {\bibfnamefont
  {G.}~\bibnamefont {Klimeck}}, \bibinfo {author} {\bibfnamefont {M.~C.}\
  \bibnamefont {Hersam}}, \bibinfo {author} {\bibfnamefont {L.~J.}\
  \bibnamefont {Lauhon}}, \bibinfo {author} {\bibfnamefont {N.~P.}\
  \bibnamefont {Stern}}, \ and\ \bibinfo {author} {\bibfnamefont
  {T.}~\bibnamefont {Kubis}},\ }\href {\doibase 10.1063/1.5005958} {\bibfield
  {journal} {\bibinfo  {journal} {Journal of Applied Physics}\ }\textbf
  {\bibinfo {volume} {122}},\ \bibinfo {pages} {224302} (\bibinfo {year}
  {2017}{\natexlab{b}})}\BibitemShut {NoStop}%
\bibitem [{\citenamefont {Berkelbach}\ \emph {et~al.}(2013)\citenamefont
  {Berkelbach}, \citenamefont {Hybertsen},\ and\ \citenamefont
  {Reichman}}]{berkelbach_variational}%
  \BibitemOpen
  \bibfield  {author} {\bibinfo {author} {\bibfnamefont {T.~C.}\ \bibnamefont
  {Berkelbach}}, \bibinfo {author} {\bibfnamefont {M.~S.}\ \bibnamefont
  {Hybertsen}}, \ and\ \bibinfo {author} {\bibfnamefont {D.~R.}\ \bibnamefont
  {Reichman}},\ }\href {\doibase 10.1103/PhysRevB.88.045318} {\bibfield
  {journal} {\bibinfo  {journal} {Phys. Rev. B}\ }\textbf {\bibinfo {volume}
  {88}},\ \bibinfo {pages} {045318} (\bibinfo {year} {2013})}\BibitemShut
  {NoStop}%
\bibitem [{\citenamefont {Kumar}\ and\ \citenamefont
  {Ahluwalia}(2012)}]{kumar_physicab_2012}%
  \BibitemOpen
  \bibfield  {author} {\bibinfo {author} {\bibfnamefont {A.}~\bibnamefont
  {Kumar}}\ and\ \bibinfo {author} {\bibfnamefont {P.}~\bibnamefont
  {Ahluwalia}},\ }\href {\doibase https://doi.org/10.1016/j.physb.2012.08.034}
  {\bibfield  {journal} {\bibinfo  {journal} {Physica B: Condens. Matter}\
  }\textbf {\bibinfo {volume} {407}},\ \bibinfo {pages} {4627 } (\bibinfo
  {year} {2012})}\BibitemShut {NoStop}%
\bibitem [{\citenamefont {Danovich}\ \emph {et~al.}(2018)\citenamefont
  {Danovich}, \citenamefont {Ruiz-Tijerina}, \citenamefont {Hunt},
  \citenamefont {Szyniszewski}, \citenamefont {Drummond},\ and\ \citenamefont
  {Fal'ko}}]{complexes2018}%
  \BibitemOpen
  \bibfield  {author} {\bibinfo {author} {\bibfnamefont {M.}~\bibnamefont
  {Danovich}}, \bibinfo {author} {\bibfnamefont {D.~A.}\ \bibnamefont
  {Ruiz-Tijerina}}, \bibinfo {author} {\bibfnamefont {R.~J.}\ \bibnamefont
  {Hunt}}, \bibinfo {author} {\bibfnamefont {M.}~\bibnamefont {Szyniszewski}},
  \bibinfo {author} {\bibfnamefont {N.~D.}\ \bibnamefont {Drummond}}, \ and\
  \bibinfo {author} {\bibfnamefont {V.~I.}\ \bibnamefont {Fal'ko}},\ }\href
  {\doibase 10.1103/PhysRevB.97.195452} {\bibfield  {journal} {\bibinfo
  {journal} {Phys. Rev. B}\ }\textbf {\bibinfo {volume} {97}},\ \bibinfo
  {pages} {195452} (\bibinfo {year} {2018})}\BibitemShut {NoStop}%
\bibitem [{\citenamefont {Heo}\ \emph {et~al.}(2015)\citenamefont {Heo},
  \citenamefont {Sung}, \citenamefont {Cha}, \citenamefont {Jang},
  \citenamefont {Kim}, \citenamefont {Jin}, \citenamefont {Lee}, \citenamefont
  {Ahn}, \citenamefont {Lee}, \citenamefont {Shim}, \citenamefont {Choi},\ and\
  \citenamefont {Jo}}]{Heo2015}%
  \BibitemOpen
  \bibfield  {author} {\bibinfo {author} {\bibfnamefont {H.}~\bibnamefont
  {Heo}}, \bibinfo {author} {\bibfnamefont {J.~H.}\ \bibnamefont {Sung}},
  \bibinfo {author} {\bibfnamefont {S.}~\bibnamefont {Cha}}, \bibinfo {author}
  {\bibfnamefont {B.-G.}\ \bibnamefont {Jang}}, \bibinfo {author}
  {\bibfnamefont {J.-Y.}\ \bibnamefont {Kim}}, \bibinfo {author} {\bibfnamefont
  {G.}~\bibnamefont {Jin}}, \bibinfo {author} {\bibfnamefont {D.}~\bibnamefont
  {Lee}}, \bibinfo {author} {\bibfnamefont {J.-H.}\ \bibnamefont {Ahn}},
  \bibinfo {author} {\bibfnamefont {M.-J.}\ \bibnamefont {Lee}}, \bibinfo
  {author} {\bibfnamefont {J.~H.}\ \bibnamefont {Shim}}, \bibinfo {author}
  {\bibfnamefont {H.}~\bibnamefont {Choi}}, \ and\ \bibinfo {author}
  {\bibfnamefont {M.-H.}\ \bibnamefont {Jo}},\ }\href {\doibase
  10.1038/ncomms8372} {\bibfield  {journal} {\bibinfo  {journal} {Nat.
  Commun.}\ }\textbf {\bibinfo {volume} {6}},\ \bibinfo {pages} {7372}
  (\bibinfo {year} {2015})}\BibitemShut {NoStop}%
\bibitem [{\citenamefont {Zhu}\ \emph {et~al.}(2017)\citenamefont {Zhu},
  \citenamefont {Wang}, \citenamefont {Gong}, \citenamefont {Kim},
  \citenamefont {Hone},\ and\ \citenamefont {Zhu}}]{charge_separation_2017}%
  \BibitemOpen
  \bibfield  {author} {\bibinfo {author} {\bibfnamefont {H.}~\bibnamefont
  {Zhu}}, \bibinfo {author} {\bibfnamefont {J.}~\bibnamefont {Wang}}, \bibinfo
  {author} {\bibfnamefont {Z.}~\bibnamefont {Gong}}, \bibinfo {author}
  {\bibfnamefont {Y.~D.}\ \bibnamefont {Kim}}, \bibinfo {author} {\bibfnamefont
  {J.}~\bibnamefont {Hone}}, \ and\ \bibinfo {author} {\bibfnamefont {X.-Y.}\
  \bibnamefont {Zhu}},\ }\href {\doibase 10.1021/acs.nanolett.7b00748}
  {\bibfield  {journal} {\bibinfo  {journal} {Nano Lett.}\ }\textbf {\bibinfo
  {volume} {17}},\ \bibinfo {pages} {3591} (\bibinfo {year}
  {2017})}\BibitemShut {NoStop}%
\bibitem [{\citenamefont {Deilmann}\ and\ \citenamefont
  {Thygesen}(2018)}]{Thygesen_large_optical_2018}%
  \BibitemOpen
  \bibfield  {author} {\bibinfo {author} {\bibfnamefont {T.}~\bibnamefont
  {Deilmann}}\ and\ \bibinfo {author} {\bibfnamefont {K.~S.}\ \bibnamefont
  {Thygesen}},\ }\href {\doibase 10.1021/acs.nanolett.8b00438} {\bibfield
  {journal} {\bibinfo  {journal} {Nano Lett.}\ }\textbf {\bibinfo {volume}
  {18}},\ \bibinfo {pages} {2984} (\bibinfo {year} {2018})}\BibitemShut
  {NoStop}%
\bibitem [{\citenamefont {Keldysh}(1979)}]{keldysh}%
  \BibitemOpen
  \bibfield  {author} {\bibinfo {author} {\bibfnamefont {L.~V.}\ \bibnamefont
  {Keldysh}},\ }\href@noop {} {\bibfield  {journal} {\bibinfo  {journal}
  {Pis'ma Zh. Eksp. Teor. Phys.}\ }\textbf {\bibinfo {volume} {29}},\ \bibinfo
  {pages} {716} (\bibinfo {year} {1979})},\ \bibinfo {note} {[JETP Lett.\
  \textbf{29}, 659 (1979)]}\BibitemShut {NoStop}%
\bibitem [{\citenamefont {Moskalenko}\ and\ \citenamefont
  {Snoke}(2000)}]{moskalenko2000bose}%
  \BibitemOpen
  \bibfield  {author} {\bibinfo {author} {\bibfnamefont {S.~A.}\ \bibnamefont
  {Moskalenko}}\ and\ \bibinfo {author} {\bibfnamefont {D.~W.}\ \bibnamefont
  {Snoke}},\ }\href@noop {} {\emph {\bibinfo {title} {Bose-{E}instein
  condensation of excitons and biexcitons and coherent nonlinear optics with
  excitons}}}\ (\bibinfo  {publisher} {Cambridge University Press},\ \bibinfo
  {year} {2000})\BibitemShut {NoStop}%
\bibitem [{\citenamefont {Yu}\ \emph {et~al.}(2015)\citenamefont {Yu},
  \citenamefont {Wang}, \citenamefont {Tong}, \citenamefont {Xu},\ and\
  \citenamefont {Yao}}]{wangyao_interlayer_X}%
  \BibitemOpen
  \bibfield  {author} {\bibinfo {author} {\bibfnamefont {H.}~\bibnamefont
  {Yu}}, \bibinfo {author} {\bibfnamefont {Y.}~\bibnamefont {Wang}}, \bibinfo
  {author} {\bibfnamefont {Q.}~\bibnamefont {Tong}}, \bibinfo {author}
  {\bibfnamefont {X.}~\bibnamefont {Xu}}, \ and\ \bibinfo {author}
  {\bibfnamefont {W.}~\bibnamefont {Yao}},\ }\href {\doibase
  10.1103/PhysRevLett.115.187002} {\bibfield  {journal} {\bibinfo  {journal}
  {Phys. Rev. Lett.}\ }\textbf {\bibinfo {volume} {115}},\ \bibinfo {pages}
  {187002} (\bibinfo {year} {2015})}\BibitemShut {NoStop}%
\bibitem [{\citenamefont {Gong}\ \emph {et~al.}(2014)\citenamefont {Gong},
  \citenamefont {Lin}, \citenamefont {Wang}, \citenamefont {Shi}, \citenamefont
  {Lei}, \citenamefont {Lin}, \citenamefont {Zou}, \citenamefont {Ye},
  \citenamefont {Vajtai}, \citenamefont {Yakobson}, \citenamefont {Terrones},
  \citenamefont {Terrones}, \citenamefont {Tay}, \citenamefont {Lou},
  \citenamefont {Pantelides}, \citenamefont {Liu}, \citenamefont {Zhou},\ and\
  \citenamefont {Ajayan}}]{WS2_MoS2_2014}%
  \BibitemOpen
  \bibfield  {author} {\bibinfo {author} {\bibfnamefont {Y.}~\bibnamefont
  {Gong}}, \bibinfo {author} {\bibfnamefont {J.}~\bibnamefont {Lin}}, \bibinfo
  {author} {\bibfnamefont {X.}~\bibnamefont {Wang}}, \bibinfo {author}
  {\bibfnamefont {G.}~\bibnamefont {Shi}}, \bibinfo {author} {\bibfnamefont
  {S.}~\bibnamefont {Lei}}, \bibinfo {author} {\bibfnamefont {Z.}~\bibnamefont
  {Lin}}, \bibinfo {author} {\bibfnamefont {X.}~\bibnamefont {Zou}}, \bibinfo
  {author} {\bibfnamefont {G.}~\bibnamefont {Ye}}, \bibinfo {author}
  {\bibfnamefont {R.}~\bibnamefont {Vajtai}}, \bibinfo {author} {\bibfnamefont
  {B.~I.}\ \bibnamefont {Yakobson}}, \bibinfo {author} {\bibfnamefont
  {H.}~\bibnamefont {Terrones}}, \bibinfo {author} {\bibfnamefont
  {M.}~\bibnamefont {Terrones}}, \bibinfo {author} {\bibfnamefont {B.~K.}\
  \bibnamefont {Tay}}, \bibinfo {author} {\bibfnamefont {J.}~\bibnamefont
  {Lou}}, \bibinfo {author} {\bibfnamefont {S.~T.}\ \bibnamefont {Pantelides}},
  \bibinfo {author} {\bibfnamefont {Z.}~\bibnamefont {Liu}}, \bibinfo {author}
  {\bibfnamefont {W.}~\bibnamefont {Zhou}}, \ and\ \bibinfo {author}
  {\bibfnamefont {P.~M.}\ \bibnamefont {Ajayan}},\ }\href
  {http://dx.doi.org/10.1038/nmat4091} {\bibfield  {journal} {\bibinfo
  {journal} {Nat. Mater.}\ }\textbf {\bibinfo {volume} {13}},\ \bibinfo {pages}
  {1135} (\bibinfo {year} {2014})}\BibitemShut {NoStop}%
\bibitem [{\citenamefont {Nagler}\ \emph {et~al.}(2017)\citenamefont {Nagler},
  \citenamefont {Plechinger}, \citenamefont {Ballottin}, \citenamefont
  {Mitioglu}, \citenamefont {Meier}, \citenamefont {Paradiso}, \citenamefont
  {Strunk}, \citenamefont {Chernikov}, \citenamefont {Christianen},
  \citenamefont {Sch\"uller},\ and\ \citenamefont {Korn}}]{MoSe2_WSe2_2017}%
  \BibitemOpen
  \bibfield  {author} {\bibinfo {author} {\bibfnamefont {P.}~\bibnamefont
  {Nagler}}, \bibinfo {author} {\bibfnamefont {G.}~\bibnamefont {Plechinger}},
  \bibinfo {author} {\bibfnamefont {M.~V.}\ \bibnamefont {Ballottin}}, \bibinfo
  {author} {\bibfnamefont {A.}~\bibnamefont {Mitioglu}}, \bibinfo {author}
  {\bibfnamefont {S.}~\bibnamefont {Meier}}, \bibinfo {author} {\bibfnamefont
  {N.}~\bibnamefont {Paradiso}}, \bibinfo {author} {\bibfnamefont
  {C.}~\bibnamefont {Strunk}}, \bibinfo {author} {\bibfnamefont
  {A.}~\bibnamefont {Chernikov}}, \bibinfo {author} {\bibfnamefont {P.~C.~M.}\
  \bibnamefont {Christianen}}, \bibinfo {author} {\bibfnamefont
  {C.}~\bibnamefont {Sch\"uller}}, \ and\ \bibinfo {author} {\bibfnamefont
  {T.}~\bibnamefont {Korn}},\ }\href
  {http://stacks.iop.org/2053-1583/4/i=2/a=025112} {\bibfield  {journal}
  {\bibinfo  {journal} {2D Mat.}\ }\textbf {\bibinfo {volume} {4}},\ \bibinfo
  {pages} {025112} (\bibinfo {year} {2017})}\BibitemShut {NoStop}%
\bibitem [{\citenamefont {Zhang}\ \emph {et~al.}(2018)\citenamefont {Zhang},
  \citenamefont {Surrente}, \citenamefont {Baranowski}, \citenamefont {Maude},
  \citenamefont {Gant}, \citenamefont {Castellanos-Gomez},\ and\ \citenamefont
  {Plochocka}}]{MoSe2_MoS2_2018}%
  \BibitemOpen
  \bibfield  {author} {\bibinfo {author} {\bibfnamefont {N.}~\bibnamefont
  {Zhang}}, \bibinfo {author} {\bibfnamefont {A.}~\bibnamefont {Surrente}},
  \bibinfo {author} {\bibfnamefont {M.}~\bibnamefont {Baranowski}}, \bibinfo
  {author} {\bibfnamefont {D.~K.}\ \bibnamefont {Maude}}, \bibinfo {author}
  {\bibfnamefont {P.}~\bibnamefont {Gant}}, \bibinfo {author} {\bibfnamefont
  {A.}~\bibnamefont {Castellanos-Gomez}}, \ and\ \bibinfo {author}
  {\bibfnamefont {P.}~\bibnamefont {Plochocka}},\ }\href {\doibase
  10.1021/acs.nanolett.8b03266} {\bibfield  {journal} {\bibinfo  {journal}
  {Nano Lett.}\ }\textbf {\bibinfo {volume} {18}},\ \bibinfo {pages} {7651}
  (\bibinfo {year} {2018})},\ \Eprint
  {http://arxiv.org/abs/https://doi.org/10.1021/acs.nanolett.8b03266}
  {https://doi.org/10.1021/acs.nanolett.8b03266} \BibitemShut {NoStop}%
\bibitem [{\citenamefont {Kozawa}\ \emph {et~al.}(2014)\citenamefont {Kozawa},
  \citenamefont {Kumar}, \citenamefont {Carvalho}, \citenamefont {Kumar~Amara},
  \citenamefont {Zhao}, \citenamefont {Wang}, \citenamefont {Toh},
  \citenamefont {Ribeiro}, \citenamefont {Castro~Neto}, \citenamefont
  {Matsuda},\ and\ \citenamefont {Eda}}]{X_in_main_four_2014}%
  \BibitemOpen
  \bibfield  {author} {\bibinfo {author} {\bibfnamefont {D.}~\bibnamefont
  {Kozawa}}, \bibinfo {author} {\bibfnamefont {R.}~\bibnamefont {Kumar}},
  \bibinfo {author} {\bibfnamefont {A.}~\bibnamefont {Carvalho}}, \bibinfo
  {author} {\bibfnamefont {K.}~\bibnamefont {Kumar~Amara}}, \bibinfo {author}
  {\bibfnamefont {W.}~\bibnamefont {Zhao}}, \bibinfo {author} {\bibfnamefont
  {S.}~\bibnamefont {Wang}}, \bibinfo {author} {\bibfnamefont {M.}~\bibnamefont
  {Toh}}, \bibinfo {author} {\bibfnamefont {R.~M.}\ \bibnamefont {Ribeiro}},
  \bibinfo {author} {\bibfnamefont {A.~H.}\ \bibnamefont {Castro~Neto}},
  \bibinfo {author} {\bibfnamefont {K.}~\bibnamefont {Matsuda}}, \ and\
  \bibinfo {author} {\bibfnamefont {G.}~\bibnamefont {Eda}},\ }\href
  {http://dx.doi.org/10.1038/ncomms5543} {\bibfield  {journal} {\bibinfo
  {journal} {Nat. Commun.}\ }\textbf {\bibinfo {volume} {5}},\ \bibinfo {pages}
  {4543} (\bibinfo {year} {2014})}\BibitemShut {NoStop}%
\bibitem [{\citenamefont {Ruppert}\ \emph {et~al.}(2014)\citenamefont
  {Ruppert}, \citenamefont {Aslan},\ and\ \citenamefont
  {Heinz}}]{X_MoTe2_2014}%
  \BibitemOpen
  \bibfield  {author} {\bibinfo {author} {\bibfnamefont {C.}~\bibnamefont
  {Ruppert}}, \bibinfo {author} {\bibfnamefont {O.~B.}\ \bibnamefont {Aslan}},
  \ and\ \bibinfo {author} {\bibfnamefont {T.~F.}\ \bibnamefont {Heinz}},\
  }\href {\doibase 10.1021/nl502557g} {\bibfield  {journal} {\bibinfo
  {journal} {Nano Lett.}\ }\textbf {\bibinfo {volume} {14}},\ \bibinfo {pages}
  {6231} (\bibinfo {year} {2014})}\BibitemShut {NoStop}%
\bibitem [{\citenamefont {Lezama}\ \emph {et~al.}(2015)\citenamefont {Lezama},
  \citenamefont {Arora}, \citenamefont {Ubaldini}, \citenamefont {Barreteau},
  \citenamefont {Giannini}, \citenamefont {Potemski},\ and\ \citenamefont
  {Morpurgo}}]{X_MoTe2_2015}%
  \BibitemOpen
  \bibfield  {author} {\bibinfo {author} {\bibfnamefont {I.~G.}\ \bibnamefont
  {Lezama}}, \bibinfo {author} {\bibfnamefont {A.}~\bibnamefont {Arora}},
  \bibinfo {author} {\bibfnamefont {A.}~\bibnamefont {Ubaldini}}, \bibinfo
  {author} {\bibfnamefont {C.}~\bibnamefont {Barreteau}}, \bibinfo {author}
  {\bibfnamefont {E.}~\bibnamefont {Giannini}}, \bibinfo {author}
  {\bibfnamefont {M.}~\bibnamefont {Potemski}}, \ and\ \bibinfo {author}
  {\bibfnamefont {A.~F.}\ \bibnamefont {Morpurgo}},\ }\href {\doibase
  10.1021/nl5045007} {\bibfield  {journal} {\bibinfo  {journal} {Nano Lett.}\
  }\textbf {\bibinfo {volume} {15}},\ \bibinfo {pages} {2336} (\bibinfo {year}
  {2015})}\BibitemShut {NoStop}%
\bibitem [{\citenamefont {Robert}\ \emph {et~al.}(2016)\citenamefont {Robert},
  \citenamefont {Picard}, \citenamefont {Lagarde}, \citenamefont {Wang},
  \citenamefont {Echeverry}, \citenamefont {Cadiz}, \citenamefont {Renucci},
  \citenamefont {H\"ogele}, \citenamefont {Amand}, \citenamefont {Marie},
  \citenamefont {Gerber},\ and\ \citenamefont {Urbaszek}}]{X_MoTe2_2016}%
  \BibitemOpen
  \bibfield  {author} {\bibinfo {author} {\bibfnamefont {C.}~\bibnamefont
  {Robert}}, \bibinfo {author} {\bibfnamefont {R.}~\bibnamefont {Picard}},
  \bibinfo {author} {\bibfnamefont {D.}~\bibnamefont {Lagarde}}, \bibinfo
  {author} {\bibfnamefont {G.}~\bibnamefont {Wang}}, \bibinfo {author}
  {\bibfnamefont {J.~P.}\ \bibnamefont {Echeverry}}, \bibinfo {author}
  {\bibfnamefont {F.}~\bibnamefont {Cadiz}}, \bibinfo {author} {\bibfnamefont
  {P.}~\bibnamefont {Renucci}}, \bibinfo {author} {\bibfnamefont
  {A.}~\bibnamefont {H\"ogele}}, \bibinfo {author} {\bibfnamefont
  {T.}~\bibnamefont {Amand}}, \bibinfo {author} {\bibfnamefont
  {X.}~\bibnamefont {Marie}}, \bibinfo {author} {\bibfnamefont {I.~C.}\
  \bibnamefont {Gerber}}, \ and\ \bibinfo {author} {\bibfnamefont
  {B.}~\bibnamefont {Urbaszek}},\ }\href {\doibase 10.1103/PhysRevB.94.155425}
  {\bibfield  {journal} {\bibinfo  {journal} {Phys. Rev. B}\ }\textbf {\bibinfo
  {volume} {94}},\ \bibinfo {pages} {155425} (\bibinfo {year}
  {2016})}\BibitemShut {NoStop}%
\bibitem [{\citenamefont {Zhao}\ \emph {et~al.}(2013)\citenamefont {Zhao},
  \citenamefont {Ghorannevis}, \citenamefont {Chu}, \citenamefont {Toh},
  \citenamefont {Kloc}, \citenamefont {Tan},\ and\ \citenamefont
  {Eda}}]{X_WS2_and_WSe2_2013}%
  \BibitemOpen
  \bibfield  {author} {\bibinfo {author} {\bibfnamefont {W.}~\bibnamefont
  {Zhao}}, \bibinfo {author} {\bibfnamefont {Z.}~\bibnamefont {Ghorannevis}},
  \bibinfo {author} {\bibfnamefont {L.}~\bibnamefont {Chu}}, \bibinfo {author}
  {\bibfnamefont {M.}~\bibnamefont {Toh}}, \bibinfo {author} {\bibfnamefont
  {C.}~\bibnamefont {Kloc}}, \bibinfo {author} {\bibfnamefont {P.-H.}\
  \bibnamefont {Tan}}, \ and\ \bibinfo {author} {\bibfnamefont
  {G.}~\bibnamefont {Eda}},\ }\href {\doibase 10.1021/nn305275h} {\bibfield
  {journal} {\bibinfo  {journal} {ACS Nano}\ }\textbf {\bibinfo {volume} {7}},\
  \bibinfo {pages} {791} (\bibinfo {year} {2013})}\BibitemShut {NoStop}%
\bibitem [{\citenamefont {Landau}\ \emph {et~al.}(2013)\citenamefont {Landau},
  \citenamefont {Lifshitz},\ and\ \citenamefont {Pitaevskii}}]{Landau_ED}%
  \BibitemOpen
  \bibfield  {author} {\bibinfo {author} {\bibfnamefont {L.~D.}\ \bibnamefont
  {Landau}}, \bibinfo {author} {\bibfnamefont {E.~M.}\ \bibnamefont
  {Lifshitz}}, \ and\ \bibinfo {author} {\bibfnamefont {L.~P.}\ \bibnamefont
  {Pitaevskii}},\ }\href@noop {} {\emph {\bibinfo {title} {Electrodynamics of
  continuous media}}},\ \bibinfo {edition} {2nd}\ ed.\ (\bibinfo  {publisher}
  {Elsevier},\ \bibinfo {year} {2013})\BibitemShut {NoStop}%
\end{thebibliography}%

\appendix

\section{Photoluminescence of hybridized excitons}\label{app:PL}
To estimate the photoluminescence (PL) intensity of a TMD heterobilayer ${\rm MX_2/M'X_2'}$, we assume a photoexcited thermal population of excitons described by the Bose-Einstein distribution
\begin{equation}
	n_{\rm B}(E,T)=\frac{1}{\exp{(E-E_{\rm gnd})/k_{\rm B}T}-1},
\end{equation}
where $E_{\rm gnd}$ is the global minimum of the exciton moir\'e band structure.

The light-matter interaction Hamiltonian is given by
\begin{equation}
\begin{split}
	&H_{\rm LM} = \frac{e\gamma}{\hbar c}\sum_{s,\eta}\sum_{\xxi,\kk}\sqrt{\frac{4\pi\hbar c}{V\xi}}c_{v\eta s}^\dagger(\kk-\xxi_\parallel)c_{c\eta s}(\kk)a_{\eta}^\dagger(\xxi)\\
	&+\frac{e\gamma'}{\hbar c}\sum_{s',\eta'}\sum_{\xxi,\kk'}\sqrt{\frac{4\pi\hbar c}{V\xi}}c_{v'\eta' s'}^\dagger(\kk'-\xxi_\parallel)c_{c'\eta' s'}(\kk')a_{\eta}^\dagger(\xxi) + {\rm H.c.},
\end{split}
\end{equation}
where the operator $a_{\tau}^\dagger(\xxi)$ creates a photon of momentum $\hbar \xxi$ and circular polarization $\eta$ ($\eta=\pm1$ corresponding to counter-clockwise and clockwise polarization, respectively); $\gamma$ and $\gamma'$ are the momentum matrix elements at the ${\rm MX_2}$- and ${\rm M'X_2'}$-layer $K$ valleys, respectively (Table \ref{tab:parameters}); and $V=SL$, with $S$ the heterostructure surface area, and $L$ the height of the optical cavity. We evaluate the radiative decay (number of photons per unit time) of hXs perturbatively, using Fermi's golden rule in its thermodynamic form
\begin{equation}
	\Gamma_i = \frac{2\pi}{\hbar}\sum_{f}\abs{\braoket{f}{H_{\rm LM}}{i}}^2\,n_{\rm B}(E,T)\delta(E_f-E_i),
\end{equation}
with single-photon final states $\ket{f}\equiv a_{\eta}^\dagger(\xxi)\ket{\Omega}$, and initial states
\begin{equation}\label{eq:hXexpanded}
\begin{split}
	\ket{i} &= \ket{{\rm hX}_s^\tau(\QQ)}_n\\
	 &\equiv \sum_{m=0}^\infty\Big[ A_{nm}^{s\tau}(\QQ)\ket{{\rm X}_{ss}^{\tau\tau}(\QQ)}_m + B_{nm}^{s\tau}(\QQ)\ket{{\rm IX}{}_{ss}^{\tau'\tau}}_m\\
	&+C_{nm}^{s\tau}(\QQ)\ket{{\rm X}'{}_{ss}^{\tau'\tau'}}_m + D_{nm}^{s\tau}(\QQ)\ket{{\rm IX}'{}_{ss}^{\tau\tau'}}_m \Big].
\end{split}
\end{equation}
The indices $m,\,n$ number the minibands, and stand for the double index $(i,j)$ introduced in Eq.\ \eqref{eq:foldings}, and $\QQ \in {\rm mBZ}$. With the definitions of Eq.\ \eqref{eq:XandY}, we obtain the matrix elements
\begin{equation}\label{eq:PLmelem}
\begin{split}
	&\braoket{\eta,\xxi}{H_{\rm R}}{{\rm hX}_s^\tau(\QQ)}_n \\
	&= \frac{e}{\hbar c}\sum_{m=0}^\infty\delta_{\xxi_{\parallel},\QQ+\bb_m}\sqrt{\frac{8\hbar c}{L\sqrt{|\QQ+\bb_m|^2+\xi_\perp^2}}}\\
	&\quad\times \left[\delta_{\eta,\tau}\frac{\gamma A_{nm}^{s\tau}(\QQ)}{ a_X}+\delta_{\eta,\tau'}\frac{\gamma' C_{nm}^{s\tau}(\QQ)}{ a_{X'}}\right].
\end{split}
\end{equation}
Note that for $\tau=\tau'$ (P stacking), Fermi's golden rule will give interference between the last two terms in Eq.\ \eqref{eq:PLmelem}, whereas no interference occurs for $\tau'=-\tau$ (AP stacking). Keeping this in mind, we will focus on the latter case, for the sake of concreteness. Fermi's golden rule gives
\begin{widetext}
\begin{equation}\label{eq:FGRPL}
\begin{split}
	&\Gamma_{n,s}^\tau(\QQ)=\frac{16\pi e^2 }{\hbar^2 c\,L}\sum_{\xi_z}n_{B}(E_{n,s}^\tau(\QQ)) \frac{\delta (\,E_{n,s}^\tau(\QQ) - hc\sqrt{|\QQ+\bb_m|^2+\xi_z^2}\,)}{\sqrt{|\QQ+\bb_m|^2+\xi_z^2}}\left[\left| \sum_{m=0}^\infty \frac{\gamma A_{nm}^{s\tau}(\QQ)}{ a_X} \right|^2 +\left|\sum_{m=0}^\infty \frac{\gamma' C_{nm}^{s\tau}(\QQ)}{ a_{X'}}\right|^2  \right].
\end{split}
\end{equation}
\end{widetext}
Given the steepness of the photon dispersion relation, all terms with $\bb_m \ne \boldsymbol{0}$ are removed by the Dirac delta function in \eqref{eq:FGRPL}. Moreover, only light-cone excitons with center-of-mass momentum $Q \le Q_{\rm LC} \approx E_{n,s}^\tau(0)/hc$ can recombine, according to energy-momentum conservation. Given the smallness of $Q_{\rm LC}$, we set $\QQ = \boldsymbol{0}$ for the exciton dispersions and wave-function coefficients in \eqref{eq:FGRPL}. After taking the continuum limit for $\xi_z$ to evaluate the sum as an integral, we find that
\begin{equation}
\begin{split}
	S^{-1}\Gamma_{n,s}^\tau \approx& \frac{2e^2\,n_{\rm B}(E_{sn}^\tau(0))}{\pi^2 \hbar^2c }\left[\left| \frac{\gamma A_{n0}^{s\tau}(0)}{a_X} \right|^2 + \left| \frac{\gamma' C_{n0}^{s\tau}(0)}{a_X'} \right|^2 \right]\\
	&\times \int_0^\infty \ud \xi_z\,\frac{\delta(\,E_{n,s}^\tau(0) -hc\sqrt{Q^2+\xi_z^2} \,)}{\sqrt{Q^2+\xi_z^2}}.
\end{split}
\end{equation}
The total PL intensity (photons per unit time per unit area) is obtained by evaluating this integral, and further integrating the resulting expression over exciton wave number within the light cone, finally giving
\begin{equation}
\begin{split}
	I_{{\rm PL},n} =& \frac{e^2}{\hbar c}\frac{2E_{n,s}^\tau(0)\,c}{4\pi^3(\hbar c)^3}\left[\left| \frac{\gamma A_{n0}^{s\tau}(0)}{a_X} + \frac{\gamma' C_{n0}^{s\tau}(0)}{a_X'} \right|^2 \right]\,\,\quad(\text{P}),\\
	I_{{\rm PL},n} =& \frac{e^2}{\hbar c}\frac{2E_{n,s}^\tau(0)\,c}{4\pi^3(\hbar c)^3}\left[\left| \frac{\gamma A_{n0}^{s\tau}(0)}{a_X} \right|^2 + \left| \frac{\gamma' C_{n0}^{s\tau}(0)}{a_X'} \right|^2 \right]\,(\text{AP}).
\end{split}
\end{equation}
A significant exciton population will only exist in the few lowest-energy minibands, so we evaluate only $I_{{\rm PL},0}$ and $I_{{\rm PL},1}$. The main PL line appears for photon energies $\hbar\omega =  E_{0,s}^\tau(0)$, and has an activation temperature given by $E_{0,s}^\tau(0)-E_{\rm gnd}\equiv k_{\rm B}T_\star$ [Figs.\ \ref{fig:Xdispersions}(c) and \ref{fig:Xdispersions}(d)]. For Fig.\ \ref{fig:AbsVsTh}, we have introduced a Lorentzian line shape
\begin{equation*}
	L(\hbar \omega) = \frac{\beta/\pi}{[\hbar \omega - E_{n,s}^\tau(0)]^2+\beta^2},
\end{equation*}
with phenomenological broadening $\beta = 5\,{\rm meV}$.

\section{Optical absorption by hybridized excitons}\label{app:Abs}
For the heterostructure's optical absorption spectrum (number of photons absorbed per unit time per unit area), we have used the $T=0$ version of Fermi's golden rule, with relaxed energy conservation (line broadening):
\begin{equation}
	\Gamma_i = \frac{2\pi}{\hbar}\sum_{f}\abs{\braoket{f}{H_{\rm LM}}{i}}^2\,\frac{\beta/\pi}{(E_f-E_i)^2+\beta^2},
\end{equation}
setting $\ket{i}=a_{\eta}^\dagger(\xxi)\ket{\Omega}$, and $\ket{f}=\ket{{\rm hX}_s^\tau(\QQ)}_n$ [see Eq.\ \eqref{eq:hXexpanded}]. After some algebra, we obtain the absorption rate for photons of momentum $\hbar \xxi$ and polarization $\eta$ given by
\begin{widetext}
\begin{equation}
\begin{split}
	\Gamma_{\eta s}(\xxi) \approx&\frac{16 \pi}{\hbar  L \xi}\frac{e^2}{\hbar c}\sum_{n}\abs{\sum_{m}\left[ \frac{\gamma A_{nm}^{s\tau}(0)}{ a_{\rm X}}+\frac{\gamma' C_{nm}^{s\tau}(0)}{ a_{\rm X'}} \right]}^2 \frac{\beta/\pi}{\left[h c \xi - E_{n,s}^\eta(0) \right]^2 + \beta^2}\quad\qquad\quad \text{(P)},\\
	\Gamma_{\eta s}(\xxi) \approx&\frac{16 \pi}{\hbar  L \xi}\frac{e^2}{\hbar c}\sum_{n} \left[\abs{\sum_{m}\frac{\gamma A_{nm}^{s\tau}(0)}{ a_{\rm X}}}^2 + \abs{\sum_m\frac{\gamma' C_{nm}^{s\tau}(0)}{ a_{\rm X'}}}^2\right] \frac{\beta/\pi}{\left[h c \xi - E_{n,s}^\eta(0) \right]^2 + \beta^2}\quad \text{(AP)}.
\end{split}
\end{equation}
\end{widetext}
The total number of absorbed photons is obtained by multiplying this expression by the number of photons states in the infinitesimal energy range $\epsilon$ to $\varepsilon +  d\varepsilon$; that is, the number of photons with wave number of magnitude between $\epsilon/hc$ and $[\varepsilon+d\varepsilon]/hc$. Since the reciprocal volume elements is $4\pi\xi^2\,d\xi$, and each volume element contains $SL/(2\pi)^3$ photon states, this number of photons is $[SL/(2\pi^2\hbar^2c^3)]\varepsilon^2d\varepsilon$. The resulting total absorption rate from exciton band $\ket{{\rm hX}_{s}^\tau}_n$ is given by
\begin{widetext}
\begin{equation}
\begin{split}
	A_{n,s}^\tau(\varepsilon) \approx& \frac{\varepsilon d\varepsilon}{\pi \hbar^2c^2}\frac{8}{\hbar}\frac{e^2}{\hbar c}\sum_n\abs{\sum_m\left[\frac{\gamma A_{nm}^{s\tau}(0)}{a_{\rm X}} + \frac{\gamma' C_{nm}^{s\tau}(0)}{a_{\rm X'}} \right]}^2\frac{\beta/\pi}{\left[h c \xi - E_{n,s}^\eta(0) \right]^2 + \beta^2}\,\quad\qquad\quad\text{(P)},\\
	A_{n,s}^\tau(\varepsilon) \approx& \frac{\varepsilon d\varepsilon}{\pi \hbar^2c^2}\frac{8}{\hbar}\frac{e^2}{\hbar c}\sum_n\left[\abs{\sum_{m}\frac{\gamma A_{nm}^{s\tau}(0)}{ a_{\rm X}}}^2 + \abs{\sum_m\frac{\gamma' C_{nm}^{s\tau}(0)}{ a_{\rm X'}}}^2\right]\frac{\beta/\pi}{\left[h c \xi - E_{n,s}^\eta(0) \right]^2 + \beta^2}\quad\text{(AP)}.
\end{split}
\end{equation}
\end{widetext}
In an experimental setup, the energy differential $d\varepsilon$ can be identified with the detector's resolution, which we set to $1\,{\rm meV}$, together with the phenomenological line broadening $\beta = 5\,{\rm meV}$, to produce the spectra of Figs.\ \ref{fig:Summary}, \ref{fig:AbsVsTh}--\ref{fig:hXAbsEz_MoSe2_WS2}.

\section{Dependence of the MoTe${}_2$/MoSe${}_2$ and MoSe${}_2$/WS${}_2$ optical spectra on the model parameters}\label{app:parametric}
\begin{figure}[t!]
\begin{center}
\includegraphics[width=0.7\columnwidth]{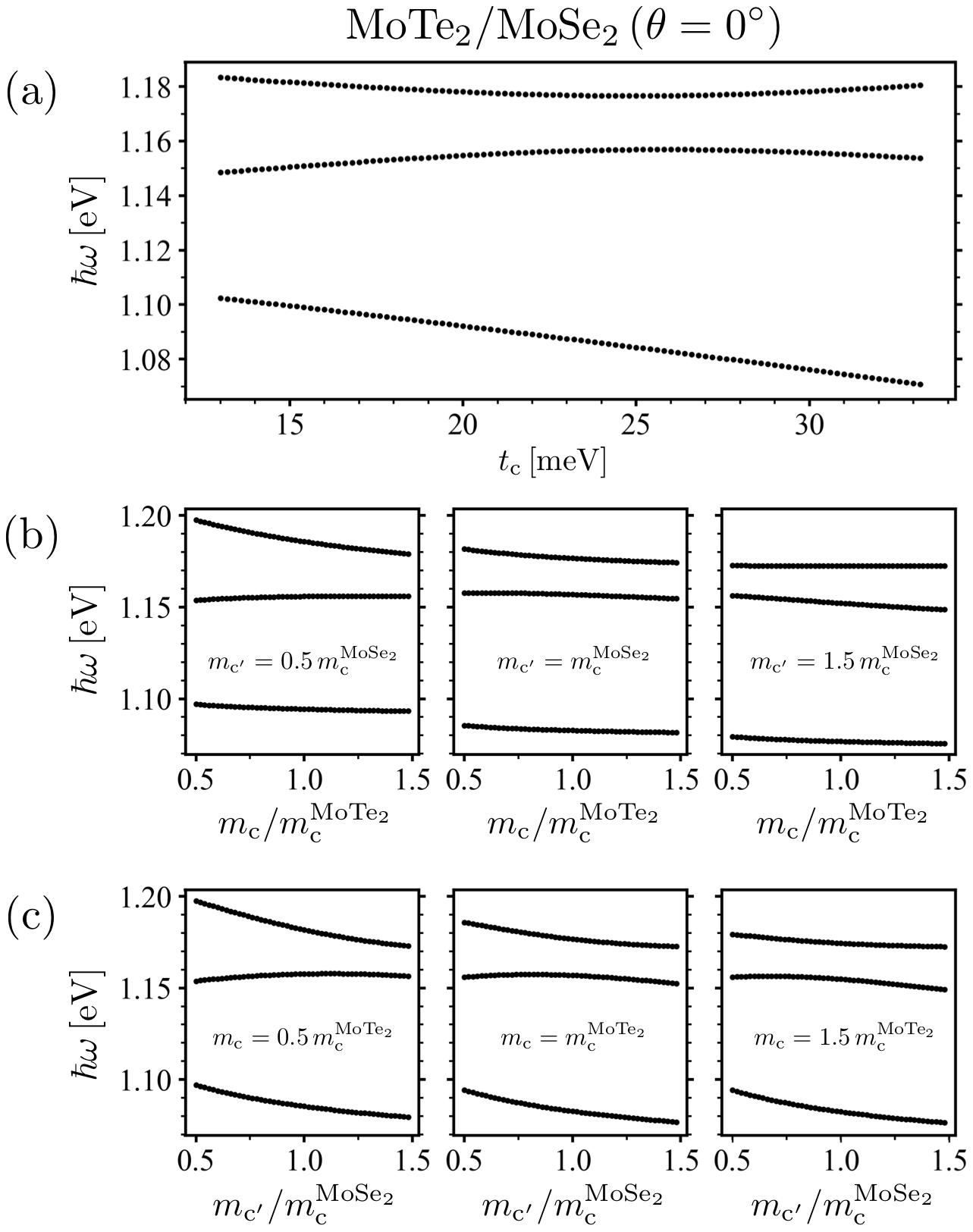}
\caption{Dependence of the low-energy three-peak structure in the absorption spectrum of perfectly aligned ($\theta=0^\circ$) MoTe${}_2$/MoSe${}_2$ on (a) the electron interlayer hopping energy $t_{\rm c}$, and (b) the MoTe${}_2$ and (c) MoSe${}_2$ conduction-band masses. Reference mass values $m_{\rm c}^{\rm MoTe_2}$ and $m_{\rm c}^{\rm MoSe_2}$ are given in Table \ref{tab:parameters}.}
\label{fig:hXvsMassMoTe2}
\end{center}
\end{figure}

\begin{figure}[t!]
\begin{center}
\includegraphics[width=0.7\columnwidth]{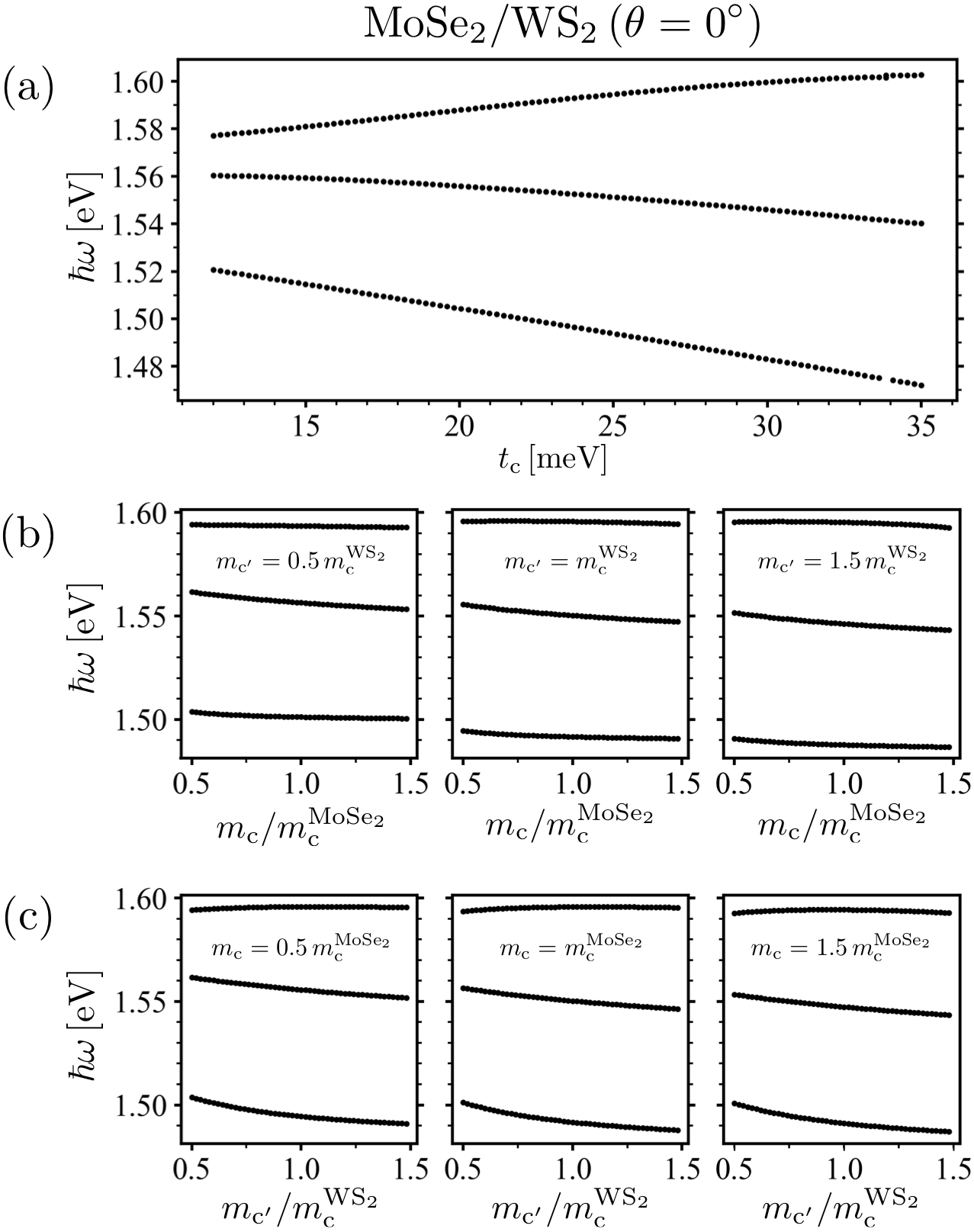}
\caption{Dependence of the low-energy three-peak structure in the absorption spectrum of perfectly aligned ($\theta=0^\circ$) MoSe${}_2$/WS${}_2$ on (a) the electron interlayer hopping energy $t_{\rm c}$, and (b) the MoSe${}_2$ and (c) WS${}_2$ conduction-band masses. Reference mass values $m_{\rm c}^{\rm MoSe_2}$ and $m_{\rm c}^{\rm WS_2}$ are given in Table \ref{tab:parameters}.}
\label{fig:hXvsMassMoSe2}
\end{center}
\end{figure}

We evaluated the low-energy absorption spectra of perfectly aligned ($\theta = 0^\circ$) MoTe${}_2$/MoSe${}_2$ and MoSe${}_2$/WS${}_2$, for different values of the relevant parameters in our theoretical model: the interlayer electron tunneling energy $t_{\rm c}$ and the conduction- and valence-band masses $m_{\rm c}$ and $m_{\rm v}$. The latter two parameters affect the intra- and interlayer exciton masses and Bohr radii, such that the intralayer-interlayer exciton mixing energies of Eq.\ \eqref{eq:Xmelem} are modified by all three parameters. Therefore, for each combination of $m_{\rm c}$ and $m_{\rm v}$, we evaluated the relevant exciton Bohr radii using the finite elements method discussed in Sec.\ \ref{sec:optical}. Figs.\ \ref{fig:hXvsMassMoTe2} and \ref{fig:hXvsMassMoSe2} show the variation of the three main hX absorption peaks discussed in the main text with varying $t_{\rm c}$, $m_{\rm c}$ and $m_{\rm v}$ within $50\%$ of their reference values. The weak dependence found for both material pairs indicates that the three-peak structure should appear for samples of different qualities, and prepared by different methods, where all three parameters may vary.


\end{document}